\documentclass[12pt,Bold,Draft,CPage]{ufthesis}
\usepackage{epsfig}
\usepackage{lscape}
\usepackage{multirow}
\usepackage{amsmath}
\SetFullName{Luis Breva-Newell}%
\SetThesisType{Dissertation}%
\SetDegreeType{Doctor of Philosophy}%
\SetGradMonth{December}%
\SetGradYear{2004}%
\SetDepartment{Physics}%
\SetChair{John Yelton}%
\SetTitle{DECAYS OF THE $\Upsilon$(1S) INTO A PHOTON AND TWO CHARGED HADRONS}
\newcommand{\Uis}{$\Upsilon$(1S)\ }
\newcommand{\gev}{\ \rm GeV}
\newcommand{\mev}{\ \rm MeV}
\newcommand{\ev}{\ \rm eV}

\newcommand{\mom}{\ \rm GeV/c}
\newcommand{\mass}{\ \rm GeV/c^2}
\newcommand{\miss}{\ \rm MeV/c^2}

\begin{document}
\frontmatter %
\pagenumbering{roman}
\maketitle %
\dedication{To my Parents}
\acknowledge{I owe my gratitude to many people without whom this work
would not have been possible. First of all, I would like to thank my
supervisor, Dr. Yelton. His office door was always open for me, and
he spent long hours of his time listening to my ideas even though only
a few of them ever worked. He gave me freedom to explore and branch off
in different directions while steadily guiding me forward at the same
time. After these years of working together I consider him more than a
mentor, and a think of him as a close friend.

Many people from CLEO have put their own time and work into this
analysis. I am grateful to all my committee members Tomas
Ferguson, John Cummings, Thomas Coan, and especially Rich Galik for all
their help and efforts to improve this work. 

During my stay at the University of Florida I have met many people
inside and outside the physics world. I would like to thank my first
year graduate teachers 
Dr. Sikivie and Dr. Woodard who were a source of inspiration for
me. My thanks also go to my fellow graduate students Vijay Potlia,
Rukshana Patel, Necula Valentin, Jennifer Sippel, Suzette Atienza,
G. Suhas, and many more for all our fun physics discussions and late
homework sessions. I would also like to thank the good friends I made
outside the physics building, James Power, Yaseen Afzal (Paki), Ramji
Kamakoti (Ramjizzle), Dan DeKee (Double Down), and Fernando Zamit
(Fernizzle) who always reminded me that there is more to life than
Physics. 

The most important person I have met during my graduate career is
my wife, Jennifer. I 
thank her for enduring all those endless days when I would answer her
questions with only grunts and nods while my attention remained fixed on
the computer screen. Her love and support are invaluable to me.

Finally, I would like to thank my parents Manuel and Charlene, my
sister Teresa, and my brother Gaizka. They have always been there for me.}
  
\tableofcontents
\begin{abstract}
Using the CLEO III detector we report on a new study of exclusive
radiative $\Upsilon(1S)$ decays into the final states
$\gamma\pi^+\pi^-$, $\gamma K^+ K^-$ and $\gamma p \bar{p} $. 
We present  branching ratios for the decay modes
$\Upsilon(1S) \rightarrow \gamma f_2(1270)$, $\Upsilon(1S) \rightarrow \gamma f_2'(1525)$, 
$\Upsilon(1S) \rightarrow \gamma f_0(1710)$ with $f_0(1710) \rightarrow K^+ K^-$
and $\Upsilon(1S) \rightarrow \gamma f_4(2050)$.
\end{abstract}
\mainmatter
\addcontentsline{toc}{extraentry}{CHAPTER}%
\pagenumbering{arabic}
\chapter{THEORY}
\section{Particle Physics}

Particle physics is the branch of physics dedicated to the study of
matter and energy at the most fundamental level. This means that the
job of a particle physicist is to identify the smallest constituents
of matter and describe how they interact with each other.

Humankind has been interested in this subject since ancient times. Two of
the first particle physicists in recorded history are the
Greek thinkers Empedocles and Democritus from the fifth century
BC. Empedocles stated that our complex world was made from combining
four fundamental elements (earth, air, fire and water) in different
proportions. Democritus on the  
other hand, believed that the apparently continuous objects in the
natural world were not really continuous, but made from voids and
indivisible particles called atoms. 

Over the last half century Particle physics has advanced
tremendously and we now have a beautiful, but incomplete, theory
firmly grounded on experiment that describes the fundamental
constituents of matter and how they interact with each other. This
theory is called ``The Standard Model.''

\subsection{The Standard Model}

According to the Standard Model, the fundamental building blocks of
matter are point like particles which interact with each other in
as many as three different ways. Each type of interaction, or force, 
is itself carried by point like particles called force carriers. The
particles which are force carriers are bosons~\footnote{Bosons are
defined as particles with integer spin in quantum mechanics.}
and are collectively called gauge bosons because they are needed for the
theory to be gauge invariant. The non-force carrying particles are
fermions.~\footnote{Fermions are defined  as particles with half odd
fractional spin.}

The three interactions described by the Standard Model are called the
electromagnetic force, the weak force, and the strong
force. The force of gravity is not included in the Standard
Model, and this is one of the reasons the Standard Model is not yet
complete. The 
electromagnetic force carriers are photons, the weak force carriers
are the $Z^0$, $W^{+}$, and $W^{-}$ particles, and finally, the strong force is
carried by eight kinds of gluons. Table~\ref{table:bosons} summarizes
the situation. 

 \begin{table}[ht]
 \begin{center}
 \caption{Gauge bosons and the force they carry.}
 \begin{tabular}{c | c |  c }
 \hline
 \hline
 Symbol & Name & Force Carried \\
 \hline 
 \hline
 $\gamma$ & Photon & Electromagnetic \\
 $Z^0$ & $Z^0$ & Weak \\
 $W^{+}$ & $W^{+}$ & Weak \\
 $W^{-}$ & $W^{-}$ & Weak \\
 $g$ & gluon & Strong \\
 \hline
 \hline
 \end{tabular}
 \label{table:bosons} 
 \end{center}
 \end{table}

Particles which interact through a particular force are said to
couple to it and to carry an associated charge. The
nomenclature is as follows, particles that interact through the
electromagnetic force have an electromagnetic charge called
electric charge, those which interact through the weak force have a
weak charge called weak isospin, and those which interact
strongly have a strong charge called color.

The rest of the particles in the Standard Model which are not force
carriers, the fermions, are subdivided into different groups depending on their
properties (See
table~\ref{table::fermions}) , reminiscent of the way chemists
organized the elements into the 
periodic table during the second half of the nineteenth century. Fermions are
divided into quarks (generically represented by the symbol q) and 
leptons. The main difference between these two groups is that quarks
interact through the strong force while leptons do not. There are six
types of leptons and six types of quarks (also called the six quark
flavors) which are grouped into three generations. Each generation
consist of two quarks and two leptons. All three generations replicate
the same set of force charges, the main difference between
generations is the mass of the particles (for example, the ratio of
the masses of $e:\mu:\tau$ is $1:200:3500$). For each fermion there is
an anti-fermion with equal mass and spin and opposite charge.

 \begin{table}[ht]
 \caption{Fermion symbols classified into quarks, leptons, and the
 three generations along with the generational common charges.}
 \begin{center}
 \begin{tabular}{c || c | c | c || c c c}
 \hline
 \hline
  & First & Second & Third & Electric & Weak & Has \\
  & Generation & Generation & Generation & Charge & Charge & Color \\
 \hline
 \hline
\multirow{2}{15mm}{Quarks} & $u$ & $c$ & $t$ & $+2/3$ & $+1/2$ & Yes \\
& $d$ & $s$ & $b$ & $-1/3$ & $-1/2$ & Yes \\
\hline
\multirow{2}{15mm}{Leptons} & $\nu_{e}$ & $\nu_{\mu}$ & $\nu_{\tau}$ & 0 &
 $+1/2$ & No \\
& $e$ & $\mu$ & $\tau$ & -1 & $-1/2$ & No \\ 
 \hline
 \hline
 \end{tabular}
 \label{table::fermions} 
 \end{center}
 \end{table}

\subsection{Quantum Chromodynamics}

Quantum chromodynamics (QCD) is the part of the Standard Model that
describes the strong force. QCD is based on local gauge invariance
and color symmetry. There are three possible color
charges for quarks called $r$ (red), $b$ (blue), and $g$
(green). Anti-quarks have opposite colors called $\bar{r}$, $\bar{b}$,
$\bar{g}$. The strong force between quarks only depends on their
colors and is independent of their flavor.

A very important characteristic of the strong force is
that gluons themselves carry a color charge and an anti-color charge
so they can interact with other quarks through the strong force and
change their color. Since there are three colors and three anti-colors
one might think that there are nine different gluons. However, there
is one linear combination of color anti-color states that has no net
color and leaves a quark unchanged. There are therefore $9-1 = 8$
gluons. The 8 individual gluon color states can be written as follows,
\begin{equation}
\begin{split}
|1> &= \frac{1}{\sqrt{2}}(r\bar{b}+b\bar{r})\\
|2> &= \frac{-i}{\sqrt{2}}(r\bar{b}-b\bar{r})\\
|3> &= \frac{1}{\sqrt{2}}(r\bar{r}-b\bar{b})\\
|4> &= \frac{1}{\sqrt{2}}(r\bar{g}+g\bar{r})\\
|5> &= \frac{-i}{\sqrt{2}}(r\bar{g}-g\bar{r})\\
|6> &= \frac{1}{\sqrt{2}}(b\bar{g}+g\bar{b})\\
|7> &= \frac{-i}{\sqrt{2}}(b\bar{g}-g\bar{b})\\
|8> &= \frac{1}{\sqrt{6}}(r\bar{r}+g\bar{g}-2b\bar{b}) \\
\end{split}
\label{equation:gluons}
\end{equation} 
and represent the eight different gluons that exist in nature. The single
and unique color state left out is called a color singlet, 
\begin{equation}
|9> = \frac{1}{\sqrt{3}}(r\bar{r}+b\bar{b}+g\bar{g}) 
\label{equation:colorsinglet}
\end{equation}
which is invariant under a redefinition of the color (a rotation in
color space). In group theory this decomposition of the color states
into an octet and a singlet is denoted by $3\otimes\bar{3} =
8\oplus1$.  It is worth noting here that a colorless sate, such as $|3>$
or $|8>$, is not
necessarily a color singlet state. 

This situation is analogous to the perhaps more familiar example of
two spin 1/2 particles. Each particle can have their spin up
($\uparrow$) or down ($\downarrow$) along the $z$ axis corresponding
to four possible 
combinations reprsented by each giving a total spin $S$ = 0
or 1 represented by $|S\ S_z>$. The S = 1 states form a triplet, 
\begin{equation}
\begin{split}
|1 \ +1> &= |\uparrow\uparrow> \\
|1 \ 0> &= \frac{1}{\sqrt{2}}(|\uparrow\downarrow> + |\downarrow\uparrow>) \\
|1 \ -1> &= |\downarrow\downarrow> \\
\end{split}
\label{equation:spin1}
\end{equation}
and there is a singlet state with S = 0,
\begin{equation}
|0 \ 0> = \frac{1}{\sqrt{2}}(|\uparrow\downarrow> - |\downarrow\uparrow>).
\label{equation:spin2}
\end{equation}

Since gluons themselves carry color they can interact with each other
through the strong force. This interaction among the force carriers is
unique to the strong force and, when included in perturbative QCD
calculations, leads to two important properties observed in nature called
``asymptotic freedom'' and ``color confinement.''  Asymptotic freedom
means that the interaction gets weaker at short distances. Color
confinement is the requirement that observed states have neutral
color, or in other words,  they must be in a color singlet state.

Confinement explains why free quarks or free gluons, which have a net
color charge, have never been observed. It also explains why no
fractional charged particles made from a $qq$ bound state have never
been observed since 
it is not possible to construct a color singlet for such a state (in
group theory terms $3\otimes3 = 6\oplus3$ where we have a sextet and a
triplet, but no singlet). On the other hand, color singlets can be
constructed for a $q\bar{q}$ or $qqq$ system. The color singlet for
$q\bar{q}$ is simply the state shown
on the right side of Equation~\ref{equation:colorsinglet}. The color singlet
for $qqq$ can be obtained from the decomposition $3\otimes3\otimes3 =
10\oplus8\oplus8\oplus1$ and is, 
\begin{equation}
|qqq>_{color\ singlet} = \frac{1}{\sqrt{6}}(rgb-grb+brg-bgr+gbr-rbg).\label{equation:singlet2}
\end{equation}

Particles that are bound states of $q\bar{q}$ are
abundant in nature and are called mesons, those that are bound states
of $qqq$ are called baryons and also abound in nature. Both groups are
collectively called hadrons. There have been
hundreds of different hadrons observed, confirming the validity of QCD
and the quark flavors. Nevertheless, QCD leaves
room for more possibilities. A bound state of two gluons $gg$ (sometimes
called a glueball) can be in a color singlet and in principle could be
observed. 

Although glueballs are allowed by QCD
there is no convincing experimental observation of one. Another
possibility are bound
states that are a mixture of the previous states, such as
$\alpha q\bar{q} + \beta gg$ with arbitrary $\alpha,\ \beta$. These
states are called ``hybrid mesons'' or simply ``hybrids''. It is
believed~\cite{Filippi:2003hb},~\cite{cleo7},~\cite{epj} that hybrids are
necessary to explain the spectrum of light mesons between 1-2 \gev.  

As with any quantum mechanical theory, the spectrum of bound states is a
fundamental test.~\footnote{A good example of this is the successful
description of the hydrogen atom's spectrum by quantum
mechanics.} Glueballs are allowed by QCD, yet there is no conclusive
experimental evidence of their observation, despite intense
experimental
searches~\cite{experiment1},~\cite{experiment2},~\cite{experiment3}
complemented by lattice QCD 
calculations~\cite{theory1},~\cite{theory2} and other theoretical
contributions like bag models~\cite{theory3}, flux-tube
models~\cite{theory4}, QCD sum rules~\cite{theory5}, weakly bound
bound-state models~\cite{maurizio}, and QCD factorization formalism
models~\cite{theory6}.  
Physicists cannot be sure they understand QCD until such states are
observed, or until they can explain why we cannot observe them.

\subsection{Introduction to the Radiative Decays of the \Uis}  

The \Uis is a meson composed of $b\bar{b}$ quarks. Mesons of
this kind, composed of a quark and an anti-quark of the same flavor,
are in general called quarkonia.
 
There is a convention behind the name of the \Uis. The
$\Upsilon$ symbol is reserved for particles composed of $b\bar{b}$
where the combined spin of the quark and anti-quark is 1. The
``1S'' symbol is borrowed from atomic spectroscopy with the ``1''
meaning that the $b\bar{b}$ pair are in the lowest-energy bound-state,
and the ``S'' meaning that the $b\bar{b}$ have a relative angular
momentum $L = 0$. A description of particle naming conventions
can be found in~\cite{textbook}.

The \Uis is unstable, existing for only about $10^{-24}$ seconds after
which it decays into daughter particles (which in turn decay
themselves if they are not stable). The term ``radiative decay'' is
reserved to any \Uis decay where one of the stable daughters is a
photon. 

The different ways the \Uis can decay must obey the
symmetries in nature. Examples of such symmetries are ``parity'' and ``charge
conjugation'', both of which will be described soon. Symmetries are very
important in the standard model. The usefulness of symmetries can be
seen in Noethers' theorem, 
which states that for every symmetry there is an associated conserved
quantity. Examples of Noethers'
theorem are the conservation of momentum connected to translational
invariance and the conservation of angular momentum associated with rotational
invariance. 

In the next sections we will use the parity and charge
conjugation symmetries to find \Uis decays which
conserve the associated symmetry constants and are allowed by nature. In
particular, we will show that the radiative decay of the \Uis through
a photon and two gluons (see Figure~\ref{Figure:Udecay.eps}) is
allowed. The key observation is that the two gluons must be in a color
singlet since both the \Uis and the radiated photon have no color and
color must be conserved. This means that the two gluons satisfy color
confinement and could form a
glueball, although more conventional meson states, or hadrons in no bound state
at all, are also 
possible outcomes. Regardless of what the gluons do, their energy
will eventually manifests itself as hadrons. Sometimes, two charged
hadrons of opposite charge will emerge. This work examines those two 
hadrons from a radiative \Uis decay to experimentally probe the $gg$ spectrum. 

\subsubsection{Parity}

The parity operator, $\hat{P}$, reverses the sign of an
object's spatial coordinates. Consider a particle $|a>$ with a wave function
$\Psi_a(\vec{x},t)$. By the definition of the parity operator,
\begin{equation}
\hat{P}\Psi(\vec{x},t) = P_a\Psi_a(-\vec{x},t)
\end{equation}
where $P_a$ is a constant phase factor. If we consider an
eigenfunction of momentum
\begin{equation}
\Psi_{\vec{p}}(\vec{x},t) = e^{i( \vec{p} \cdot \vec{x}-Et)}
\end{equation}
then
\begin{equation}
\hat{P}\Psi_{\vec{p}}(\vec{x},t) = P_a\Psi_{\vec{p}}(-\vec{x},t) =
P_a\Psi_{-\vec{p}}(\vec{x},t),
\end{equation}
so that any particle at rest, with $\vec{p}=0$, remains unchanged up
to a multiplicative number, $P_a$, under the parity operator. States with this
property are called eigenstates with eigenvalue $P_a$. $P_a$ is also
called the intrinsic parity of particle $a$, or more usually just the
parity of particle $a$, with the words at rest left implicit. Since
two successive parity transformations leave the system unchanged,
$P_a^2 = 1$, implying that the possible values for the parity
eigenvalue are $P_a=\pm1$.

In addition to a particle at rest, a particle with definite orbital
angular momentum is also en eigenstate of parity. The wave function
for such a particle in spherical coordinates is,
\begin{equation}
\Psi_{nlm}(\vec{x},t) = R_{nl}(r)Y_l^m(\theta,\phi),
\end{equation}

where $(r,\theta, \phi)$ are spherical polar coordinates, $R_{nl}(r)$
is a function of the radial variable $r$ only, and the $Y_l^m(\theta,
\phi)$ is a spherical harmonic.

The spherical harmonics are well known functions which have the
following property,
\begin{equation}
Y_l^m(\theta,\phi) = (-1)^lY_l^m(\pi-\theta,\pi+\phi).
\end{equation}
Hence 
\begin{equation}
\hat{P}\Psi_{nlm}(\vec{x},t) = P_a\Psi_{nlm}(-\vec{x},t) =
P_a(-1)^l\Psi_{nlm}(\vec{x},t)  
\end{equation}
proving that a particle with a definite orbital angular momentum $l$
is indeed an eigenstate of the parity operator with eigenvalue $P_a(-1)^l$.

The parities of the fundamental fermions cannot be measured or
derived. All that nature requires is that the parity of a fermion be
opposite to that of an anti-fermion. As a matter of convention fermions are
assigned $P = +1$ and anti-fermions are assigned $P = -1$. In contrast,
the parities of the photon and gluon can be derived by applying
$\hat{P}$ to the field equations resulting in $P_{\gamma} = -1$ and
$P_{g} = -1$.

The \Uis has $P = P_{b}P_{\bar{b}}(-1)^{L} = -1$ since $L = 0$.  

Parity is a good quantum number because it is a symmetry of the strong
and electromagnetic force. This means, that in any reaction involving
these forces, parity must be conserved.

\subsubsection{Charge Conjugation}

Charge conjugation is simply the  
operation which replaces all particles by their anti-particles. In
quantum mechanics the charge conjugation operator is represented by
$\hat{C}$. For any particle $|a>$ we can write,
\begin{equation}
\hat{C}|a> = c_a|\bar{a}>
\end{equation}
where $c_a$ is a phase factor. If we let the $\hat{C}$ operator act twice to
recover the original state $|a>$,
\begin{equation}
|a> = \hat{C}^2|a> = \hat{C}(c_a|\bar{a}>) = c_a\hat{C}|\bar{a}> = c_ac_{\bar{a}}|a>
\end{equation}
which shows that $c_ac_{\bar{a}} = 1$. If (and only
if) $a$ is its own anti-particle, it is an eigenstate of $\hat{C}$.
The possible eigenvalues are limited to $C = c_a = c_{\bar{a}} = \pm1$. 

All systems composed of a the same fermion and an anti-fermion pair
are eigenstates of $\hat{C}$ with eigenvalue $C = (-1)^{(L+S)}$. This
factor can be understood because of the need to exchange both
particles' position and 
spin to recover the original state after the charge conjugation
operator is applied. Exchanging the particles' position gives a factor of
$(-1)^L$ as was shown in the previous section, exchanging the particles
spin gives a factor of $(-1)^{S+1}$ as can be verified by 
inspecting Equations~\ref{equation:spin1} and~\ref{equation:spin2}, and a factor
of (-1) which arises in quantum field theory whenever fermions and
anti-fermions are interchanged. With this result we can calculate the
charge conjugation eigenvalue for the \Uis and obtain $C = -1$ since
$L+S = 1$.  

The photon is an eigenstate of $\hat{C}$ since
it is its own anti-particle. The $C$ eigenvalue for the photon can be
derived by inserting $\hat{C}$ into the field equations and is
$C_{\gamma} = -1$.

Finally, we consider a system composed of two gluons that are in a
color singlet. The two gluons are bosons and they must have a 
symmetric wave function, $\Psi_G$ under a $g_1 \leftrightarrow g_2$
exchange. Under this exchange, the orbital angular momentum part of the wave
function contributes with a factor $(-1)^L$, the spin part of the wave
function for two spin 1
particles contributes with a factor of $(-1)^{S}$, and the
color singlet part of the wave function 
contributes with a factor of $+1$ since it is symmetric. To ensure
that $\Psi_G$ is symmetric we need $L+S$ to be even. This implies that
$C = (-1)^{L+S} = +1$. The $L = 0$ and $L = 1$ possible
$gg$ bound states are shown in Table~\ref{table:possiblegluons}. The
$J^{PC} = 1^{-+}$ is peculiar because it is impossible for a $q\bar{q}$
system to have these quantum numbers. If this state is ever observed,
it must be a glueball. Experimental searches for such a state have
been done, for example, in~\cite{exotic}. 

 \begin{table}[ht]
 \begin{center}
 \caption{Possible $gg$ bound states with $L = 0$ or $L = 1$. The
 possible quantum numbers are 
 limited by the condition that $J = L + S $ be even, which is needed
 to ensure a symmetric wave function for the two gauge bosons.}
 \begin{tabular}{c | c |  c}
 \hline
 \hline
 L & S & $J^{PC}$ \\
 \hline 
 \hline
 0 & 0 & $0^{++}$ \\
 0 & 2 & $2^{++}$ \\
 1 & 1 & $0^{-+},1^{-+},2^{-+}$ \\
 \hline
 \hline
 \end{tabular}
 \label{table:possiblegluons} 
 \end{center}
 \end{table}

Charge conjugation is a symmetry of the strong and electromagnetic force.
For those particles that are eigenstates of $\hat{C}$, $C$ is a good
quantum number because in any reaction involving these
forces $C$ must be conserved. 
   
The $P$ and $C$ values of various particles used in this analysis are
shown in Table~\ref{table:pcvalues} along with their quark
composition, orbital angular momentum, and internal spin.

 \begin{table}[ht]
 \begin{center}
 \caption{Symbol, name, quark composition, angular momentum (L),
 internal spin (S), parity (P), and charge conjugation eigenvalues (C)
 for a
 few of the particles used in this analysis.}
 \begin{tabular}{c | c |  c | c | c | c | c}
 \hline
 \hline
 Symbol & Name & Quark Composition & L & S & P & C \\
 \hline 
 \hline
 \Uis & Upsilon(1S) & $b\bar{b}$ & 0 & 1 & -1 & -1 \\
 $\pi^+$ & Pion & $u\bar{d}$ & 0 & 0 & -1 & x \\
 $\pi^-$ & Pion & $d\bar{u}$ & 0 & 0 & -1 & x \\
 $K^+$ & Kaon & $u\bar{s}$ & 0 & 0 & -1 & x \\
 $K^-$ & Kaon & $s\bar{u}$ & 0 & 0 & -1 & x \\
 $p$ & Proton & $uud$ & 0 & 1/2 & +1 & x \\
 $\bar{p}$ & Anti-proton & $\bar{u}\bar{u}\bar{d}$ & 0 & 1/2 & -1 & x \\
 $\gamma$ & Photon & x & x & 1 & -1 & -1 \\
 \hline
 \hline
 \end{tabular}
 \label{table:pcvalues} 
 \end{center}
 \end{table}

\subsubsection{Possible Decays of the \Uis}

At this point we can understand the different possible ways the \Uis
can decay. The Possible decays are limited because the strong and
electromagnetic force must conserve color, $C$ and $P$. 

The simplest possibility is for the $b\bar{b}$ pair to interact
electro-magnetically and annihilate into one virtual
photon~\footnote{Such a photon is called virtual because it cannot
conserve the 4-momentum of $b\bar{b}$ and is unstable, only existing
for a brief period of time, as allowed by the uncertainty principle, after
which it decays.}. This is allowed by parity, charge conjugation
symmetry and color conservation. The decay of the \Uis to one gluon is
not allowed by color conservation and is therefore forbidden. \Uis
decays to two photons are forbidden by charge conjugation. \Uis decays to two
gluons are also forbidden by charge conjugation. \Uis decays
to 3 gluons are allowed (3 gluons can form a  
color singlet). \Uis decays to 3 photons are also allowed, but are
largely suppressed by decays to one photon since 3 successive
electromagnetic interactions are much less likely to occur than a
single one. Finally, \Uis decays two one photon and two gluons are
also allowed under the condition that the two gluons be in a color
singlet state. 

\begin{figure}[ht]
\begin{center}
\epsfig{bbllx=66,bblly=275,bburx=580,bbury=625,width=4.5in,clip=,file=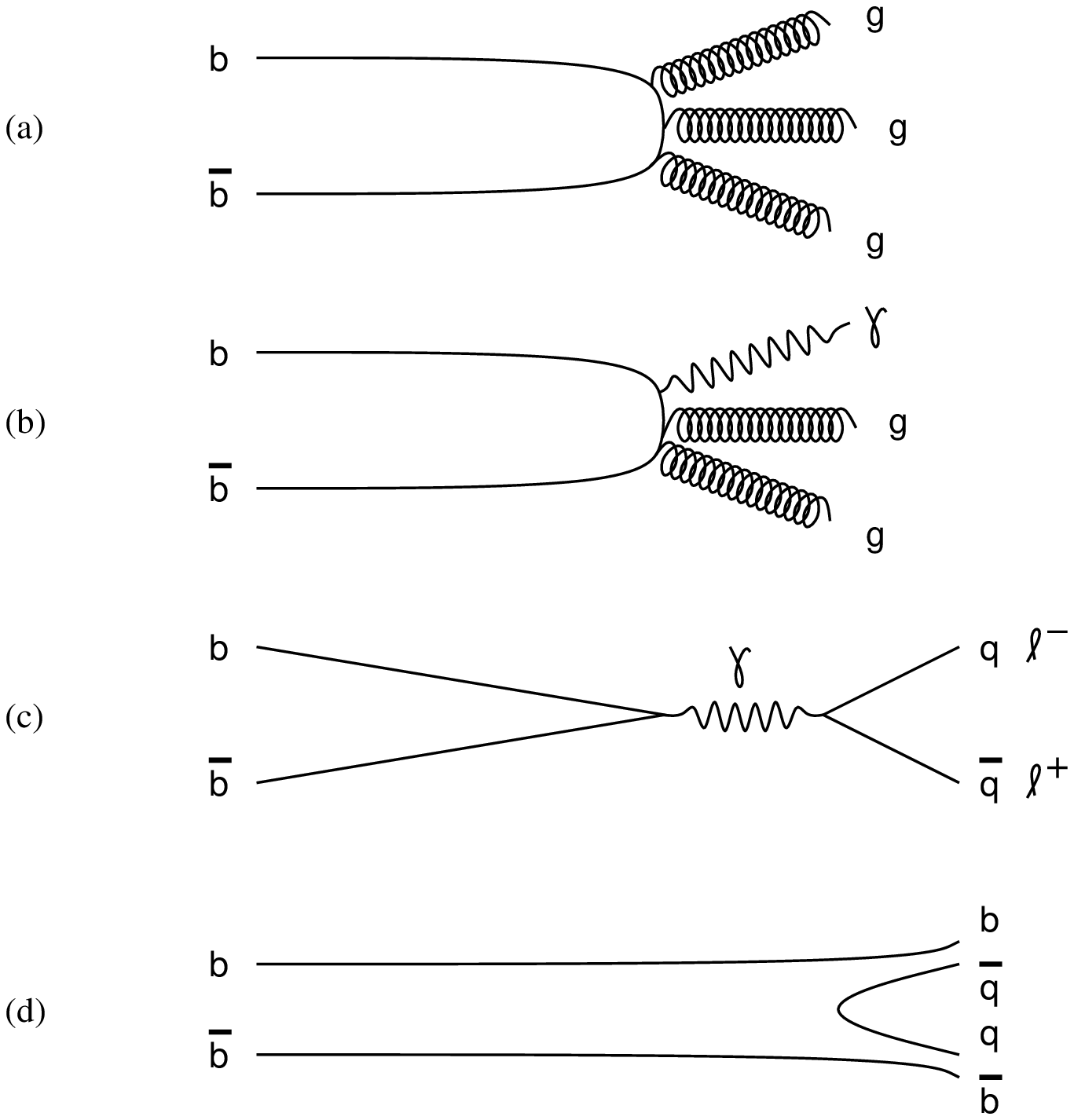} 
\end{center}
\caption{Lowest order decays of the \Uis allowed by color conservation, charge
conjugation symmetry, and parity. (a) Shows the decay into three
gluons, (b) shows a radiative decay, and (c) shows the electromagnetic
decay through a virtual photon that in turn decays electromagnetically
into a pair of charged fundamental particles, such as quarks or charged leptons
(the charged leptons are represented by the symbol
$l$).}\label{Figure:Udecay.eps} 
\end{figure}

The three different possible \Uis decays with least amount of interactions
(also called lowest order decays) are shown in Figure~\ref{Figure:Udecay.eps}.

\subsubsection{Observable Resonances}

In this work we search for a resonance $X$ produced in a  radiative
\Uis decay, 
\begin{equation}
\Upsilon(1S) \to \gamma X.
\end{equation}
Using charge conjugation symmetry on both sides, 
\begin{equation}
-1 = -1C_X. 
\end{equation}
Therefore, $C_X = +1$.

In order for us to observe $X$ it must decay into two charged hadrons,
\begin{equation}
X \to h^+ h^-,\label{equation:xhh}
\end{equation}
where $h = \pi, K, p$ are the hadrons whose momentum we are going to
measure. Applying charge conjugation to this last decay, 
\begin{equation}
+1 = (-1)^{L+S}, 
\end{equation}
where $L$ and $S$ are respectively the angular momentum and spin of
the $h^+h^-$ system. This last equation implies that $L+S$ must be
even.
This has consequences for the possible $X$ parities we can
observe. By parity conservation in~\ref{equation:xhh}, 
\begin{equation}
P_X = (-1)^{L}(-1)(-1) = (-1)^L.
\end{equation}
For $h = \pi,\ K$, $S = 0$ and $L=J$ must be even, which implies that $P_X
= +1$. For $h = p$, $S$ can be 0 or 1 and particles with both
positive and negative parities can be detected.

 \begin{table}[ht]
 \begin{center}
 \caption{Possible $S$, $L$, $J$, $P$ and $C$ values for $X$ from the
 radiative decay 
 $\Upsilon(1S) \to \gamma X$ 
 reconstructed in different decay modes.}
 \begin{tabular}{c | c | c | c | c | c }
 \hline
 \hline
 Decay Mode & $S$ & $L$ & $J$ & $P$ & $C$ \\
 \hline 
 \hline
 $X \to \pi^+ \pi^-$ & 0 & even & even & +1 & +1 \\
 $X \to K^+ K^-$ & 0 & even & even & +1 & +1 \\
 $X \to p \bar{p}$ & 0 & even & even & +1 & +1 \\
 $X \to p \bar{p}$ & 1 & odd & even and odd & $-1$ & +1 \\
 \hline
 \hline
 \end{tabular}
 \label{table:possibleX} 
 \end{center}
 \end{table}

\section{Radiative Decays of Quarkonia Overview}

\begin{figure}[ht]
\begin{center}
\epsfig{bbllx=0,bblly=75,bburx=580,bbury=615,width=4.5in,clip=,file=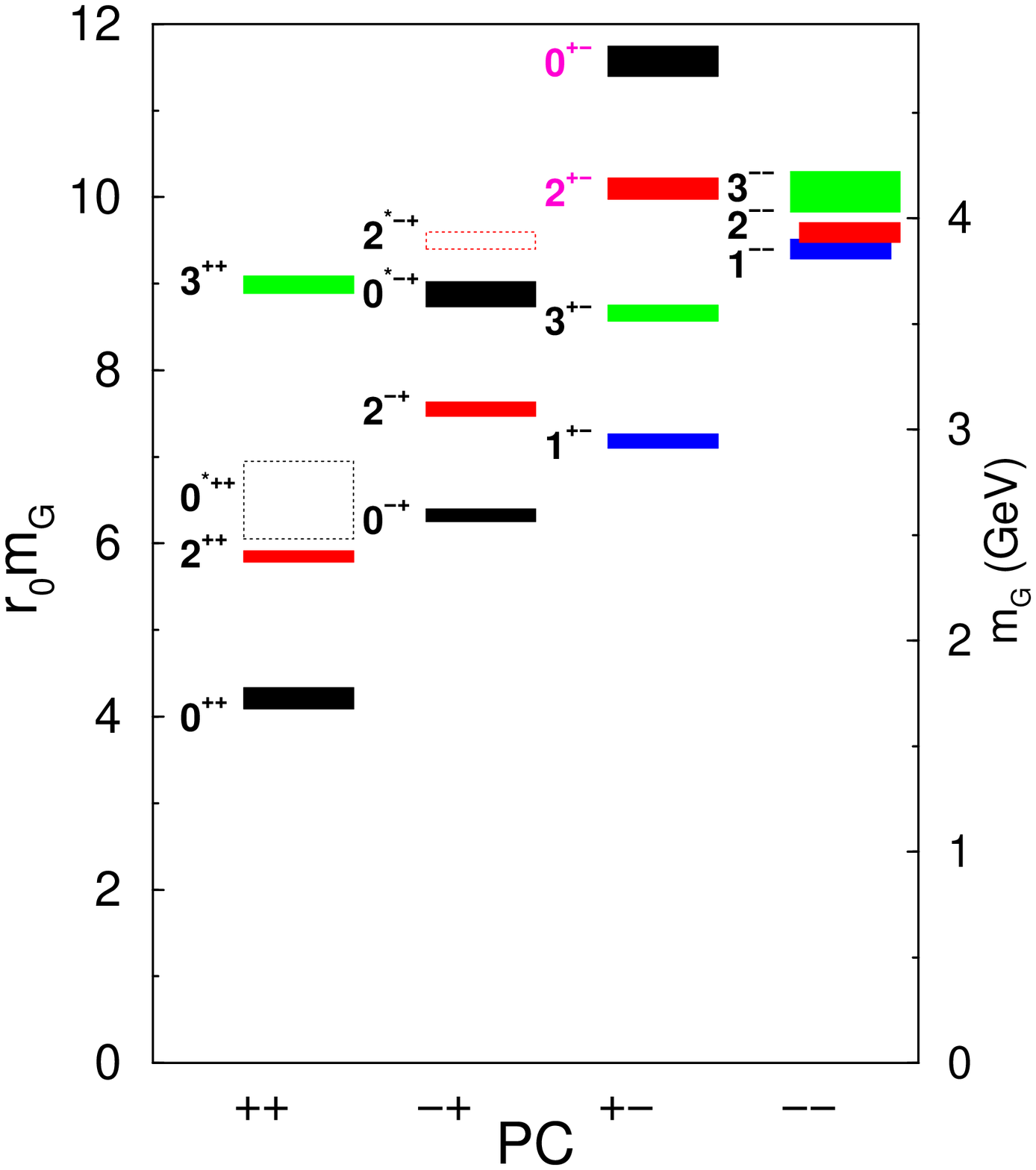} 
\end{center}
\caption{Quenched lattice calculation result for the glueball spectrum
for different $P$ and $C$ values. The mass scale is shown in
terms of a scaling parameter from the QCD lattice calculation named
$r_0$ on the left and in absolute terms on the right
by taking $r_0^{-1}$ = 410 \mev.}\label{Figure:glueballs.eps}   
\end{figure}

Theoretical models exist for glueball production in quarkonia
decay~\cite{maurizio} and for the glueball spectrum. For
example, a quenched lattice calculation~\cite{theory1}
predicts a $J^{PC}=2^{++}$ glueball~\footnote{Here $J$ stands for the
internal angular momentum (spin) of the glueball, $J = L + S$.}
in the 2.2$\mass$ mass 
region (see Figure~\ref{Figure:glueballs.eps}). According to
Table~\ref{table:possibleX}, in the charged pion
and kaon modes, we are limited to detect glueballs in the 
leftmost column where $P=C=+1$ of Figure~\ref{Figure:glueballs.eps},
while in the proton mode we are restricted to the two left most columns.

\begin{figure}[ht]
\begin{center}
\epsfig{bbllx=0,bblly=0,bburx=580,bbury=550,width=4.5in,clip=,file=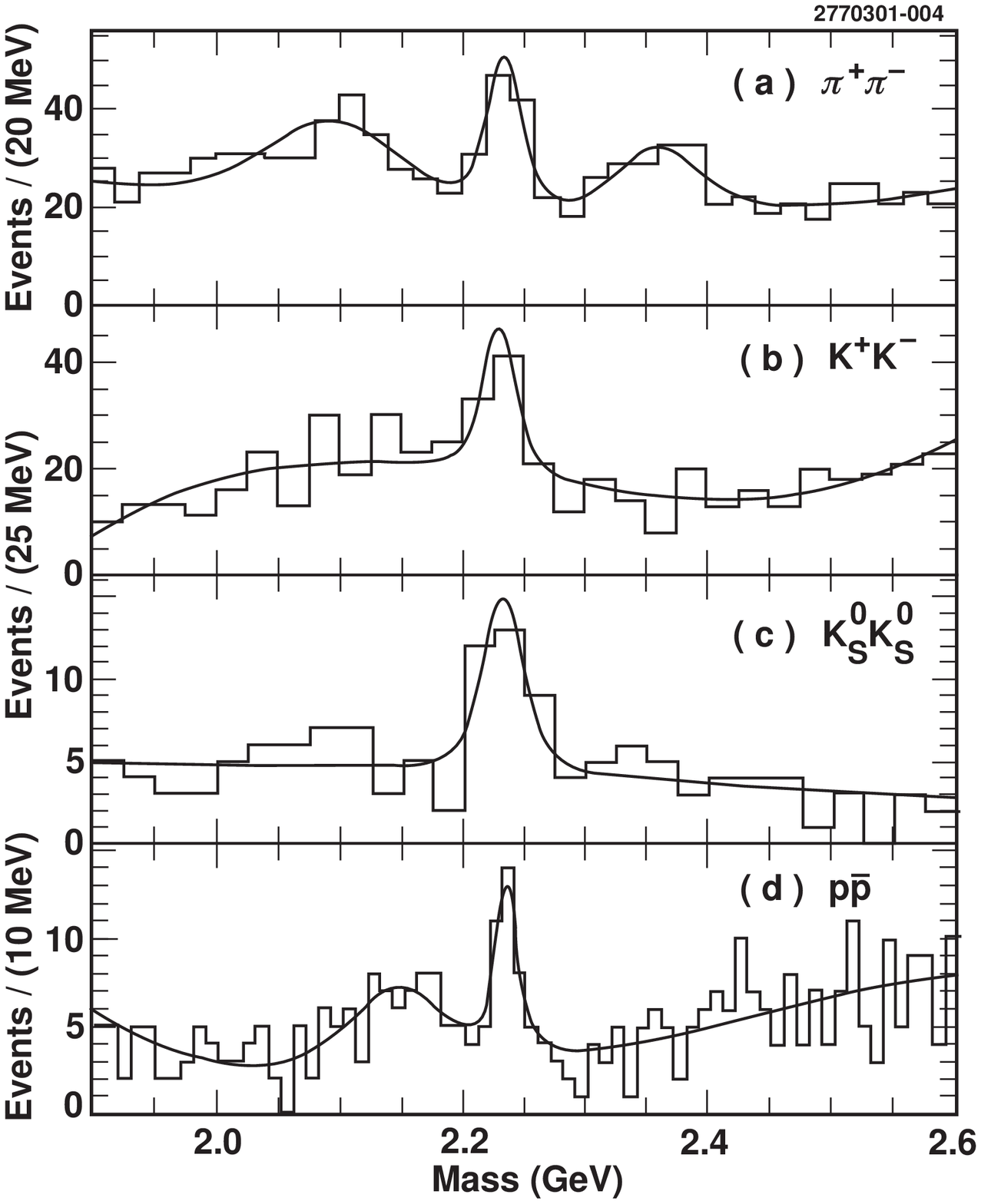} 
\end{center}
\caption{The mass spectrum obtained by the BES collaboration in
radiative $J/\psi$ decays into different hadronic
modes.}\label{Figure:fj2220.eps}
\end{figure}

For some time bound states not considered to be pure glueballs,
such as $f_2(1270)$, $f_4(2050)$, $\eta$, and $\eta^\prime$ have been
observed in $J/\psi$ radiative decays at the $10^{-3}$ 
production level~\cite{pdg}~\footnote{The $J/\psi$ particle is a $c\bar{c}$
bound state. Since the strong force is flavor blind the situation
in $J/\psi$ decays is in principle similar to that in \Uis decays.}.
In 1996, the BES collaboration claimed the observation
of a resonance they called the $f_J(2220)$ particle in 
the radiative $J/\psi$ system at the $10^{-5}$
level~\cite{glueball} (see Figure~\ref{Figure:fj2220.eps}). A lot of
excitement was generated at the time 
because it is possible to interpret the $f_J(2220)$ as a
glueball. However, this result has not been confirmed.

Key to identifying a particle as a
glueball are (a) suppressed production in two-photon collisions
(unlike quarks, gluons don't carry electric charge and do not couple to
photons), and (b) flavor symmetric decays, since a pure glueball has
no valence quarks.  

CLEO has already done several studies of radiative decays of the
$\Upsilon(1S)$. Naively, one expects these types of decays to be
suppressed by a factor 
\begin{equation}
[(q_b/q_c)(m_c/m_b)]^2 \approx 0.025
\end{equation}
with respect to $J/\psi$ radiative decays. This comes from noticing
that the quark-photon 
coupling is proportional to the electric charge and the quark
propagator is approximately $1/m$ for low momentum
quarks. In 1999, CLEO made the first observation of a
two-body $\Upsilon(1S)$ radiative decay~\cite{cleo1}. The spin of the
observed resonance 
could not be measured, but its mass and width where consistent with
the $f_2(1270)$ particle. Under this assumption,
comparing the measured branching ratio of \Uis $\to \gamma f_2(2220)$
to the measured branching fraction of
the $J/\psi \to \gamma f_2(1270)$, a suppression
factor of $0.06 \pm 0.03$ was obtained. After the BES result for the
$f_J(2220)$ in radiative $J/\psi$ decays, a corresponding search was performed
by CLEO in the radiative $\Upsilon(1S)$ system~\cite{cleo2}. This
analysis put limits on the $f_J(2220)$
production in radiative $\Upsilon$(1S) decays.

In this work we are privileged to have available the largest collection
of radiative $\Upsilon$(1S) decays in the world. With it, we can
study the structure of color singlet $gg$ hadronization, and
shed more light on the $f_J(2220)$ result from BES.

\chapter{EXPERIMENTAL APPARATUS}
To carry out our study of the di-hadron spectrum we need to first produce the
\Uis resonance and secondly observe its daughter particles
flying away at relativistic speeds. These two tasks are respectively
accomplished by the Cornell Electron Storage Ring and the CLEO III detector.

\section{The Cornell Electron Storage Ring}\label{luminositydef}

The Cornell Electron Storage Ring (CESR), located at Cornell
University, is a circular particle accelerator that produces $e^+ e^-$
collisions. 

\begin{figure}[ht]
\begin{center}
\epsfig{bbllx=66,bblly=100,bburx=580,bbury=700,width=4.5in,clip=,file=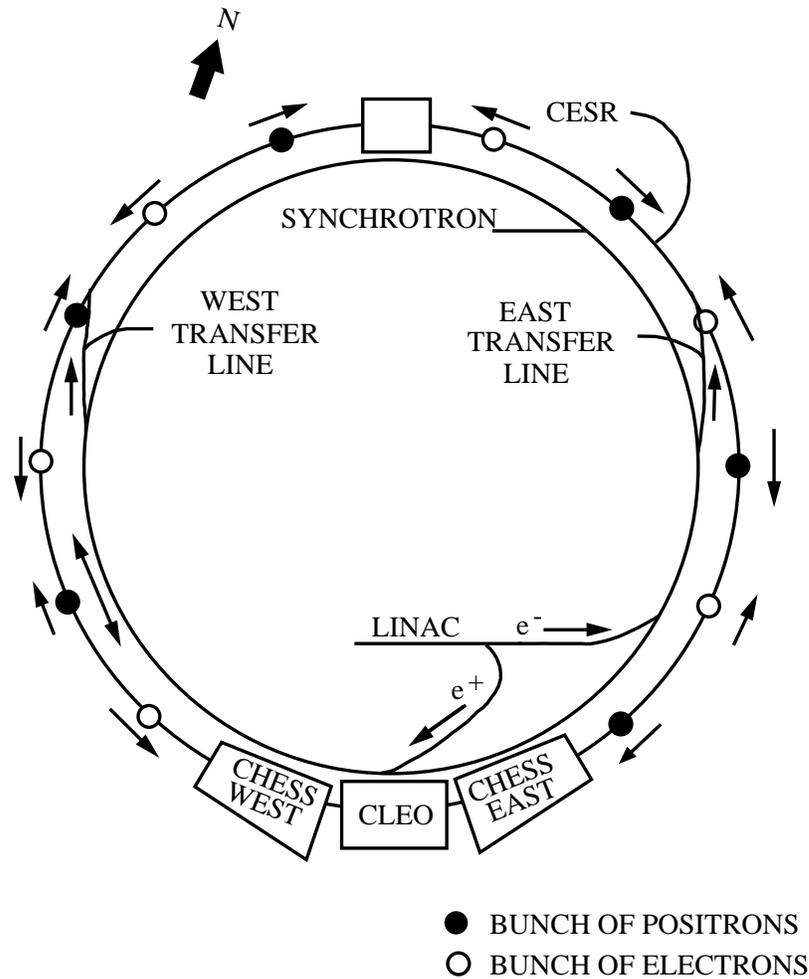}
\end{center}
\caption{The Wilson Laboratory accelerator located about 40 feet
beneath Cornell University's Alumni Fields.}\label{Figure:cesr.ps}
\end{figure}

In order to produce such collisions electrons and
positrons need to be created, accelerated and stored. CESR's different
components, shown in Figure~\ref{Figure:cesr.ps}, have been carrying out
this task since 1979.

A typical CESR run begins at the linear accelerator (LINAC) where
electrons and positrons are produced. To create positrons, electrons
are evaporated off a filament and linearly accelerated by
electromagnetic fields towards a tungsten target. The collision
creates a spray of electrons, positrons and photons. The electrons are
cleared away with magnetic fields and the positrons are introduced
into the synchrotron. The filling procedure is identical, except that
the tungsten target is removed.

Once the electron and positron beams are introduced into the
synchrotron, they are accelerated to the operating energy. In hour case they
are accelerated to the point where their combined energy is the \Uis
mass, 9.46 \gev. 

\begin{figure}[ht]
\begin{center}
\epsfig{bbllx=0,bblly=0,bburx=580,bbury=625,width=4.5in,clip=,file=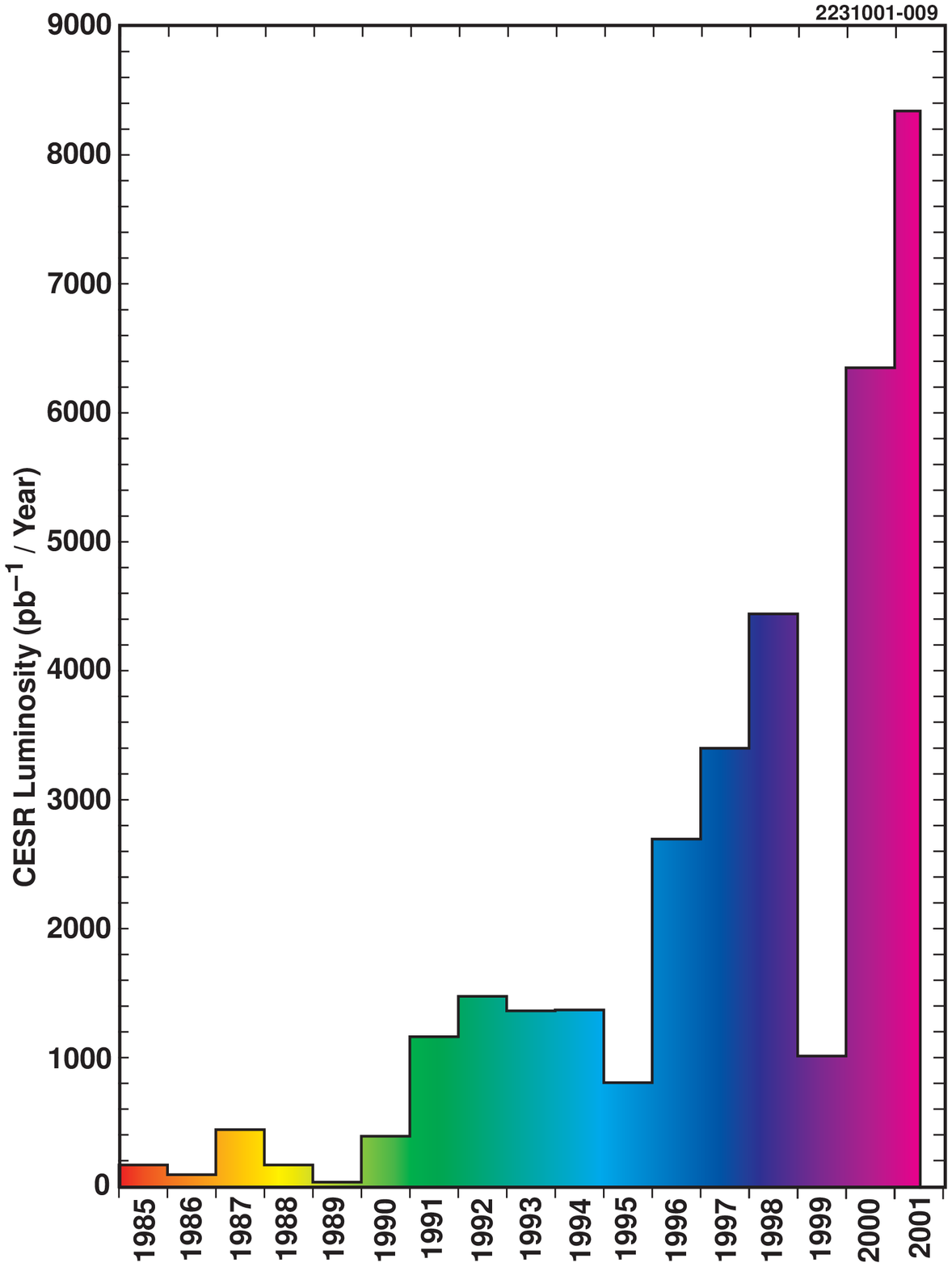}
\end{center}
\caption{CESR yearly luminosity. The gaps in 1995 and 1999
correspond to down times when the CLEO II.V and CLEO III detectors
where being installed.}\label{Figure:cesrluminosity.eps}
\end{figure}

Once the beams are at the desired energy they are transfered to the
storage ring, where they will remain for about an our. At one point in
the storage ring the beams are forced to cross paths. This is the
point where $e^+e^-$ collisions occur~\footnote{This point is
not fixed in space, but varies from event to event inside a small
volume of space called the interaction region (IR).} and
where the center of the CLEO III detector is located.
If the accelerator is performing well a high collision rate results. A
high collision rate is crucial for the success of an accelerator and
the experiments it serves. The important figure is the number of
possible collisions per second per unit area; this is called the
luminosity. In order to maximize the luminosity, the beams are focused
as small as possible at the IR. During the CLEO III installation several magnetic quadrupoles were
added to CESR to improve the beam focus. CESR has consistently outdone
itself while collecting luminosity over the years (see Figure~\ref{Figure:cesrluminosity.eps}).

The CLEO III detector measures the time integrated luminosity over a period
of time by counting how many times a benchmark process occurs during a certain
time interval at the IR. For redundancy, there are two benchmark processes that
are used, one where the $e^+e^-$ particles interact to produce a new
$e^+e-$ pair, and
the other one where the $e^+e^-$ annihilate and produce two
photons. Using the known cross-section for each process, the number of
events is converted to a luminosity called the Bhabha integrated luminosity for
the first process, and the $\gamma\gamma$ integrated luminosity for the second
one. The term ``integrated'' is sometimes left out and the total
luminosity is referred to as simply the Bhabha or
$\gamma\gamma$ luminosity, with the time integration left implicit.


\section{The CLEO III Detector}

When the $e^+e^-$ collision occurs, the two particles are
annihilated we enter the world of particle physics. Nature decides
what to do with the energy from the annihilation. We have no chance of directly observing
what is happening at the annihilation point, but eventually long lived
semi-stable particles are created that fly off at relativistic
speeds. These particles carry information about what happened after the $ e^+e^-$ collision, and can tell us what nature did. The
CLEO III detector has the important mission of detecting and measuring
such particles.  

\begin{figure}[ht]
\begin{center}
\epsfig{bbllx=0,bblly=0,bburx=700,bbury=550,width=4.5in,clip=,file=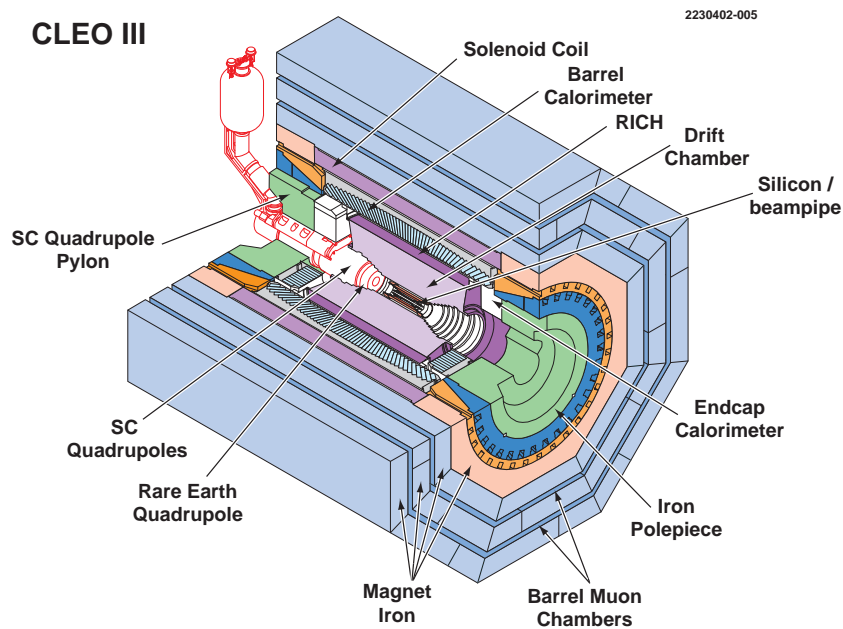}
\end{center}
\caption{The CLEO III detector.}\label{Figure:cleo3.ps} 
\end{figure}

As one can see in Figure~\ref{Figure:cleo3.ps}, the CLEO III detector  
is a composite of many detector elements. These sub-detectors are
typically arranged as concentric cylinders. The entire detector is
approximately cube shaped, with one side measuring about 6 meters, and
weighs over 500 thousand kilograms.

As implied by the name, CLEO III is not the only CLEO detector. CLEO
III was preceded by CLEO II.V, CLEO II, CLEO I.V and the original CLEO
detectors. The CLEO III detector was a major upgrade compared with the
previous version of CLEO~\cite{cleoiiiupgrade},~\cite{cleo2detector},
and has an improved particle identification system together with a new
drift chamber and a new silicon vertex detector. 

\subsection{Superconducting Coil}

All the detector subsystems except for the muon chambers are located
inside a superconduction coil. The coil remains unchanged since CLEO
II. It is kept in a superconducting
state by liquid helium. The purpose of the coil is to provide a
1.5 Tesla magnetic field, which is uniform to 0.2\%, to bend the paths
of charged particles in the detector. By measuring how much much a charged
particle bends, experimenters can measure the momentum of the particle.

The coil inner radius is 1.45m and its outer radius is 1.55 m, with a
radial thickness of 0.10 m. The total length of the coil in z is 3.50 m.
It is wound from a 5mm x 16 mm superconducting cable (Al surrounding
Cu-NbTi strands). It is wound in 2 layers, with 650 turns per layer, on
an aluminum shell. When in operation a current of 3300 amps flows
through the coil.

\subsection{Tracking System}

After particles from the interaction point pass through the beam pipe,
they begin to encounter the active detector elements of the tracking
system. There are two sub-detectors responsible for tracking the
curving path of charged particles. The first one encountered by
particles is the silicon vertex detector, and the second one is the
central drift chamber. The CLEO III tracking system is responsible for
tracking a charged particle's path and measuring its momentum. Typical
momentum resolution is 0.3\% (1\%) for 1\gev~(5\gev) tracks. The
tracking system also measures ionization energy losses with an
accuracy of about 6\%. 

\subsubsection{Silicon Vertex Detector}

\begin{figure}[ht]
\begin{center}
\epsfig{bbllx=130,bblly=385,bburx=500,bbury=750,width=4.0in,clip=,file=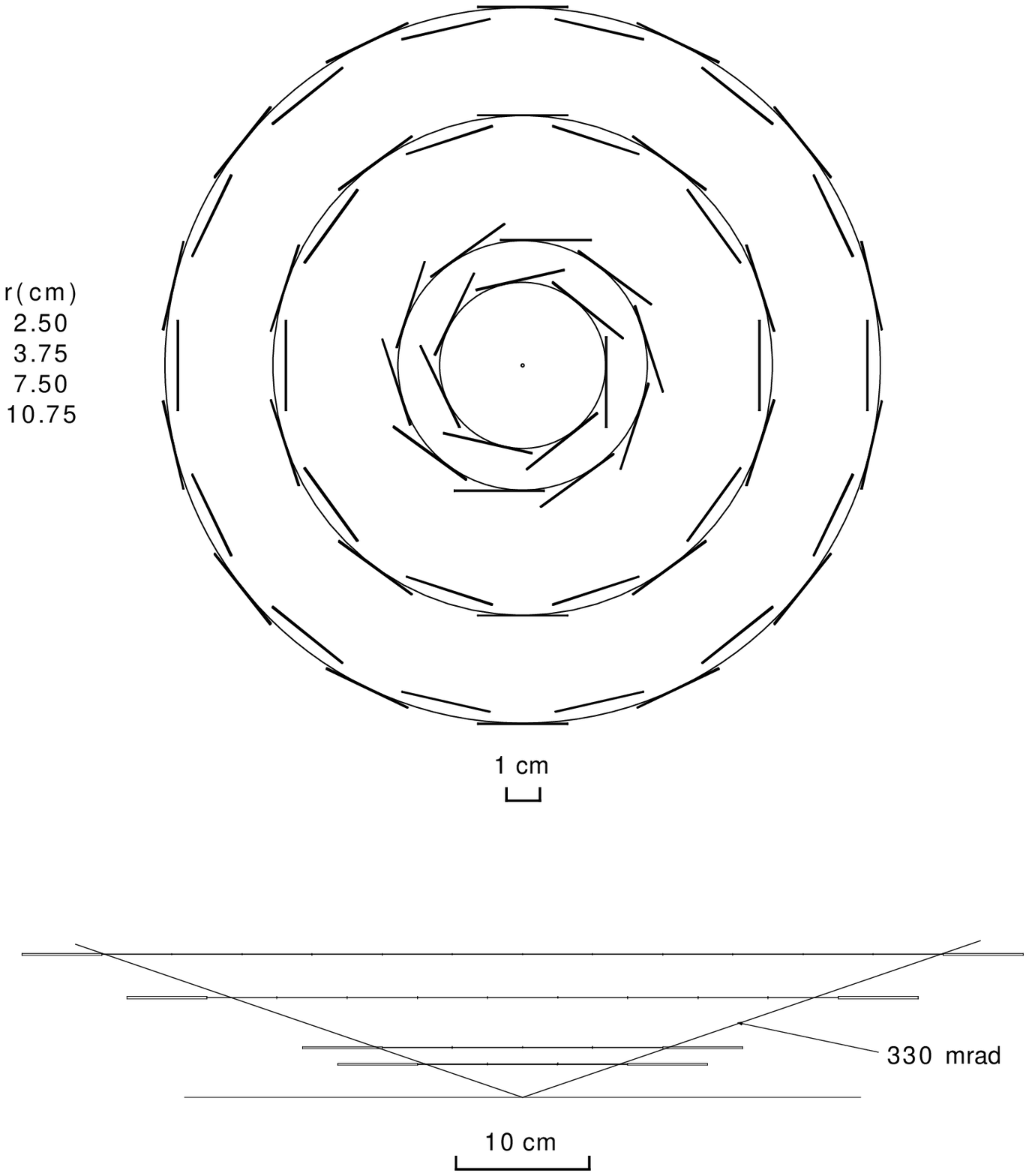} 
\end{center}
\caption{View of the SVD III along the beampipe.}\label{Figure:svd.ps} 
\end{figure}

The silicon vertex detector in CLEO III~\cite{sv3}, also called SVD III, is a
four-layer barrel-only structure with no endcaps that surrounds the
beam pipe. This detector (see Figure~\ref{Figure:svd.ps})
provides four $\phi$ and four $z$ measurements covering over over 93\% of the
solid angle. The average radius of inner surface of the four layers is
25 mm, 37.5 mm, 72 mm, 102 mm. The detector is constructed from 447
identical double-sided silicon wafers, each 27.0 mm in $\phi$, 52.6 mm in
z and 0.3 mm thick. The wafers are instrumented and read out on both
sides. The instrumentation on each side consists of an array of
aluminum strips on the wafer surface. These strips are connected to
preamplifiers at the end of the detector. The inner side has 512
strips in the $z$ direction and the outer side has 512 in the $\phi$
direction. Therefore each wafer contains 512+512 sensors. The 447 wafers
are arranged in the 4 layers, as follows: 7 sections in $\phi$, each
with 3 wafers in $z$, total = 21 wafers in the first layer; 10 sections in $\phi$, each with 4 wafers in $z$, total
= 40 wafers  in the second layer; 18 sections in $\phi$, each with 7
wafers in $z$, total  = 126 wafers in the third layer; 26 sections in
phi, each with 10 in Z, total = 260 wafers in the fourth layer.

Charged particles traversing the wafer lose energy and create electron
hole pairs. Approximately 3.6 \ev~ is required to create a single
electron-hole pair. The electrons and holes then travel in opposite
directions in the electric field applied to the surfaces of the
wafers until they end up on the aluminum strips, and the detector registers a
``hit''. When combined together, the hit on the inner side of a wafer
and the hit on the outer side give a measurement of the ($z$,
$\phi$). The wafer position itself determines $r$.

\subsubsection{The Central Drift Chamber}

The CLEO III central drift chamber (DR III) is full of a gas mixture
with 60\% Helium and 40\% propane held at about 270 K and at a
pressure slightly above one atmosphere. The drift chamber is strung
with array of anode wires of gold-plated tungsten of 20 $\mu$m in
diameter and cathode wires of gold-plated aluminum tubes of 130 $\mu$m
in diameter. The anodes are kept at a positive voltage (about 2000 V),
and the cathodes are kept grounded, which provides an electric field
between the anode and the cathode wires. Anode and cathode wires are
often called ``sense'' and ``field'' wires respectively. 

As a charged particle passes through the DR III, it interacts
electromagnetically with the gas molecules giving energy to the outer
electrons which become free in a process called ionization. The free
electrons from the ionized gas molecules drift in the electric field
toward the nearest anode wire. As the electrons get close to the
anode, the electric field becomes very strong which causes an
avalanche as further ionization is induced. The result of the
avalanche is a large number of electrons collapsing upon the sense wire in a
very short amount of time (less than one nanosecond). When this
happens to a sense wire, we say that there is a ``hit''.

The current on the anode wire from the avalanche is amplified and
collected at the end of the anode wire. Both the amount of charge
and the time it takes it to move to the end of the detector are
measured. A calibration of the drift chamber is used to convert the
amount of charge to a specific ionization measurement of the incident
particle. A calibrated drift chamber can also convert the time to roughly
measure the position along the sense wire where the charge was
deposited. 

The wires are strung along the $z$ direction. About 2/3 of the outer
part of the drift chamber (the farthest part from the interaction
point) is strung in with a slight angle (about 25 miliradians) with respect
to the $z$ direction to help with the z measurement. Wires strung in
the z direction are called ``axial'' wires, while those that are strung at
a slight angle are called ``stereo'' wires.

The DR III consists of an inner stepped section with 16 axial layers,
and an outer part with conical endplates and 31 small angle stereo
layers. There are 3 field wires per sense wire and
they approximately form a 1.4 cm side square. The drift resolution is
around 150 $\mu$m in $r-\phi$ and about 6 mm in $z$. All wires are
held at sufficient tension to have only a 50 $\mu$m gravitational
sag at the center ($z$ = 0). There are 1696 axial sense wires and 8100
stereo sense wires, a 9796 total.

\subsection{Ring Imaging Cherenkov Detector}

The Ring Imaging Cherenkov (RICH) detector~\cite{rich} is a new
detector subsystem for CLEO III. It replaces the CLEO II.V time of
flight system designed to measure particles' velocities.

Cherenkov radiation occurs when a particle travels faster than the speed of light in a
certain medium,
\begin{equation}
v > c/n.
\end{equation}
Where $v$ is the velocity of the particle, $c$ is the speed of light
in vacuum, and $n$ is the index of refraction of the medium the particle is
traveling in. The charged particle polarizes the molecules of the medium,
which then turn back rapidly to their 
ground state, emitting radiation. The emitted light forms a
coherent wavefront if $ v> c/n$ and Cherenkov light is emitted 
under a constant Cherenkov angle, $\delta$, with the particle
trajectory forming a cone of light. The cone half-angle is given by
the Cherenkov angle which is,
\begin{equation}
\cos\delta = \frac{c}{vn} = \frac{1}{\beta n}.
\end{equation}
If the radiation angle, $\delta$, is measured, the speed of the
incident particle is known. This measurement, combined with the
momentum measurement from the tracking system, gives a
measurement of the particles mass, and can be used in particle
identification. 

The threshold velocity at which Cherenkov radiation is
emitted is $v_{min} = \frac{c}{n}$. When a particle traveling at the
threshold velocity transverses the medium a very small cone with
$\delta \approx 0$ is produced. The maximum emission angle
occurs when $v_{max} = c$ and is given by 
\begin{equation}
\cos\delta_{max} = \frac{1}{n}.
\end{equation}
 
\begin{figure}[ht]
\begin{center}
\epsfig{bbllx=66,bblly=225,bburx=580,bbury=575,width=4.5in,clip=,file=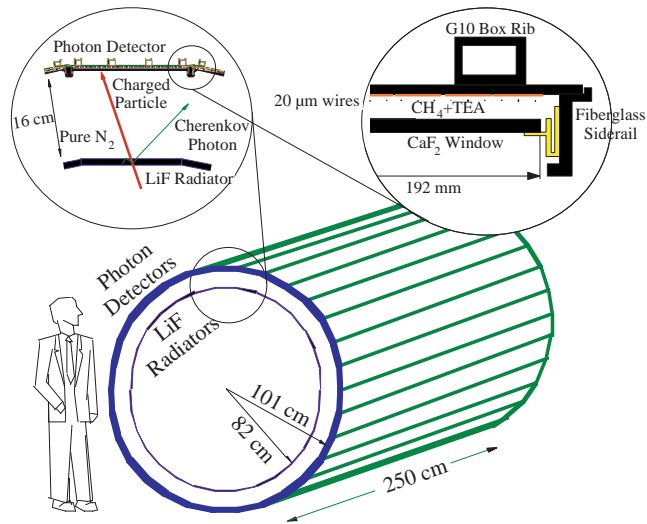}
\end{center}
\caption{The RICH detector subsystem.}\label{Figure:rich}
\end{figure}

\begin{figure}[ht]
\begin{center}
\epsfig{bbllx=0,bblly=0,bburx=300,bbury=175,clip=,file=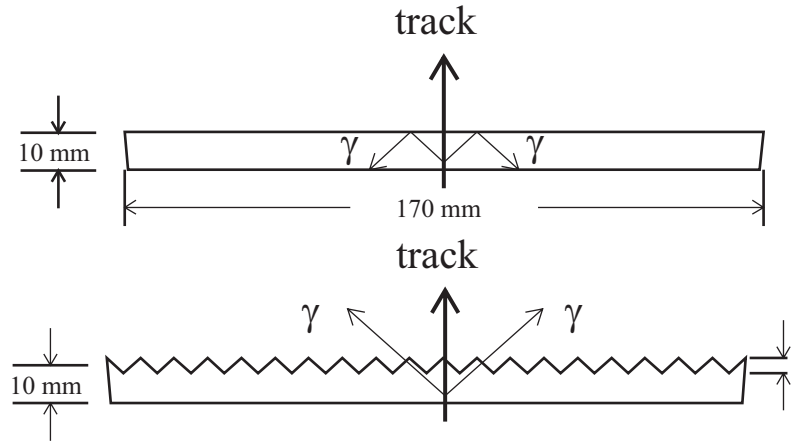}
\end{center}
\caption{The two kinds of RICH LiF radiators. For normal incidence
particles ($z$ $\approx$ 0) a sawtooth radiator is necessary to avoid
internal reflection.}\label{Figure:richradiators}
\end{figure}

The RICH (see Figure~\ref{Figure:rich}) consists of 30 modules in phi,
0.192 m wide and 2.5 m long. The detector starts at a radius of 0.80 m
and extends to 0.90 m. Each module has 14 panes of solid crystal LiF
radiator at approximately 0.82 m radius, 0.192 m wide, 0.17 m long, 1
cm thick. Inner separation between radiators is typically $50
\mu$m. The LiF index of refraction is $n =
1.5$. The radiators closest to z = 0 in each module have a 45 degree
sawtooth outer face, to reduce total internal reflection of the
Cherenkov light for normal incident particles (see
Figure~\ref{Figure:richradiators}). The radiators are 
followed by a 15.7 cm (radial) drift 
space filled with pure N2. The drift space is followed by the photodetector,
a thin-gap multiwire photosensitive proportional chamber.


\begin{figure}[ht]
\begin{center}
\epsfig{bbllx=66,bblly=100,bburx=580,bbury=600,width=4.5in,clip=,file=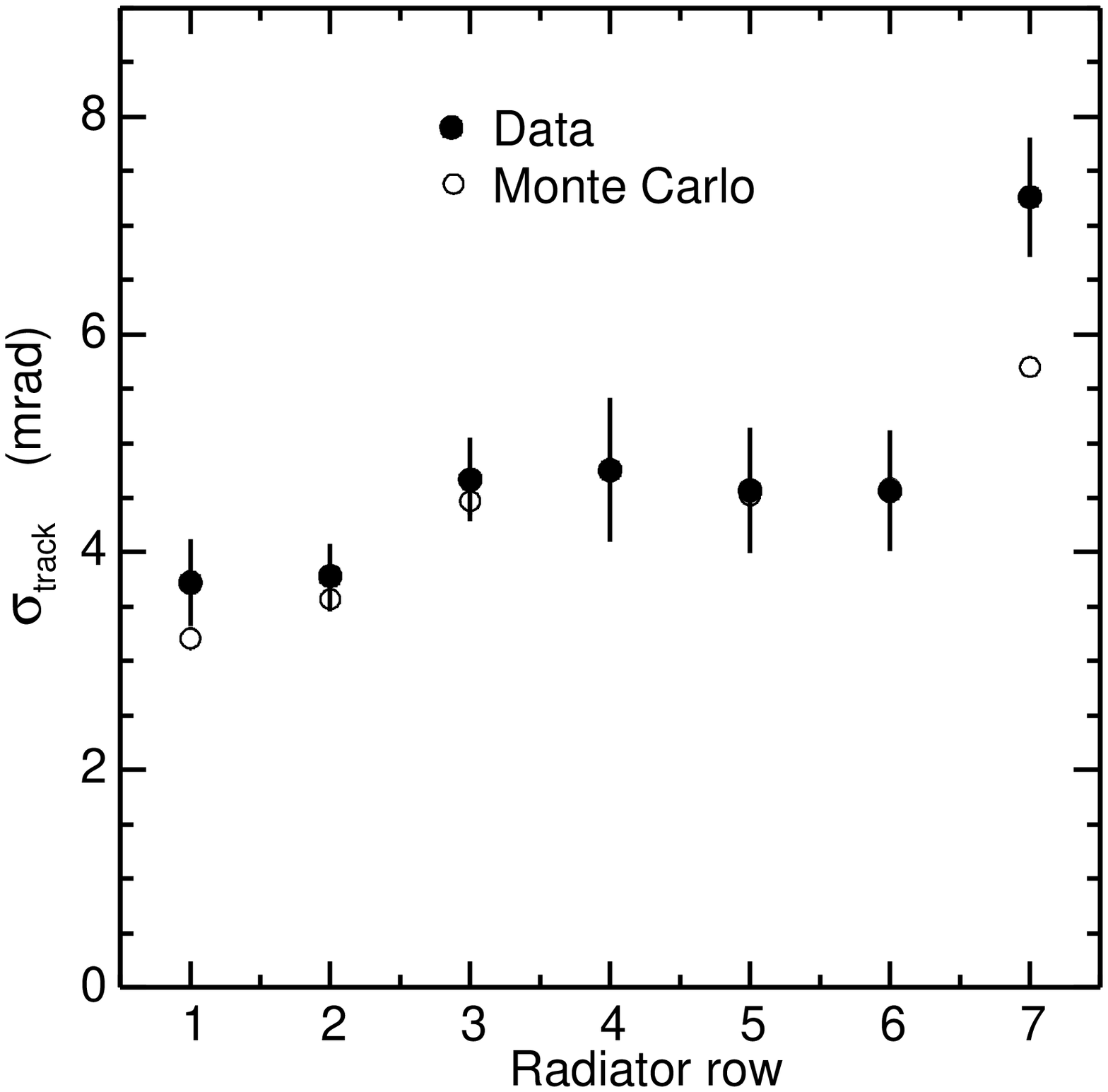}
\end{center}
\caption{ Cherenkov angle resolutions per track
as a function of radiator row for Bhabha events. Row 1 corresponds to the
two rows closest to $z = 0$, etc.}\label{Figure:richresolution}
\end{figure}

\begin{figure}[ht]
\begin{center}
\epsfig{bbllx=66,bblly=100,bburx=580,bbury=600,width=4.5in,clip=,file=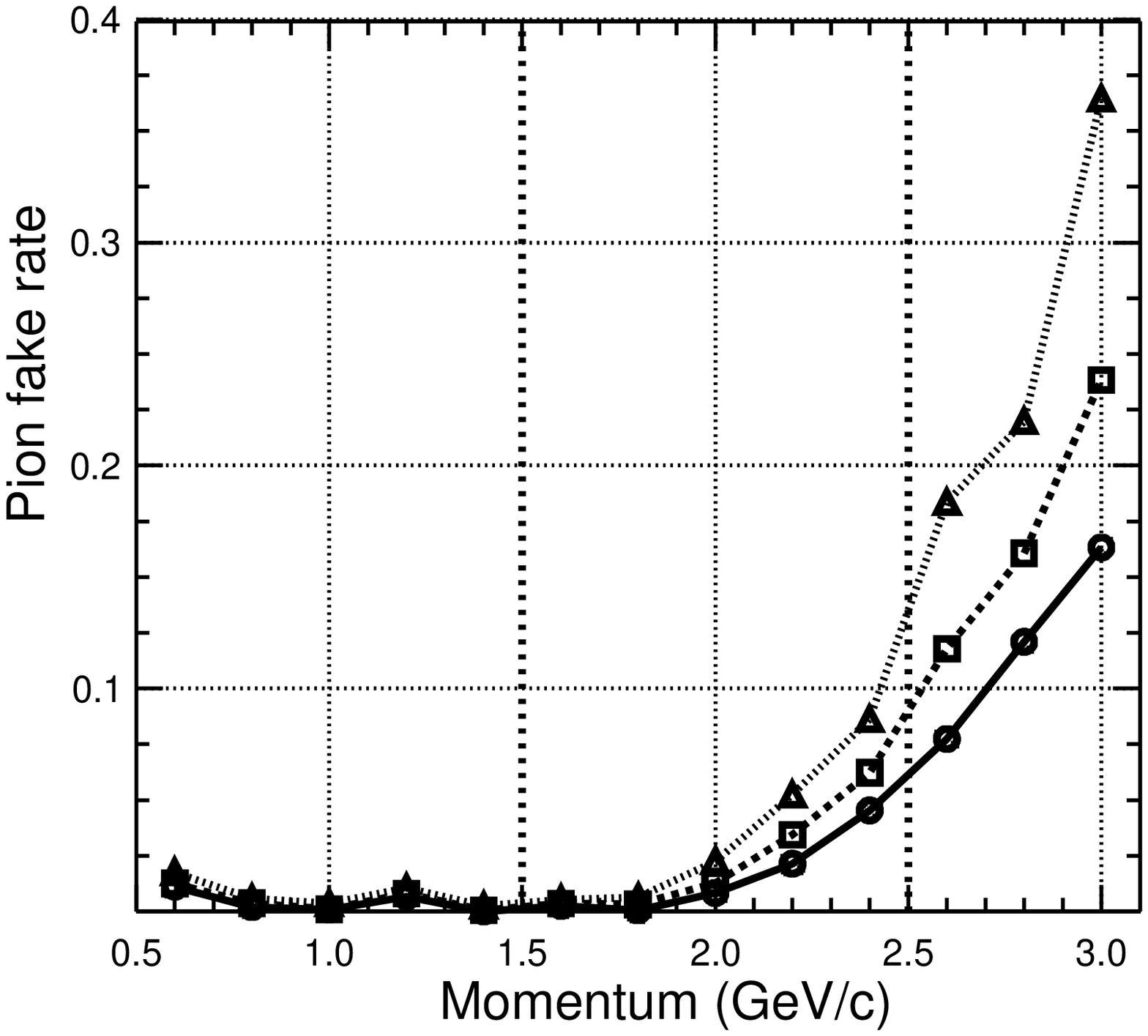}
\end{center}
\caption{Pion fake rate as a function of particle momentum for
      kaon efficiency of 80\%\ (circles), 85\%\ (squares) and
      90\%\ (triangles).}\label{Figure:richseparation}
\end{figure}

With this index, particles in the LiF radiator with beta = 1 produce
Cherenkov cones of half-angle $\cos^{-1}$(1/n) = 0.84 radians. With a 16 cm
drift space, this produces a circle of radius 13 cm. The RICH is
capable of measuring the Cherenkov angle with a resolution of a few
miliradians (see Figure~\ref{Figure:richresolution}). This great
resolution allows for good separation between pions and kaons up to
about 3 GeV as Figure~\ref{Figure:richseparation} shows.  

\subsection{Crystal Calorimeter}

The CLEO Crystal Calorimeter (CC) is composed of 7784 thallium-doped
CsI crystals. Each crystal is 30 cm long (16.2 radiation lengths)
with 5cm x 5cm square front face. The crystals absorb any incoming
electron or photon which cascades into a series
of electromagnetic showers. The electronic system composed of 4
photo-diodes present at the back
of each crystal are calibrated to measure the energy deposited by the
incoming particle. Other incoming particles other than
photons and electrons are partially, and sometimes fully, absorbed by
the crystal giving an energy reading.

The CC is arranged into a barrel section and two
endcaps, together covering 95\% of the solid angle.
The CC barrel section is unchanged since CLEO II; the endcaps have
been rebuilt for CLEO III to accommodate the new CESR interaction
region quadrupoles. The barrel detector consists of an array of 6144
crystals, 128 in $\phi$ and 48 in $z$, arranged in an
almost-projective barrel geometry. That is, the crystals are tilted in
$z$ to point to a few cm away from the interaction point, and there is
also a small tilt in $\phi$.  The CC barrel inner radius is 1.02 m,
outer radius is 1.32 m, and the length in $z$ at the inner radius is
3.26 m. It covers the polar angle range from 32 to 148 degrees. The
barrel crystals are tapered towards the front face (there are 24
slightly different tapered shapes), the endcap crystals are
rectangular, but shaved near the outer radius to fit in the
container. The CC endcaps consist of two identical end plugs, each
containing 820 crystals of square cross-section, aligned parallel to
the beam line (not projective). There are 60 crystals in the
"fixed" portion of the "keystone" piece of the endcap. 760 in the part
that slides. The keystone is made up of two parts, one on
top that has 12 crystals that for mechanical removal reasons is
separate from a container holding 48 crystals. 
The endcap extends from 0.434 m to 0.958 m in $r$. The front faces
are $z$ = $\pm$1.308 m from the interaction point. It covers the polar
angle region from 18 to 34 degrees in +$z$, and 146 to 162 in -$z$. 

The photon energy resolution in the barrel (endcap) is 1.5\% (2.5\%)
for 5 \gev photons, and goes down to 3.8\% (5.0\%) for 0.1 \gev photons.

\subsection{Muon Detectors}

The muon detectors (MU) are the most external subsystem of the CLEO
III detector. They remain unchanged from CLEO II, and are composed of
plastic proportional tubes embedded in the magnet
iron return yoke. They cover 85\% of the 4$\pi$ solid angle (roughly
30-150 degrees in polar angle). If a series of hits is detected in the
muon chamber layers they most likely correspond to muons because other
particles are blocked by the iron. Besides detecting muons, the heavy
iron of the return yoke protects the inner subsystems of the CLEO III
detector from cosmic ray background (except for cosmic ray muons of course). 

There are three planes of chambers in the barrel section, arranged in
8 octants in $\phi$. The plastic barrel planes lie at depths of
36, 72, and 108 cm of iron (at normal incidence), corresponding to
roughly 3, 5, and 7 hadronic interaction lengths (16.8 cm in
iron) referred to as DPTHMU. There is one plane of chambers in each of
the two endcap regions, arranged in 4 rough quadrants in $\phi$. They lie
at $z$ =$\pm$ 2.7 m, roughly covering the region $0.80 <
|\cos(\theta)| < 0.85$. The planar tracking chambers use plastic
proportional counters at about 2500 V with drift gas of 60\% He, 40\%
propane, identical to (and supplied by the same system as) the drift
chamber gas. Individual counters are 5 m long and 8.3 m wide, with a
space resolution (along the wire, using charge division) of 2.4
cm. The tracking chambers are made of extruded plastic, 8cm wide by 1
cm thick by 5 m long, containing eight tubes, coated on 3 sides with
graphite to form a cathode, with 50 $\mu$m silver-plated Cu-Be anode
wires held at 2500 V. The orthogonal coordinate is provided by 8cm
copper strips running perpendicular to the tubes on the side not
covered by graphite.

\chapter{ANALYSIS OF THE DATA}
This work builds on the techniques developed by previous CLEO
radiative $\Upsilon(1$S$)$ analyses ~\cite{cleo1},~\cite{cleo2},
modified to be used with CLEO III data. A new technique based on
kinematic fitting is developed to, together with the new RICH
detector, improve efficiency and particle identification. 

We search for radiative $\Upsilon(1$S$)$
decays in the modes $\Upsilon(1$S$)$ $\to \gamma \pi^+ \pi^-, \gamma
K^+K^-$, and $\gamma p 
\bar{p}$. The $e^+e^-$ collision data has both resonant events, where
the $e^+e^-$ annihilate to give a \Uis, and continuum 
events, where the $e^+e^-$ collision does not give a \Uis. To be sure
we are observing \Uis and not a continuum process, the continuum must
be subtracted by using a pure source of correctly scaled continuum
events. Pure continuum data can be obtained by operating CESR at an
energy different form the \Uis mass. After subtracting the underlying 
continuum, we examine the di-hadron invariant mass spectrum in search
of resonances. We determine the spin and production helicity (the projection of
the spin on the momentum vector at production time) of any found
resonances by examining the photon and hadron angular distributions.

\section{Data Sample}

The analysis presented here is based on CLEO III data. Throughout this
document, and unless otherwise stated we use $\gamma\gamma$
luminosity (the luminosity types used by CLEO are defined at the end
of Section 2.1). We prefer to use the 
$\gamma\gamma$ over 
the Bhabha luminosity because the resonant process $\Upsilon(1$S$) \to
e^+e^-$ artificially increases the reported Bhabha luminosity
by about 3\% in $\Upsilon(1$S$)$ data. This extra contribution would need to
be accounted for when doing a continuum subtraction. Choosing the
$\gamma\gamma$ luminosity avoids this complication.

The CLEO III data is divided into numbered sets. Sets 18 and 19 have a
luminosity of $1.13\pm0.02 fb^{-1}$ in the beam
energy range 4.727-4.734$\gev$. This data, which we call the
$\Upsilon(1$S$)$ data (or simply the 1S data),
has both resonance $e^+e^- \to \Upsilon(1$S$)$ and continuum events. We
take the number of resonant events
from~\cite{cleo4}, $N_{\Upsilon(1S)} = (2.1\pm0.1)\times  
10^7$. This number contrasts with the previous
generation measurement of CLEO II, where $N_{\Upsilon(1S)} \approx 0.15
\times 10^7$ were available. 

In datasets and 18 and 19 there are also $0.192\pm0.004 fb^{-1}$ taken
below the $\Upsilon(1$S$)$ beam energy (4.714-4.724$\gev$). This
data, which we call the $\Upsilon(1$S$)$-off data (or simply the
1S-off data), has relatively low statistics
and corresponds to purely continuum events.

To improve our continuum statistics we use $3.49\pm0.07 fb^{-1}$
from datasets 9, 10, 12, 13, and 14 of data taken near
the center-of-mass energy of the $\Upsilon(4$S$)$, which for our
purposes is defined as data with beam energy in the
5.270-5.300$\gev$ range. This set of data, which we call the $\Upsilon(4$S$)$
data (or simply the 4S data), is a
source of pure continuum because no $\Upsilon(4$S$) \to B \bar{B}$
resonant event can survive our ``cuts''~\footnote{Cuts are simply
conditions that an event must satisfy to be considered in the
analysis. Cuts are necessary to eliminate background that would
otherwise make a measurement difficult or even impossible.} (the cuts are presented in Section 3.3).

\subsection{Continuum Subtracted Distributions}

We use the continuum data taken at the $\Upsilon(4$S$)$ energy to
subtract the underlying continuum present in the $\Upsilon(1$S$)$
data. This is important because 
continuum background processes like $e^+e^- \to \gamma
\rho$ with $\rho \to  \pi^+ \pi^-$, $e^+e^- \to \gamma
\phi$ with $\phi \to  K^+ K^-$, and direct $e^+e^- \to \gamma
h^+ h^-$ (we will use the convention $h = \pi,\ K,\ p$ from now
on), look like the signal events we are searching for. To first order, the
cross section of these continuum process scales like
$1/s$, where $s$ is the square of the center of mass energy of the
$e^+e^-$ system. Taking the
luminosity-weighted average beam-energies of each interval, and the
$\gamma\gamma$
luminosities (see Table ~\ref{Table:datasets}), we
calculate that the $\Upsilon(4$S$)$ data scaled down by a factor of
0.404~\footnote{If we are to be mathematically strict we should
calculate the 
scale factor as 
\begin{equation*}
{\sum_{1S\ runs}\frac{\mathcal{L}_{1S\ run}}{s_{1S\ run}}}\ \Big{/}{\sum_{4S\
runs}\frac{\mathcal{L}_{4S\ run}}{s_{4S\ run}}}\end{equation*} with
obvious notation. This is equivalent to redefining the average energy as
\begin{equation*}
\frac{1}{\bar{E}^2} = \frac{1}{\mathcal{L}} \sum_{runs}\frac{\mathcal{L}_{run}}{E_{run}^{2}}.
\end{equation*}
However, our energy intervals are sufficiently narrow and this
calculation does not change the last significant digit of the original
scale factor. Similarly, taking into account second order terms in the
energy dependence of the cross section, like the
one that appears as $m_{\rho}^2/s$ in the explicit formula
for the $e^+e^- \to \gamma \rho$ cross section, also has an insignificant
effect on the scale factor.} represents the underlying continuum in
the $\Upsilon(1$S$)$ data. This is true up to differences in momentum
distributions and phase space. The error in the
 continuum scale factor is unknown because the luminosity ratio is
 expected to have a small but undetermined systematic
 error. We make the somewhat arbitrary decision to retain three
 significant digits in the continuum scale factor because it is
 sufficient for our purposes and there are 0.5\% effects from second
order terms in the cross section formulae.

To eliminate the contribution of continuum events from a $\Upsilon(1$S$)$ data
variable distribution (e.g., the invariant mass of two tracks, the photon
angular distribution) we proceed as follows,

$\bullet$ Obtain the $\Upsilon(4$S$)$ data distribution for the same
variable.  

$\bullet$ Efficiency correct both the $\Upsilon(1$S$)$ and 
$\Upsilon(4$S$)$ distributions, using a GEANT~\cite{geant} based Monte
Carlo (MC) simulation of the detector. Examples of MC efficiency
distributions are shown in Figures 
\ref{Figure:efficiencies} and ~\ref{Figure:systematics_helicity.ps}.

$\bullet$ Subtract the $\Upsilon(4$S$)$ distribution from the
$\Upsilon(1$S$)$ distribution using a $\Upsilon(4$S$)$ scale 
factor of 0.404.

We call this set of steps ``continuum subtraction'' by definition. For
compactness, we call such a distribution the
``continuum subtracted $<$variable$>$ distribution/plot'' or ``$<$variable$>$
continuum subtracted distribution/plot''. Except for
statistical fluctuations and phase space effects, the resulting 1S
distribution should not have any contribution from continuum processes
that scale as 1/s.
 
It is important to notice that any continuum subtracted
distribution is efficiency corrected. This means that a fit to a
continuum subtracted distribution (for example, the continuum
subtracted invariant mass distribution) gives the efficiency corrected
number of events directly. Strictly speaking, this number of events is
only correct if all the other
variables we cut on have the same initial distribution in data and
MC, or if the efficiency does not depend on them. In this
note we use ``flat MC'', defined as MC that is generated 
with a flat distribution in the mass and the helicity angles
$\theta_{\gamma}$, $\theta_{\pi}$ (these angles are defined in the appendix). 

For example, the number
of events obtained from the continuum subtracted invariant mass fit needs to be
corrected to account for the fact that the helicity angle
distributions are not flat in data (see for example 
Figure ~\ref{Figure:pipiplot_ct_cg_2.ps}, and the efficiency is highly 
dependent on these variables (see
Figure~\ref{Figure:systematics_helicity.ps}). This correction is done
in Section 7.2. 


A summary of the results from this section is shown in
Table~\ref{Table:datasets}.

 \begin{table}[ht]
 \begin{center}
 \caption{Summary of the data used in this analysis. The continuum scale factor
 is obtained using $\gamma\gamma$ luminosities because the Bhabha
 luminosity is artificially high during $\Upsilon(1$S$)$ running due to
 the process $\Upsilon(1$S$) \to e^+ e^-$.}
 \begin{tabular}{c c c c}
 \hline
 \hline
  & $\Upsilon(1$S$)$ & $\Upsilon(4$S$)$ &$\Upsilon(1$S$)$-off \\
 \hline
 \hline
Dataset & 18, 19 & 9, 10, 12, 13, 14 & 18, 19 \\
Average $E_{beam}(\rm GeV)$ & 4.730 & 5.286 & 4.717 \\
Range of $E_{beam}(\rm GeV)$ & 4.727-4.734 & $5.270-5.300$ & $4.714-4.724$ \\
$\mathcal{L} (e^+e^-) (fb^{-1})$ & $1.20\pm0.02$ & $3.56\pm0.07$ &
 $0.201 \pm 0.004$ \\
$\mathcal{L} (\gamma\gamma) (fb^{-1})$ & $1.13\pm0.02$ & $3.49\pm0.07$ &
 $0.192\pm0.004$ \\
$\Upsilon(1$S$)$ continuum scale factor & 1 & $0.404$ & $5.84$ \\
 \hline
 \hline
 \end{tabular}
\label{Table:datasets} 
 \end{center}
 \end{table}

\section{Event Selection}

Event selection for $\Upsilon(1S) \to \gamma h^+ h^-$ is
straightforward and can be 
thought of in terms
of three major stages. 

First we skim the data,
keeping only those events that contain exactly one high-energy photon and two
tracks. Next, we require that the total 4-momentum of
these three elements be consistent with the colliding $e^+e^-$
4-momentum. Finally, in the
third stage, we project the surviving data onto the three different
hadronic modes via hadron separation and QED suppression cuts. 

However,
checking the 4-momentum involves using the tracks masses. This
means that the information from stage 2 should somehow be useful in
stage 3. This is indeed the case, and the details of how we do it are
revealed in this section.

\subsection{Skim Cuts}

We skim over  the data in the ``hardGam'' subcollection. The hardGam
subcollection was developed with this type of analysis in
mind. For an event that passes the
triggers~\footnote{Triggers are basic criteria that an event must satisfy
to recorded during the data
collection processes. Triggers are designed to get rid of trash and
noise and reduce the size of the data sample while keeping all of the
important information.} to be classified as hardGam it must pass the
following cuts,

$\bullet$ eGam1  $>$ 0.5 

$\bullet$ eSh2    $<$ 0.7 

$\bullet$ eOverP1 $<$ 0.85 

$\bullet$ eVis    $>$ 0.4 

$\bullet$ aCosTh  $<$ 0.95
where ``eGam1'' is the highest isolated shower energy relative to the beam
energy, ``eSh2'' is the energy of the second highest shower relative to
the beam energy, ``eOverP1'' is the matched shower energy relative to
the momentum of the track with highest momentum, ``eVis'' is the total energy
detected (charged tracks are assumed to be pions) relative to the
center of mass energy, and ``aCosTh'' is the z component of the unitary total
momentum vector.

Monte-Carlo predicts that about 75\% of the generated $\Upsilon(1S)
\to \gamma h^+ h^-$ signal passes the
hardware and software triggers and gets classified as hardGam.

As mentioned above, we use this data to make our skim. To write an
event from the hardGam subcollection into our skim we require the
following topological cuts,

$\bullet$ There are exactly two ``good tracks''; there can be any
  number of tracks that are not ``good tracks'' but these are not
  used in the analysis. We define a ``good
  track'' as a track that satisfies the following cuts; drift chamber track
  ionization energy loss ($dE/dX$)
  information is available, the ratio of number of wire hits to those
  expected is between 0.5 and 1.5, the pion fit has
  $\chi^2/d.o.f. < 20 $ (here d.o.f. stands for degrees of freedom), and the
  distance of closest 
  approach to the beam spot in the x-y plane (called DBCD) is less
  than $5 - 3.8P$ 
  (mm) if $P < 1$ GeV/c (where $P$ is the tracks momentum in GeV/c) and
  less than 1.2 mm for tracks with $P > 1$ GeV/c. This DBCD cut is
  common in the more sophisticated
  CLEO II/II.5 analyses. It performs better than a simple DBCD $<$
  5 mm cut, 
  because it takes into account the fact that traks with higher
  momentum have a better measurement of DBCD since they scatter less.

$\bullet$ There is exactly one ``good shower'', there can be any
number of showers that are not ``good showers'' but these are not
used in the analysis. We define a ``good
shower'' as an unmatched shower with energy $> 4$ GeV.

These topological cuts are  about 85\% efficient for generated signal
events that have passed the triggers and have been classified as
hardGam.

The overall skim efficiency is between 60-65\%, depending on the mode
(see Table~\ref{Table:table_eff}).

\subsection{Analysis Cuts}

After our skim we call any cuts we make ``analysis cuts''. These cuts are
done at analysis time and are mode dependent. As a convention, and
unless otherwise stated, efficiencies 
for individual analysis cuts are reported relative to the 
events in the skim (not relative to the events generated).

\subsubsection{4-momentum Cut}
All fully reconstructed events should have the 4-momentum of the
$e^+ e^-$ system. This constraint is usually implemented with a
simple two-dimensional $\Delta$E-p box cut, where $\Delta$E is the
difference between the reconstructed energy for the event and the
colliding $e^+ e^-$ energy ($E_{CM}$), and p is the magnitude of the
reconstructed total momentum for the
event. Typical values for these cuts are $-0.03 < \Delta
E/E_{CM} <0.02$ and $p < 150 {\ \rm MeV/c}$ (taken from ~\cite{cleo2}).

\begin{figure}[ht]
\begin{center}
\epsfig{bbllx=66,bblly=150,bburx=580,bbury=600,width=4.5in,clip=,file=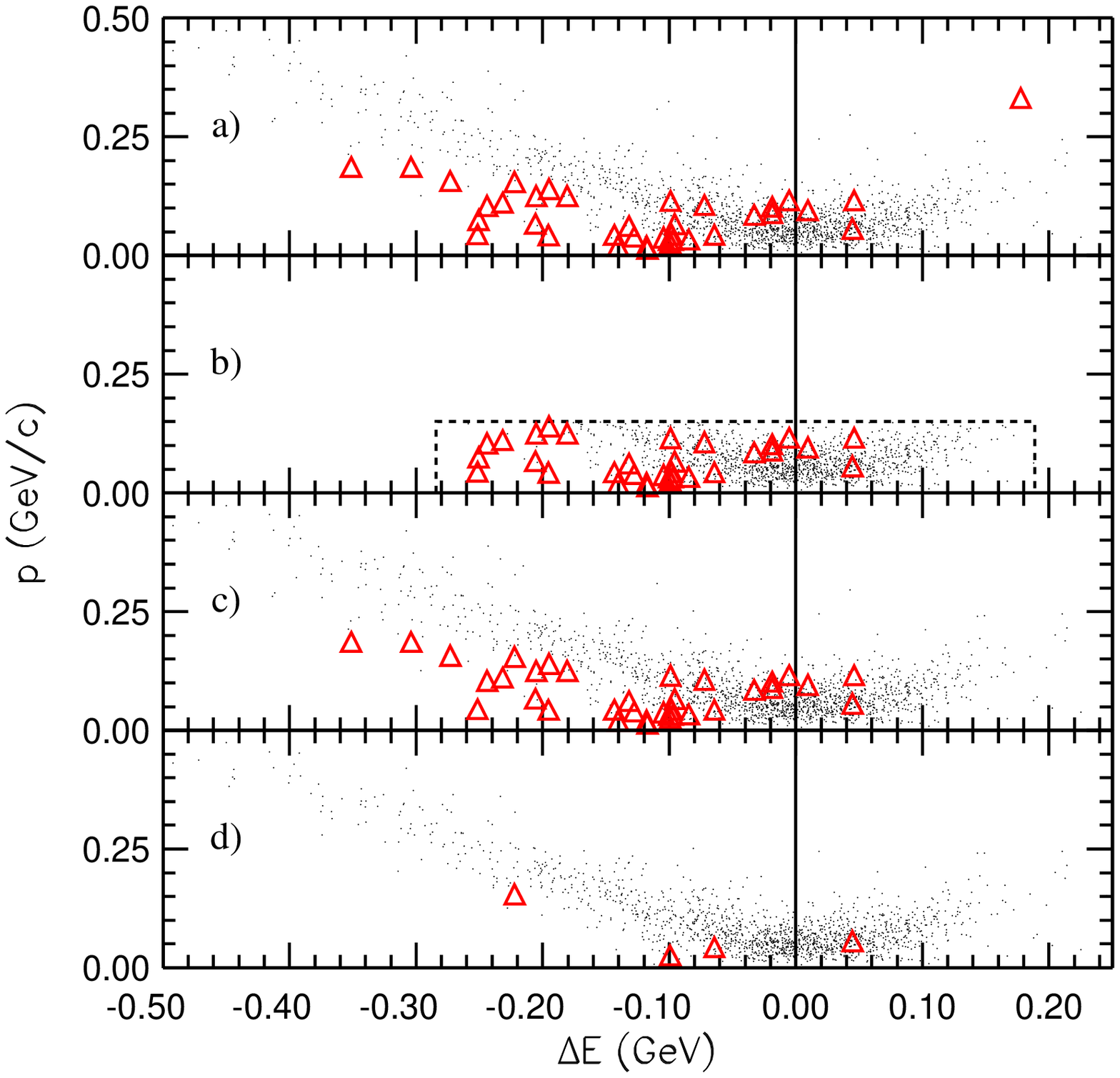}
\end{center}
\caption{Distributions for different 4-momentum cuts. Signal MC events ($\Upsilon(1S) \to \gamma
\pi^+ \pi^-$) are represented by the
black dots, and ``background'' MC events ($e^+e^- \to \gamma
\phi,$ $\phi \to  K^+ K^-$) appear as (red) triangles. Plot a) has no
cuts. Plot b) has the old 4-momentum box-cut. Plot c) has
the new 4-momentum cuts. Plot d) has the new
4-momentum cuts, and also a cut defined as
$\chi^2_{E-p}(\pi)-\chi^2_{E-p}(K) < 0$. The particle ID
potential of the newly available $\chi^2_{E-p}(h)$ is
evident.}\label{Figure:CompareEp}
\end{figure}

The traditional $\Delta$E-p box cut is somewhat useful. However, it
does not take into account the correlation
between the measured energy and momentum. Indeed, the signal lies in
diagonal bands in the $\Delta$E-p plane, making a box-shaped cut not
optimal (see Figures ~\ref{Figure:CompareEp}a and ~\ref{Figure:CompareEp}b).

We use an alternative approach to the 4-momentum cut. After a simple
substitution, $E_{\gamma} = p_{\gamma} = E_{CM}-E_{h^+h^-}$ (where
$E_{\gamma}$ is the photon's energy, $p_{\gamma}$ is the magnitude of
the photons momentum, $E_{CM}$ is the energy of the $e^+e^-$ system,
and  $E_{h^+h^-}$ is the energy of the hadron pair), we can write
the E-p conservation equations as:
\begin{equation}
	\vec{p}_{h^+h^-} + (E_{CM}-E_{h^+h^-})\widehat{p}_{\gamma} 
	= \vec{p}_{CM}\label{constraint}
\end{equation}

\noindent
where, $\vec{p}_{h^+h^-}$ is the di-hadron momentum,
$\widehat{p}_{\gamma}$ is the photon's momentum unit vector, and
$\vec{p}_{CM}$ is the momentum of the $e^+e^-$ system (which is a few
MeV because of the crossing angle). Equation~\ref{constraint} is a
3-constraint subset of the 4-momentum 
constraint and has the convenient property of avoiding the use of the
measured photon's energy, which has non-Gaussian asymmetric errors.
It is important to notice that Equation~\ref{constraint} contains the
di-hadron energy, therefore, it can help discriminate between the various
particle hypotheses.

We proceed as follows. After vertexing the hadron pair using the beam
spot with its error matrix, we
calculate the photon's direction from the hadron pair's vertex and the
shower position. We then fit the event to the 3
constraints expressed in Equation (1) using the techniques outlined
in~\cite{fitting} and cut on the $\chi^2$ of the
3-constraint fit, $\chi^2_{E-p}(h) < 100$. To complete the 4-momentum
requirement, we calculate $\Delta E(h) = E_{h^+h^-} + E_\gamma -
E_{CM}$, where $ E_{h^+h^-}$ is the updated di-hadron energy after
the constraint, and $E_\gamma$ is the measured photon's energy, and
require $-0.050*E_{CM} < \Delta E(h) < 0.025*E_{CM}$. Furthermore, we now
have available  $\chi^2_{E-p}(h)$ differences between different particle
hypotheses, which help in particle identification (ID). This is discussed in more detail in
the next section. Figure \ref{Figure:CompareEp} compares the
performance of the old and new approaches to the E-p cut.

At this point it is a good idea to check that the 4-momentum cut
rejects background events that make it through our skim cuts (see
Section 3.1). These events typically have one high energy shower, two
tracks, and (an) additional element(s). One such background is
$\Upsilon(1S) \to 
\gamma\pi^+\pi^-\pi^0$. Out of 25000 $\Upsilon(1S) \to \gamma
\eta$, with $\eta \to \pi^+\pi^-\pi^0$ MC events~\footnote{Thanks to
Vijay Potlia for generating these events} only 4 survive our
4-momentum cut. We conclude that our 
4-momentum cut is good at rejecting background events that pass the
skim cuts but have additional elements such as an extra photon,
$\pi^0$, pair of tracks, etc.

\subsubsection{Hadron Separation}

We define $\Delta \chi^2_{ID}(h_1-h_2)$ for the particle
hypotheses $h_1$ and $h_2$ of our charged track pair (e.g. $\Delta
\chi^2_{ID}(\pi - K)$) as follows,
\begin{equation}
\begin{tabular}{ll}
$\Delta \chi^2_{ID}(h_1-h_2) = $ & $
\sigma^2_{dE/dX}(h_1^+)-\sigma^2_{dE/dX}(h_2^+)+\sigma^2_{dE/dX}(h_1^-)-\sigma^2_{dE/dX}(h_2^-)$\\
&$-2\log($$\cal {L}$$_{RICH}(h_1^+))+2\log($$\cal{L}$$_{RICH}(h_2^+))$\\
&$-2\log($$\cal {L}$$_{RICH}(h_1^-))+2\log($$\cal{L}$
$_{RICH}(h_2^-))$.
\end{tabular}
\end{equation}
where the idea is to combine the dE/dX and RICH information into one
number. Pairs of tracks with  $\Delta \chi^2_{ID}(h_1-h_2) < 0 $ are
more likely to be of type $h_1$ than of type $h_2$.

In practice, we only add a tracks RICH
information if its momentum is above the Cherenkov radiation threshold
for both mass hypotheses and there are at least 3 photons within
$3\sigma$ of the Cherenkov angle for at least one of the mass
hypothesis. We also require that both hypotheses were actually
analyzed by RICH during pass2.

In addition to RICH and dE/dX, and as hinted in the previous section,
the difference in $\chi^2_{E-p}(h)$ from 
the constraint expressed in Equation~\ref{constraint} can help the particle ID
(see Figures~\ref{Figure:CompareEp}c and~\ref{Figure:CompareEp}d). We define,
\begin{equation}
\Delta \chi^2_{E-p}(h_1-h_2) = \chi^2_{E-p}(h_1)-\chi^2_{E-p}(h_2).
\end{equation}

\noindent
Events with $\Delta \chi^2_{E-p}(h_1-h_2) < 0$ are more likely to be of
type $h_1$ than of type $h_2$.

In this analysis, to select $h_1$ and reject $h_2$ the
default cut is simply $\Delta \chi^2_{ID}(h_1-h_2) < 0$. This simple
cut is highly efficient, has low fake rates, and is sometimes
sufficient. However, out of the six possible cases when one hadron
fakes another, there are three important cases where it pays off to
also use  $\Delta\chi^2_{E-p}(h_1-h_2)$ together with  $\Delta
\chi^2_{ID}(h_1-h_2)$ in an optimal way;

\begin{enumerate}
\item $\pi$ background to $K$. This background comes from the
    continuum process  $e^+e^- \to \gamma \rho,$ $\rho \to  \pi^+
    \pi^-$.

\item $\pi$ background to $p$. Again, this background comes from the
    continuum process $e^+e^- \to \gamma \rho,$ $\rho \to  \pi^+
    \pi^-$.

\item $K$ background to $p$. This background comes from the continuum
    process $e^+e^- \to \gamma\phi,$ $\phi \to  K^+ K^-$.
\end{enumerate}

In other words, the important cases occur when the lighter mass hypothesis
fakes the heavier mass hypothesis.

Mathematically, one would expect that simply adding both $\Delta\chi^2_{ID}$
and $\Delta\chi^2_{E-p}$
together (like we just did when combining RICH and dE/dX), and cutting on
the grand $\Delta\chi^2$ is the way to 
go. Unfortunately, this simple approach fails because of large
non-mathematical tails in the individual $\chi^2_{E-p}$ distributions.

\begin{figure}[ht]
\begin{center}
\epsfig{bbllx=0,bblly=150,bburx=580,bbury=650,width=4.5in,clip=,file=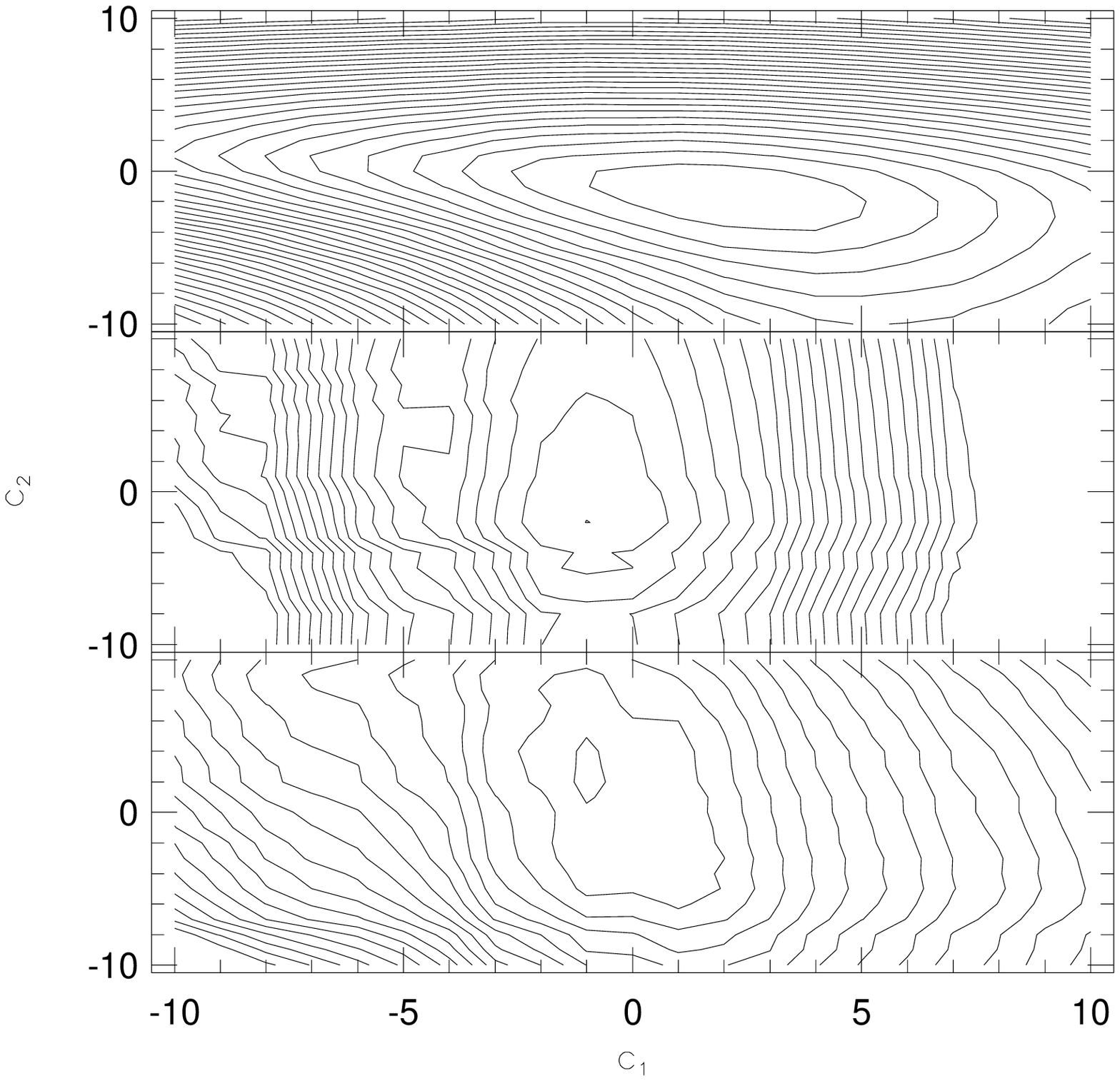}
\end{center}
\caption{Contour plots of $F(c_1,c_2)$ as defined in
Equation~\ref{Equation:F} for
different signals and backgrounds from flat MC. Top has a $K$ signal
and a $\pi$ 
background. Center has a $p$ signal and a $\pi$ background. Bottom has
a $p$ signal and a $K$ background.}\label{Figure:OptimizingID}
\end{figure}

Instead, for each of these three cases we define the best cut values
$(c_1,c_2)$ in $\Delta \chi^2_{ID} < c_1$ and $\Delta\chi^2_{E-p} <
c_2$ as those that maximize
\begin{equation}
F(c_1,c_2) = \frac{R_e}{\sqrt{R_e + W * R_f}},
\label{Equation:F}
\end{equation}
where $R_e$ ($R_f$) is the efficiency (fake rate) of the particle ID cuts
and $W$ is the rough ratio of the background to signal in the
data sample for each case. Each $W$ can't be known \textit{a priori},
but a rough idea of its value can be obtained by doing a first
iteration of the analysis with, for example, $c_1 = c_2 = 0$. We
use $W = 20$, $W = 60$, and $W = 30$ for cases 1-3
respectively. Figure~\ref{Figure:OptimizingID} shows $F(c_1,c_2)$ for
each case. Table~\ref{Table:table_c1c2} shows the optimized cut values
and their effect on particle ID.

\begin{landscape}
\vspace*{\fill}
 \begin{table}
  \caption {Cut values, efficiencies, and fake rates for $\Delta
\chi^2_{ID} < c_1$ and $\Delta \chi^2_{E-p} < c_2$ in flat MC. The different
$c_1,c_2$ values are chosen so $F$ from Equation~\ref{Equation:F} 
 is maximun. Efficiency and fake rates of each cut are reported relative to events in the skim. Statistical errors in the efficiencies
 are 0.1\% or less. Errors in the fake rates are statisticall and are shown for
completeness only.}
 \begin{center}
 \begin{tabular}{c||cc||ccc||ccc}
\hline
\hline
 & \multicolumn{2}{c||}{Cut Value} & \multicolumn{3}{c||}{Signal efficiency (\%)} & \multicolumn{3}{|c}{Fake rate (\%)} \\
 & $c_1$ & $c_2$ & $\Delta\chi^2_{ID}$ & $\Delta\chi^2_{E-p}$ & Both & $\Delta\chi^2_{ID}$ & $\Delta\chi^2_{E-p}$ & Both \\
 \hline
 \hline
 $\pi$ faking $K$ & 3 & -2 &         94.2 &         95.2  &         89.9 & $        1.73\pm        0.09$ & $        14.5\pm         0.3$ & $       0.31\pm       0.04$ \\
 $\pi$ faking $p$ & -1 & -2 &         99.0 &         99.8 &         98.9 & $       0.89\pm       0.06$ & $       1.03\pm       0.07$ & $       0.03\pm       0.01$ \\
 $K$ faking $p$   & -1 & 3 &         98.8 &         99.3  &         98.1 & $       1.89\pm       0.09$ & $       1.09\pm       0.07$ & $       0.10\pm       0.02$ \\
 \hline
 \hline
 \end{tabular}
 \label{Table:table_c1c2}
 \end{center}
 \end{table}
\vfill
\end{landscape}

\subsubsection{Quantum Electrodynamic Suppression}

Quantum Electrodynamic (QED) background in this analysis comes from
the abundant processes $e^+e^- \to \gamma e^+e^-$ and $e^+e^- \to
\gamma \mu^+\mu^-$. 

To reject $e^+e^- \to \gamma e^+e^-$ we require both tracks to have a matched
shower with energy E such that $|E/p(\pi)-0.95| > 0.1$, and to have $\Delta
\chi^2_{ID}(h-e) < 0$.

To reject $e^+e^- \to \gamma \mu^+\mu^-$ we simply require $DPTHMU <
5$ ($DPTHMU$ was defined in Section 2.2.5) for both tracks in the $K$ and $p$ modes because particle ID cuts
make the pion (and therefore the muon) fake rate small (see Tables
\ref{Table:table_c1c2} and \ref{Table:table_ef}). For the
$\pi$ mode we cannot use particle ID in a practical way because 
muons and pions have similar masses. Instead, to separate 
pions and muons we use a much stronger cut requiring that
both tracks be within the barrel part of the muon chamber
($|\cos(\theta)| < 0.7$), both have $P > 1\mom$ and both have $DPTHMU
< 5$\footnote{Other analyses (for example~\cite{cleo1}) typically
use $DPTHMU < 3$, our CLEO III MC has too many pion
tracks with  $3 < DPTHMU < 5$ and two few tracks with $DPTHMU = 0$.}.

To improve the overall muon suppression cut efficiency with
virtually no increase in muon fakes, we flag an event as ``not
muonic'' if any of the tracks deposit more than 600 MeV in the
CC. This increases the cut efficiency by about 90\% in the $\pi$ mode
and makes the detector more hermetic.

\subsection{Cut Summary, Efficiencies, and Fake Rates}

\begin{figure}
\begin{center}
\epsfig{bbllx=66,bblly=150,bburx=580,bbury=600,
        width=4.5in,clip=,file=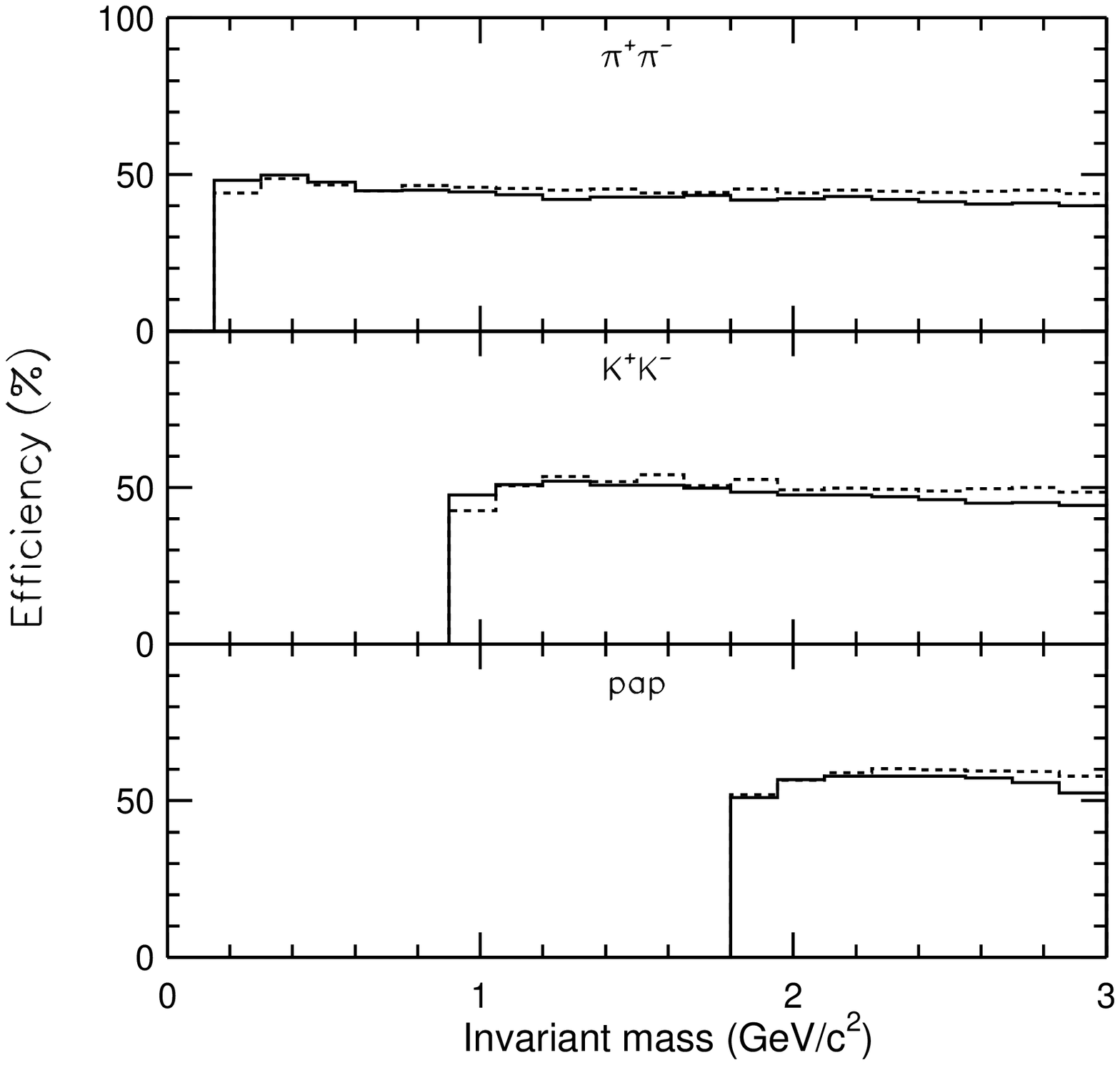}
\end{center}
\caption{Final efficiencies for each mode as a function of invariant
mass for the 1S (solid) and 4S (dashed) Monte Carlo data. The MC was
generated with a flat angular distribution.}\label{Figure:efficiencies}
\end{figure}

\begin{figure}
\begin{center}
\epsfig{bbllx=66,bblly=150,bburx=580,bbury=600,width=4.5in,clip=,file=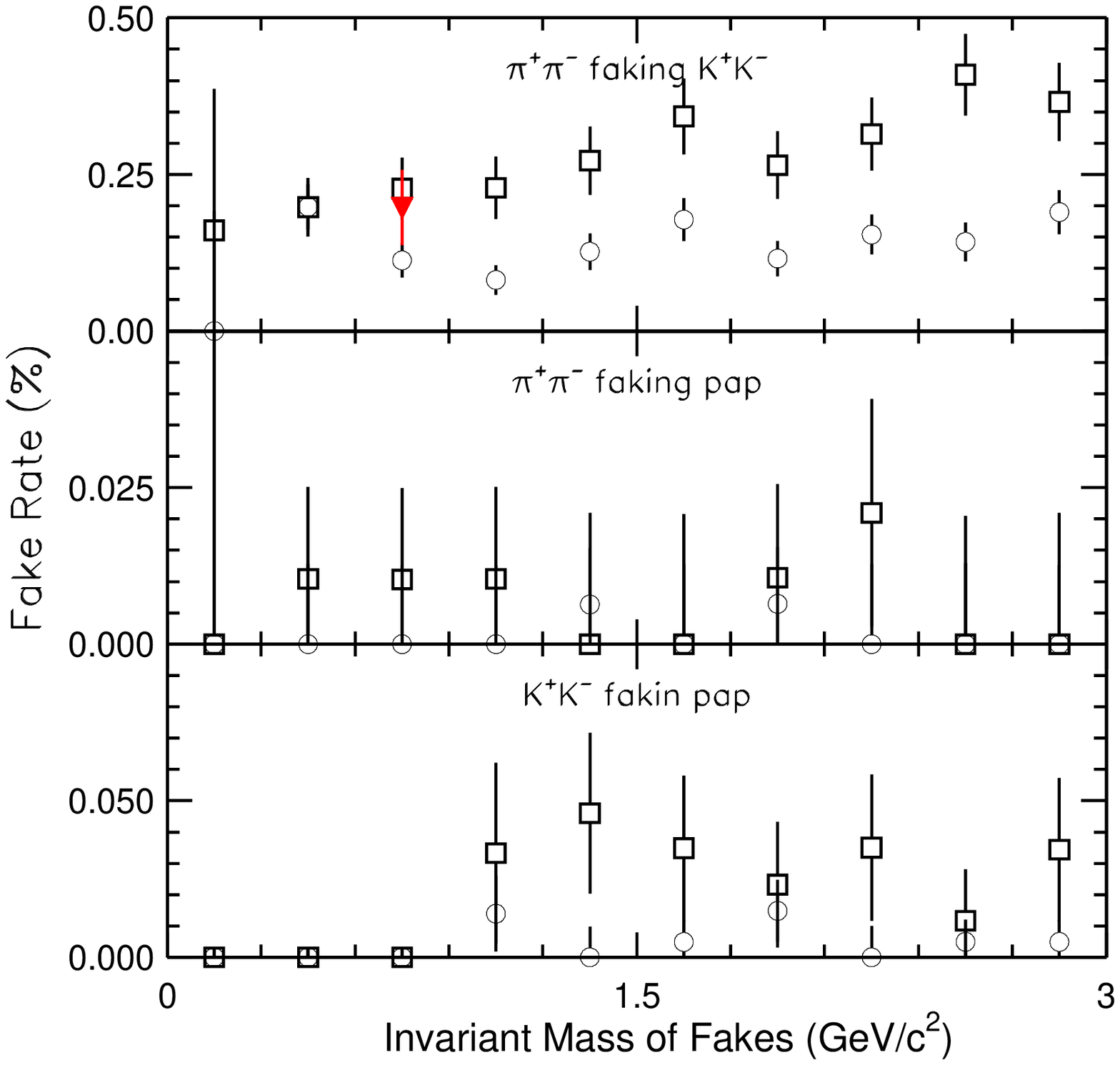}
\end{center}
\caption{Fake rates for 1S (hollow circles) and 4S (hollow squares)
according to flat MC. The (red) downward pointing triangle is obtained
using the $\rho$ peak in $4S$ data.}\label{Figure:fakes}
\end{figure}

Table \ref{Table:cuts} summarizes the cuts used in this analysis.
Figure \ref{Figure:efficiencies} shows the overall Monte Carlo efficiency after
all cuts.
Figure \ref{Figure:fakes} shows the fake rates according to the MC
and the data for different particle ID cuts. The data fake rates and
limit fake rates, which are measured for pions and kaons faking other hadrons,
are calculated from the $\rho$ and $\phi$ peaks in the
continuum (see Figures~\ref{Figure:plot_mass_pipi.ps} and
~\ref{Figure:plot_mass_kk.ps}). Tables~\ref{Table:table_eff} through \ref{Table:table_ef} summarize
the results of 
Figures \ref{Figure:efficiencies} and \ref{Figure:fakes}.

\begin{landscape}
\vspace*{\fill}
\begin{table}
\caption{Cuts used in this analysis.}\label{Table:cuts}
\begin{center}
\begin{tabular}{p{5cm}|p{11cm}}
\hline
\hline
Motivation& Definition for $\Upsilon(1S) \to \gamma h^+ h^-$ (where
$h = \pi, K$, or $p$) \\
\hline
\hline
\multicolumn{2}{c}{Skim cuts}\\
\hline
Data acquisition& Event must pass hardware (Level1) and software
triggers (pass2) and be of type hardGam \\ 
\hline
Topological & There are only two good tracks and only one isolated
shower with $E > 4\gev$ \\
\hline
\hline
\multicolumn{2}{c}{Analysis cuts}\\
\hline
Reconstructed event must have 4-momentum of the center-of-mass system &
$\chi^2_{E-p}(h) < 100.0$  and $-0.050 < \Delta E(h)/E_{CM} <
0.025$ \\
\hline
Hadron separation & Default is, $\Delta \chi^2_{ID} < 0$. The three
cases with a large fake population because of continuum processes use
a simultaneously optimized cut on $\Delta \chi^2_{ID}$ and $\Delta
\chi^2_{E-p}$, and are summarized in Table~\ref{Table:table_c1c2}\\
\hline
QED background \newline $e^+e^- \to \gamma e^+e^-$ & Both tracks have
a matched shower energy E that satisfies
$|E/p(\pi)-0.95| > 0.1$, and $\Delta\chi^2_{ID}(h-e) < 0$ \\
\hline
QED background \newline $e^+e^- \to \gamma \mu^+\mu^-$ &  For $h = K,
p$ both tracks have $DPTHMU < 5$. For $h = \pi$
(At least one track has a matched shower energy $>$ 600 MeV) or ((
Both tracks have $cos(\Theta) < 0.7$ and $P > 1\mom$) and (both tracks
have $DPTHMU < 5$))\\
\hline
\hline
\end{tabular}
\end{center}
\end{table}
\vfill
\end{landscape}

 \begin{table}
 \caption{Efficiencies in \% for cuts
 (as outlined in Table \ref{Table:cuts})
 for flat signal MC. Efficiencies in the second group are
 reported relative to the number of candidates that make it to the skim.
 The third part of the table shows the overall reconstruction efficiency.
 Statisticall errors are 0.1\% or less.}
 \begin{center}
 \begin{tabular}{c|c|c|c}
 \hline
 \hline
 Skim cuts &$\pi$&$K$&$p$ \\
 \hline
 Hardware Trigger &     89.7 &     89.0 &     90.9 \\
 Software Trigger &     96.5 &     96.2 &     96.6 \\
 hardGam        &     78.4 &     79.9 &     76.5  \\
 Topological $(n_{tracks} = 2\ \& \ E_{\gamma} >  4 GeV)$
 &     73.3  &     67.6  &     72.3 \\
 \hline
 Overall Skim efficiency &     64.3 &     60.3 &     62.4 \\
 \hline
 \hline
 Analysis cuts &$\pi$&$K$&$p$ \\
 \hline
 4-momentum &     98.6 &     98.5 &      99.0 \\
 QED $e^+e^- \to \gamma e^+e^-$ suppresion &     93.9 &      87.4 &      93.1 \\
 QED $\mu^+\mu^- \to \gamma \mu^+\mu^-$ suppresion  &     74.7 &     93.0 &     98.3 \\
 Hadron separation &     97.1 &      89.0 &      97.8\\
 \hline
 Overall analysis efficiency &      66.9 &      79.1 &      89.1 \\
 \hline
 \hline
 Overall reconstruction efficiency &      43.0 &      47.6 &      55.6 \\
 \hline
 \end{tabular}
 \end{center}
 \label{Table:table_eff}
 \end{table}

 \begin{table}
 \caption{Efficiencies in \% for cuts
 (as outlined in Table \ref{Table:cuts})
 for flat 4S MC. Efficiencies in the second group are
 reported relative to the number of candidates that make it to the skim.
 The third part of the table shows the overall reconstruction efficiency.
 Statisticall errors are 0.1\% or less.}
 \begin{center}
 \begin{tabular}{c|c|c|c}
 \hline
 \hline
 Skim cuts &$\pi$&$K$&$p$ \\
 \hline
 Hardware Trigger &     88.4 &     88.0 &     89.0 \\
 Software Trigger &     94.6 &     94.6 &     94.8 \\
 hardGam        &     79.2 &     80.5 &     77.9  \\
 Topological $(n_{tracks} = 2\ \& \ E_{\gamma} >  4 GeV)$
 &     75.2  &     69.5  &     75.8 \\
 \hline
 Overall Skim efficiency &     66.9 &     62.7 &     66.6 \\
 \hline
 \hline
 Analysis cuts &$\pi$&$K$&$p$ \\
 \hline
 4-momentum &     97.1 &     97.0 &      97.4 \\
 QED $e^+e^- \to \gamma e^+e^-$ suppresion &     93.4 &      89.1 &      93.2 \\
 QED $\mu^+\mu^- \to \gamma \mu^+\mu^-$ suppresion  &     75.8 &     92.5 &     98.1 \\
 Hadron separation &     98.0 &      91.4 &      96.6\\
 \hline
 Overall analysis efficiency &      67.4 &      80.2 &      87.3 \\
 \hline
 \hline
 Overall reconstruction efficiency &      45.0 &      50.3 &      58.2 \\
 \hline
 \end{tabular}
 \label{Table:table_eff4}
 \end{center}
 \end{table}

 \begin{table}
 \begin{center}
 \caption{Final efficiencies and fake rates after all cuts in \%.
 Statistical errors for efficiencies are 0.1\% or less.
 Statistical errors for fake MC rates are shown for completeness only.
 The 1S DATA corresponds to the 1S off resonance data sample.
 The MC was generated flat.}
 \begin{tabular}{c|c|c|c}
 \hline
 \hline
 & $\pi$ cuts & $K$ cuts & $p$ cuts \\
 \hline
 \hline
 1S $\pi$ MC &     43.0 & $        0.14\pm        0.01$ & $ <        0.007 $ \\
 4S $\pi$ MC &     45.0 & $        0.29\pm        0.02$ & $ <         0.01$ \\
 1S $\rho$ DATA & -- & $<$          1.2  & $<$          0.2 \\
 4S $\rho$ DATA & -- & $        0.20\pm        0.06$  & $<$        0.06 \\
 \hline
 1S $K$ MC & $        1.27\pm        0.03$ &     47.6  & $<$ 0.02 \\
 4S $K$ MC & $        0.92\pm        0.04$ &     50.3  & $<$ 0.06 \\
 1S $\phi$ DATA & $<$          3.8 & --  & $<$          2.0 \\
 4S $\phi$ DATA & $        4.14 \pm         0.69$ & --  & $<$          0.4 \\
 \hline
 1S $p$ MC & $        0.08\pm        0.02$ & $        0.34\pm        0.04$ &     55.6 \\
 4S $p$ MC & $        0.05\pm        0.01$ & $        0.37\pm        0.01$  &     58.2 \\
 \hline
 \hline
 \end{tabular}
 \label{Table:table_ef} 
 \end{center}
 \end{table}

\clearpage
\chapter{EXCLUSIVE RADIATIVE DECAY $\Upsilon(1S) \rightarrow \gamma
\pi^+\pi^-$}
\begin{figure}[ht]
\begin{center}
\epsfig{bbllx=66,bblly=100,bburx=580,bbury=775,width=4.5in,clip=,file=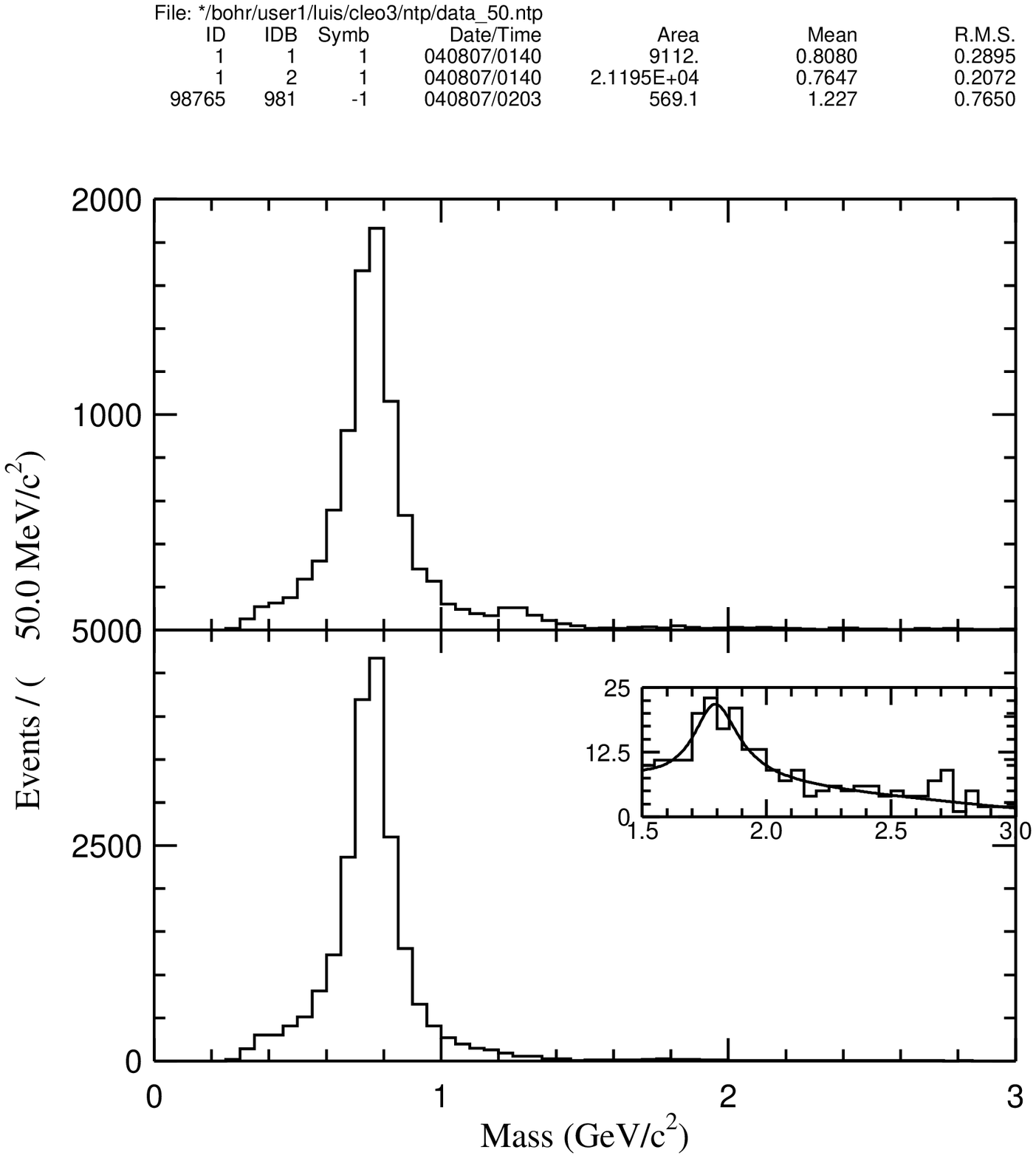}
\end{center}
\caption{Invariant mass of $\pi^+ \pi^-$ for 1S (top) and 4S
(bottom) data. For the 4S data we show a blow-up of the mass region
$1.5-3\mass$ where the $\rho^{*}$ can be
seen.}\label{Figure:plot_mass_pipi.ps} 
\end{figure}

In Figure \ref{Figure:plot_mass_pipi.ps} the $\pi^+\pi^-$ invariant
mass plot is shown for both
1S and 4S data. Figure
\ref{Figure:pipiplot_1.ps} shows the continuum subtracted $\pi^+\pi^-$
invariant mass distribution (as defined in Section 3.2.1) with the most
likely statistical fit overlayed (which is described in the next
section). The number of
events within $1\Gamma$ of the
$\rho$ region ($0.62-0.92\mass$) left after the continuum
subtraction is $200\pm300$, and 50 of these belong to the $f_0(980)$
low-mass tail. 

\section{Robustness of The Mass Distribution}

In~\cite{cleo6} the decay $\Upsilon(1$S$) \to \gamma \pi^0\pi^0$ is
analyzed, and it is shown how the analysis stream warps the shape of the
reconstructed resonance. This effect, which arises
because of the particular $\pi^0$ behavior, raises problems when
fitting the invariant mass distribution.

In Section 3.2.1 we claimed that if the data and MC had the same
$\theta_{\gamma}$ and $\theta_{\pi}$ distributions, the fit to the continuum
subtracted invariant mass distribution automatically gives the
correct efficiency corrected number of events. 

Here we test this claim. To this end, we generate 10000
$\Upsilon(1$S$) \to \gamma f_2(1270)$, with $f_2(1270) \to
\pi^+\pi^-$ with flat $\theta_{\gamma}$ and $\theta_{\pi}$
distributions. We treat
this MC as data and carry out the first two steps of the continuum
subtraction process. The resulting peak has a mass of $1.278 \pm
0.002\mass$ and a width of $0.193\pm0.006\mass$, consistent with the
generated mass and width of $1.275\mass$ and $0.185\mass$. More
importantly, the number of reconstructed events from the fit is
$10040\pm180$, which is consistent with the number of generated events.
 
\begin{figure}[h]
\begin{center}
\epsfig{bbllx=66,bblly=100,bburx=580,bbury=650,width=2.5in,clip=,file=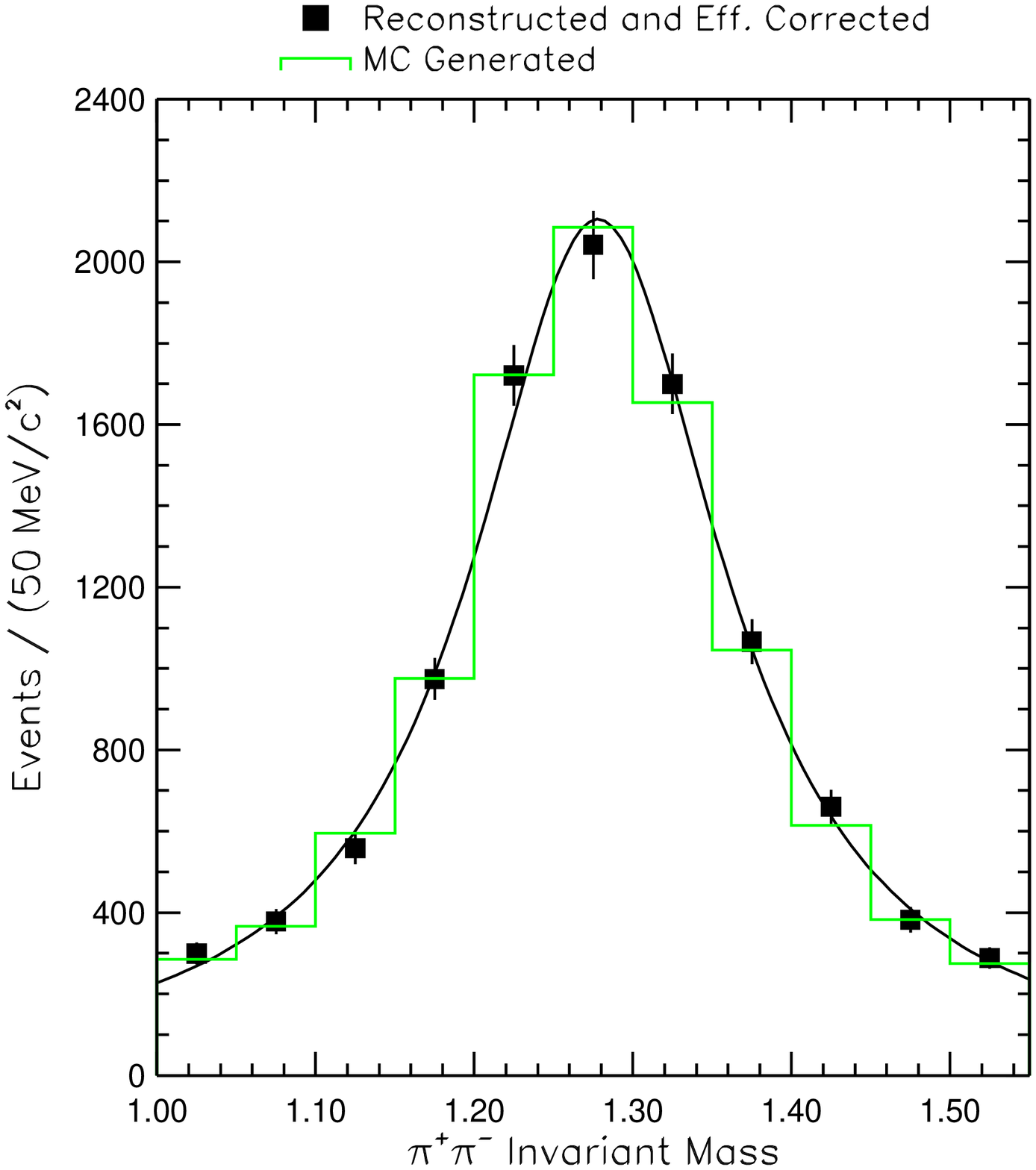}
\end{center}
\caption{Reconstructed events and efficiency corrected events, a fit
to them, and the generated events for $\Upsilon(1$S$) \to \gamma
f_2(1270)$, with $f_2(1270) \to \pi^+\pi^-$ with flat
$\theta_{\gamma}$ and $\theta_{\pi}$ distributions.}\label{Figure:ferguson3.ps}
\end{figure}

Figure~\ref{Figure:ferguson3.ps} shows the reconstructed and
efficiency corrected events, a fit to them, and the generated events.

We conclude that there is no warping of the mass distribution, and
that the analysis stream behaves like we expect when obtaining the
efficiency corrected number of events from data.

\section{Statistical Fit of the Invariant Mass Distribution}

The results of this section are summarized in
Table~\ref{Table:table_pipi}.

Figure \ref{Figure:pipiplot_1.ps} shows possible signals for for
$\Upsilon(1S) \rightarrow \gamma 
f_0(980)$, $\Upsilon(1S) \rightarrow
\gamma f_2(1270)$ and  
$\Upsilon(1S) \rightarrow \gamma f_4(2050)$. To determine the number
of events in each signal
we fit the invariant $\pi^+\pi^-$
mass continuum subtracted distribution with three spin-dependent, relativistic
Breit-Wigner line shapes. The masses and widths are allowed to float,
except for the width of the $f_4(2050)$ which has a very large
error if allowed to float and is set to its PDG value. The PDG values\cite{pdg} for the mass and
width of the three resonances are
$m_{f_0(980)} = 980 \pm 10\miss$, $\Gamma_{f_0(980)} = 70 \pm
15\miss$, 
 $m_{f_2(1270)} = 1275.4 \pm 1.2\miss$, $\Gamma_{f_2(1270)} =
185.1^{+ 3.4}_{- 2.6}\miss$,
$m_{f_4(2050)} = 2034 \pm 11\miss$, and $\Gamma_{f_4(2050)} = 222 \pm
19 \miss$. 

To measure the statistical significance of each signal we do
multiple $\chi^2$ fits fixing the signal area to different values
while letting the mass and width of the signal 
whose significance is being measured to float within
$2\sigma$ of the PDG values. At the same time the rest of the fit
parameters are fixed to the values that originally minimized the $\chi^2$.
We assign each of these multiple fits a probability proportional to
$e^{-\chi^2/2}$ and then normalize. We calculate the chance of the
signal being due to a random fluctuation by adding the normalized
probabilities for the fits with a 
negative or 0 signal. This method fails for the highly significant $f_2(1270)$
signal because the $e^{-\chi^2/2}$ value of fits with negative or 0 $f_2(1270)$ signal is of the
order of $10^{-54}$ and our computing software can only handle
numbers as small as $10^{-45}$. For completeness we state that the
significance of this signal is $< 10^{-45}$.

\begin{figure}[ht]
\begin{center}
\epsfig{bbllx=66,bblly=100,bburx=580,bbury=775,width=4.5in,clip=,file=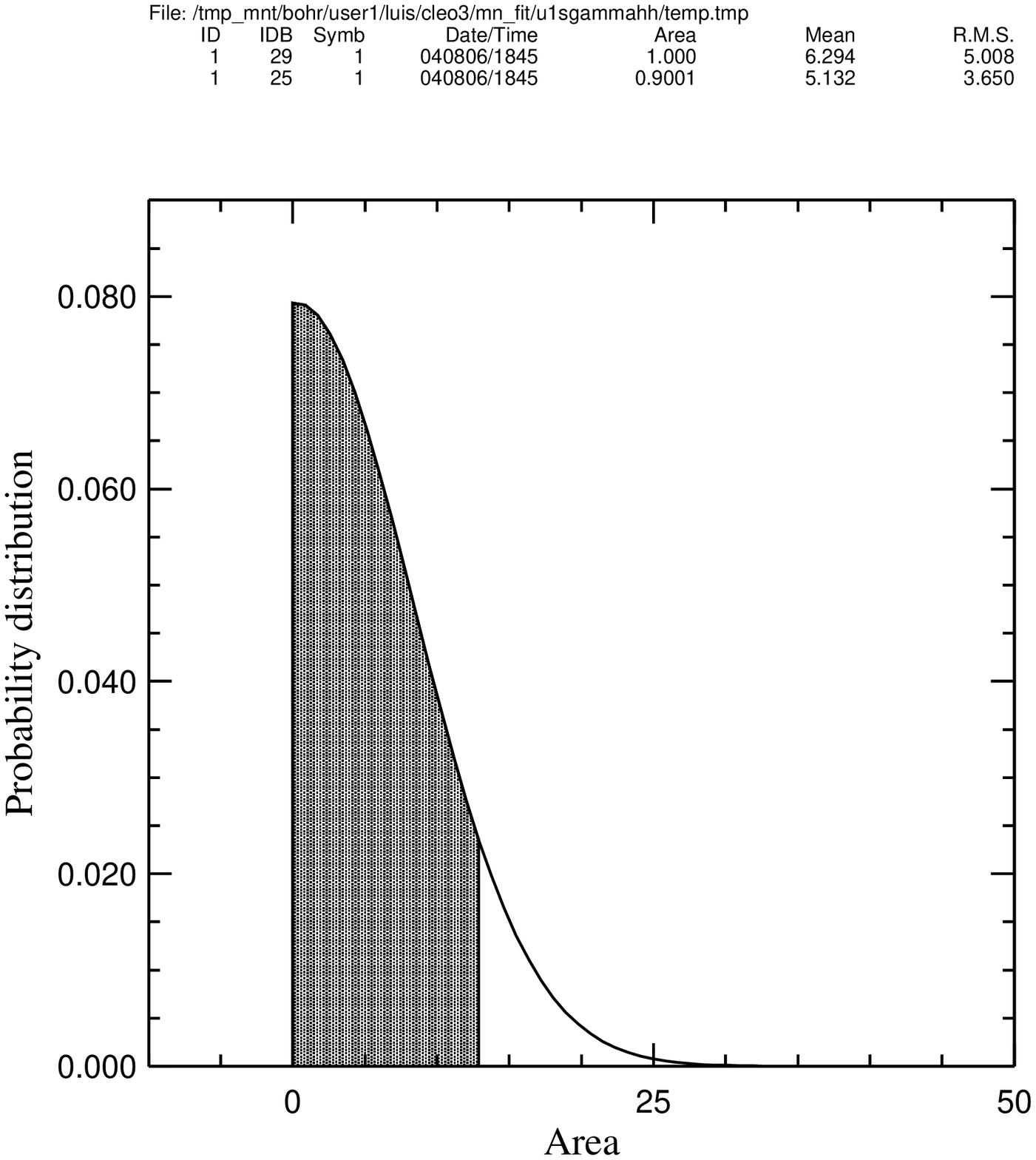}
\end{center}
\caption{Normalized probability distribution for different $f_J(2220)
\to \pi^+ \pi^-$
signal areas. The shaded area spans 90\% of the
probability.}\label{Figure:fj_prob.ps} 
\end{figure}

To measure the upper limit for the $f_J(2220)$ we also do multiple
$\chi^2$ fits for different fixed signal values, while keeping its mass and
width constant at $m_{f_J(2220)} = 2.234\mass$ and $\Gamma_{f_J(2220)}
= 17\ \rm  MeV/c^2$ as
in~\cite{glueball}. The resulting probability plot is shown in 
Figure~\ref{Figure:fj_prob.ps}.

We find clear evidence for the $f_2(1270)$,
evidence for the $f_0(980)$ and weak evidence for the $f_4(2050)$. We
also put a 90\% confidence level upper limit on $f_J(2220)$
production. Fit results are shown in Table~\ref{Table:table_pipi}.

 \begin{table}[ht]
 \begin{center}
 \caption{Results for $\Upsilon(1S) \rightarrow \gamma
 \pi^+\pi^-$. The branching fractions of $f_2(1270)$ and $f_4(2050)$ are
 taken from the PDG~\cite{pdg}. Errors shown are satistical only.}
 \begin{tabular}{c|c|c|c}
 \hline
 \hline
 Mode & Area & B.F. $(10^{{-5}})$ & Significance \\
 \hline
 $\gamma f_0(980), f_0(980) \rightarrow \pi^+\pi^-$& $    340^{+
 140}_{    -130}$  & $    1.6^{+    0.7}_{    -0.6}$ &  $8.3\times10^{-6}$
 (4.3$\sigma$) \\ 
 $\gamma f_2(1270)$& $   1230\pm     100$  & $    10.4\pm    0.8$&
 $<10^{-45}$\ \ ($> 14\sigma$)\\
 $\gamma f_4(2050)$& $     85\pm     30$ & $     3.6\pm     1.3$ &
 $5.2\times10^{-3}$ (2.6$\sigma$)\\
 $\gamma f_J(2220), f_J(2220) \rightarrow \pi^+\pi^-$& $<      13$ & $ < 6.2\times10^{-2}$ & - \\
 \hline
 \end{tabular}
 \label{Table:table_pipi} 
 \end{center}
 \end{table}

\section{Angular Distribution of The Signal}

The helicity angle distributions of $\theta_{\pi}$ and
$\theta{\gamma}$ are defined and described in the appendix. In this
section we first obtain the helicity angular distributions of the $f_0(980)$,
$f_2(980)$, and $f_4(2050)$ and then fit them to the predictions of the
helicity formalism.

In practical terms, obtaining the helicity angle distribution of a
particular resonance from data consists of two steps. First, we choose an
invariant mass interval around the resonance peak to select events
from the resonance and obtain a helicity angular distribution which
has both signal and background events. Second, we subtract the
contribution to the helicity angular distribution of the background
events in the chosen mass interval to obtain what we want; the helicity
angle distribution of the resonance.

Choosing the mass interval is not a trivial thing. If its too wide
there will be too much background, and if its to narrow there will
not be enough signal. To choose the optimum mass interval we need to know
how much signal and background we are selecting. Therefore, the two
steps described in the previous paragraph are related. How we deal with
this is revealed in the next two sections.

\subsection{Optimum Mass Interval}
 
As described above, the first step in obtaining the helicity angle
distribution for a resonance is choosing an invariant mass interval to select  events from
such a resonance. A standard $1
\Gamma$ (which corresponds to $1.6\sigma$ for a spin 0 Breit-Wigner) 
cut around the mean mass of the resonance can be chosen as a
``standard'' interval. We could 
proceed this way, but in our case because of
the large subtractions involved when obtaining the angular
distribution, a considerable increase statistical significance of each
bin in the helicity angular
distribution can be achieved by choosing the mass interval carefully
(see the last column of Table~\ref{Table:intervalresults}).

Let's consider for example the $f_0(980)$ angular distribution. We
begin with the $\Upsilon(1S)$ and the $\Upsilon(4S)$ efficiency
corrected distributions. Before
the continuum subtraction each bin in the $f_0(980)$ angular
distribution has contributions from the high end 
mass tail of the $\rho$ and the low mass tail of the
$f_2(1270)$. In order to get the final angular distribution, both of
these contributions are taken away by first doing a continuum
subtraction using the scale factor $\alpha = 0.404$, and then by
subtracting the $f_2(1270)$ distribution outside the mass interval
scaled by an appropriate scale factor $\beta$ equal to the ratio of
$f_2(1270)$ inside the mass interval and outside of
it~\footnote{Strictly speaking this is correct only to first
order. The $f_2(1270)$ distribution itself has a small contamination
from $f_0(980)$ and $f_4(2050)$. This effect, which we call
cross-contamination, is ignored
in Equation~\ref{Equation:capitalN}. Later, in section 4.2.2, we will
show how to eliminate this small cross-contamination using all the
resonances. The difference in how we actually get the helicity
distributions and Equation~\ref{Equation:capitalN} has an
insignificant effect when calculating the optimal mass
interval}. After all this,
the contributions to the bins in the 
final angular distribution are, 
\begin{equation}
N_{f_0(980)} = n_{f_0(980)} + n_{\rho(\Upsilon(1S))} + n_{f_2(1270)} -
\alpha n_{\rho(\Upsilon(4S))}-\beta n'_{f_2(1270)} \approx  n_{f_0(980)}\label{Equation:capitalN}
\end{equation}
where  $n_{f_0(980)} (n_{f_2(1270)})$ is the number of $f_0(980)
(f_2(1270))$ obtained by integrating the fitted spin-dependent, relativistic
Breit-Wigner function inside the mass interval,
$n_{\rho(\Upsilon(1S))}(n_{\rho(\Upsilon(4S))})$ is the number of $\rho$'s
inside the mass interval from 
continuum events at the $\Upsilon(1S)$($\Upsilon(4S)$) energy, and
$n'_{f_2(1270)}$ is the number of $f_2(1270)$ outside the mass 
interval being used to subtract the contribution of
$f_2(1270)$ to the $f_0(980)$ angular distribution. Following this last
definition, $\beta = n_{f_2(1270)}/n'_{f_2(1270)}$. 

\begin{figure}[ht]
\begin{center}
\epsfig{bbllx=0,bblly=100,bburx=580,bbury=775,width=4.5in,clip=,file=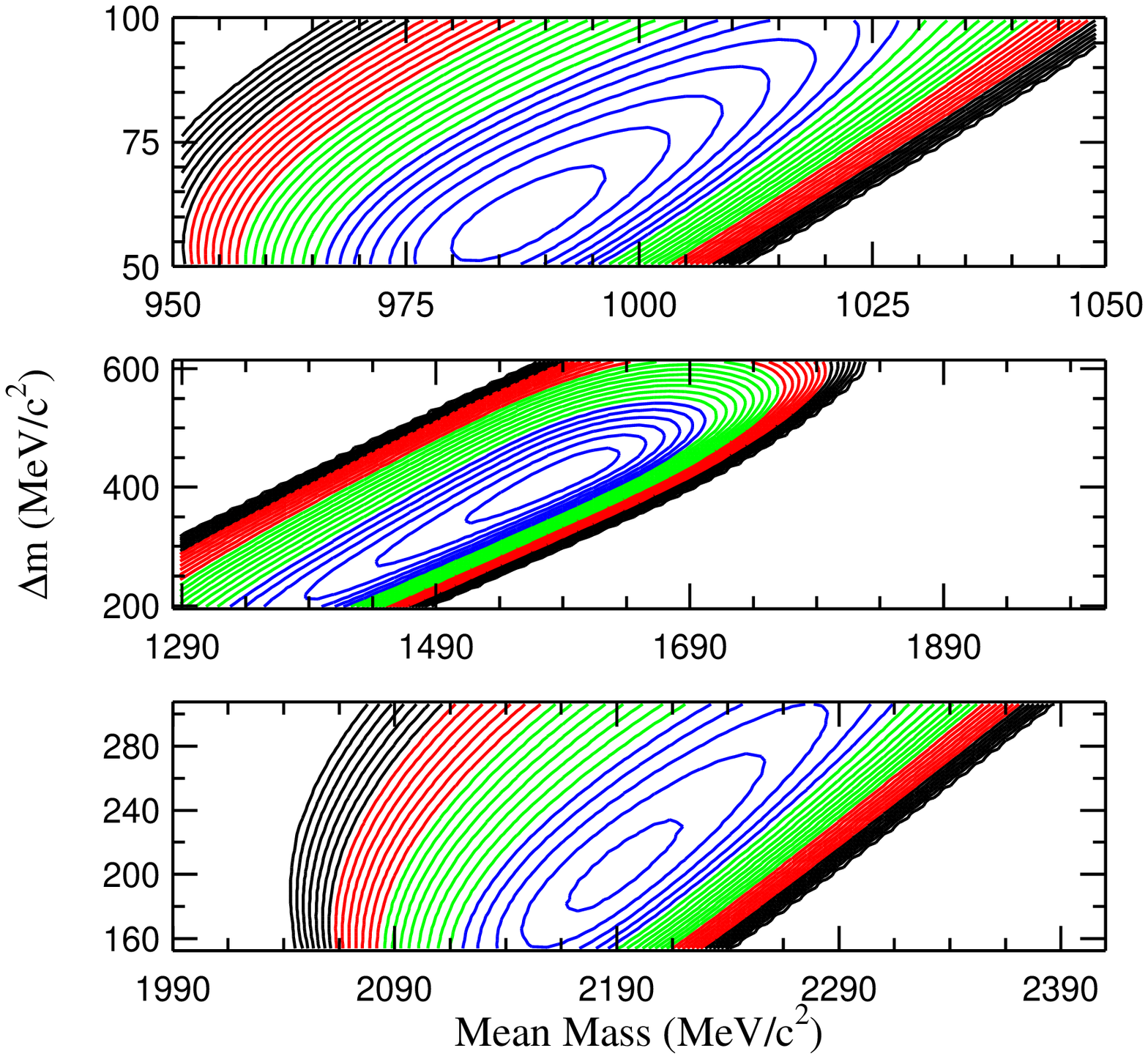}
\end{center}
\caption{Contour plot of the inverse
of the average relative bin-error from background subtractions in the
$f_0(980)$ (top), $f_2(1270)$ (middle), and $f_4(2050)$ (bottom)
angular distribution.}\label{Figure:f0980Interval.ps}
\end{figure}

Each of these terms has an associated error. Assuming that $\Delta n =
\sqrt{n}$, that the efficiency correction has infinite statistics, and
ignoring the errors on the continuum scale factor $\alpha$, the
overall error is, 
\begin{equation*}
\Delta N_{f_0(980)} = \sqrt{n_{f_0(980)} + 
(\alpha+\alpha^2) n_{\rho(\Upsilon(4S))} + (2+\beta+\beta^2) n_{f_2(1270)}}
\end{equation*}
We arrive to the conclusion that the mass interval
$(\bar{m}-{\Delta m},\bar{m}+{\Delta m})$ which
produces the helicity angle distribution with smallest relative
bin-errors is the one that maximizes, 
\begin{equation}
F_{f_0(980)}(\bar{m},\Delta m) = \frac{n_{f_0(980)}}{\sqrt{n_{f_0(980)} + 
(\alpha+\alpha^2) n_{\rho(\Upsilon(4S))} + (2+\beta+\beta^2)
n_{f_2(1270)}}}.
\end{equation}
The plot of $F_{f_0(980)}(\bar{m},\Delta m)$ is shown in
figure~\ref{Figure:f0980Interval.ps}.

This same technique can be applied to the $f_2(1270)$ and $f_4(2050)$
resonances. Results are shown in Table ~\ref{Table:intervalresults}.

\begin{landscape}
\vspace*{\fill}
 \begin{table}[ht]
 \begin{center}
 \caption{Mean masses, widths in$\mass$ and inverse of the average bin
 relative error (F) from background subtractions for the angular
 distribution of different resonances. Standard mean masses and 
 widths, corresponding to 1$\Gamma$, are taken from the fit in
 Figure~\ref{Figure:pipiplot_1.ps}, and are labeled with the 
 subscript ``s'', while those that maximize
 F are labeled with the 
 subscript ``m''. The last column shows the factor by which the
 effective statistics increase.} 
 \begin{tabular}{c||c|c||c|c||c}
 \hline
 \hline	
 Resonance & $(\bar{m}_s,\Delta m_s)$ & $F(m_s,\Delta m_s)$ &
 $(\bar{m}_m,\Delta  m_m)$ & $F(m_m,\Delta m_m)$ &
 $\frac{F^2(m_m,\Delta m_m)}{F^2(m_s,\Delta m_s)}$ \\ 
 \hline
$f_0(980)$ & $(0.970,0.070)$ & 5.7 & $(0.985,0.060)$ & 6.0 & 1.11 \\ 
$f_2(1270)$ & $(1.270,0.120)$ & 18.3 & $(1.590,0.420)$ & 20.6 & 1.27  \\ 
$f_4(2050)$ & $(2.120,0.220)$ & 3.2 & $(2.240,0.250)$ & 3.4 & 1.13 \\ 
 \hline
 \end{tabular}
\label{Table:intervalresults} 
 \end{center}
 \end{table}
\vfill
\end{landscape}

\subsection{Background Subtraction}

The continuum subtraction of the helicity angle distribution is
defined in Section 2.1. The subtraction of the tails from other
resonances requires a closer look.
Let us call the continuum subtracted helicity angle distribution of the
events in the
$f_0(980)$, $f_2(1270)$, and $f_4(2050)$ mass intervals $c_{f_0(980)}$,
$c_{f_2(1270)}$, and $c_{f_4(2050)}$ respectively. Let us call the helicity
angle distribution of the events that come exclusively from the
resonance we are 
trying to select in the same mass intervals (that is, the true helicity
angle distribution of the resonance) $t_{f_0(980)}$,
$t_{f_2(1270)}$, and $t_{f_4(2050)}$ respectively. The continuum subtracted
helicity angle 
distribution of a resonance is being contaminated by the tails of the
other resonances. Keeping this in mind we write~\footnote{Here the
cross contamination between the $f_0(980)$ and the $f_4(2050)$ is
ignored. There is no mathematical problem in including these
contamination terms, but the system of equations would be more
complicated than it needs to be since this type of cross contamination is
negligible.}
\begin{equation} 
\begin{cases}
c_{f_0(980)} = t_{f_0(980)} + \beta t_{f_2(1270)}  \\
c_{f_2(1270)} = t_{f_0(1270)} + \gamma t_{f_0(980)} + \delta t_{f_4(2050)}  \\
c_{f_4(2050)} = t_{f_0(2050)} + \epsilon t_{f_2(1270)}
\end{cases}
\label{Equation:systemlala}
\end{equation}

\begin{figure}[ht]
\begin{center}
\epsfig{bbllx=66,bblly=100,bburx=580,bbury=775,width=4.5in,clip=,file=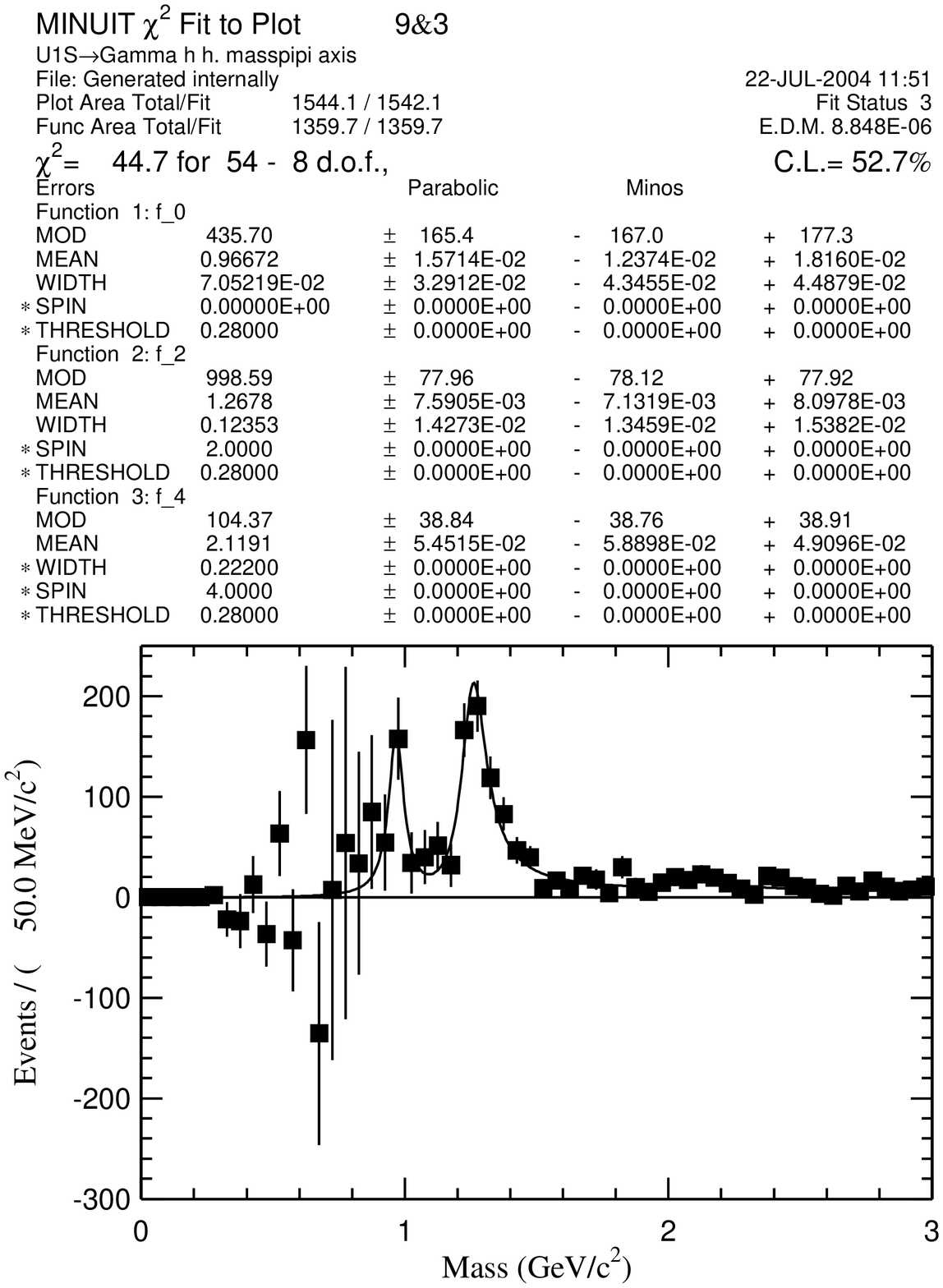}
\end{center}
\caption{Continuum subtracted (as defined in Section 3.2.1) invariant
mass of $\pi^+ \pi^-$ from $\Upsilon(1S) \rightarrow \gamma
\pi^+\pi^-$.}\label{Figure:pipiplot_1.ps}
\end{figure}

Where the small numbers $ \beta,\ \gamma,\ \delta$, and $\epsilon$ are
the ratios of the number of events from a resonance in the mass
interval where the contamination is taking place to the number of
events from the same resonance in the mass interval used to select
it. Using the $(m_m, \Delta m_m)$ values in
Table~\ref{Table:intervalresults} and the fit in
Figure~\ref{Figure:pipiplot_1.ps} we obtain $\beta = 8.5\times10^{-3}$,
$\gamma = 6.0\times10^{-2}$, $\delta = 9.7\times10^{-2}$, and
$\epsilon = 0.13$.

To obtain the background subtracted helicity angular distributions we
simply invert the system of equations expressed
in Equation~\ref{Equation:systemlala}. The solution can be conveniently
expressed as,
\begin{equation} 
\begin{cases}
t_{f_2(1270)} = \frac{1}{1-\beta \gamma - \delta \epsilon}
(c_{f_0(1270)} - \gamma c_{f_0(980)} - \delta c_{f_4(2050)})  \\ 
t_{f_0(980)} = c_{f_0(980)} - \beta t_{f_2(1270)}  \\
t_{f_4(2050)} = c_{f_0(2050)} - \epsilon t_{f_2(1270)}\label{Equation:solutionlala}
\end{cases}
\end{equation}

As a check, if we ignore the second order terms the previous solution becomes,
\begin{equation*} 
\begin{cases}
t_{f_2(1270)} = c_{f_0(1270)} - \gamma c_{f_0(980)} - \delta
c_{f_4(2050)}  \\  
t_{f_0(980)} = c_{f_0(980)} - \beta c_{f_2(1270)}  \\
t_{f_4(2050)} = c_{f_0(2050)} - \epsilon c_{f_2(1270)}
\end{cases}
\end{equation*}
which is indeed is the solution when cross-contamination is ignored.

\subsection{Statistical Fit of the Helicity Angular Distributions}

\begin{figure}[ht]
\begin{center}
\epsfig{bbllx=66,bblly=100,bburx=580,bbury=775,width=4.5in,clip=,file=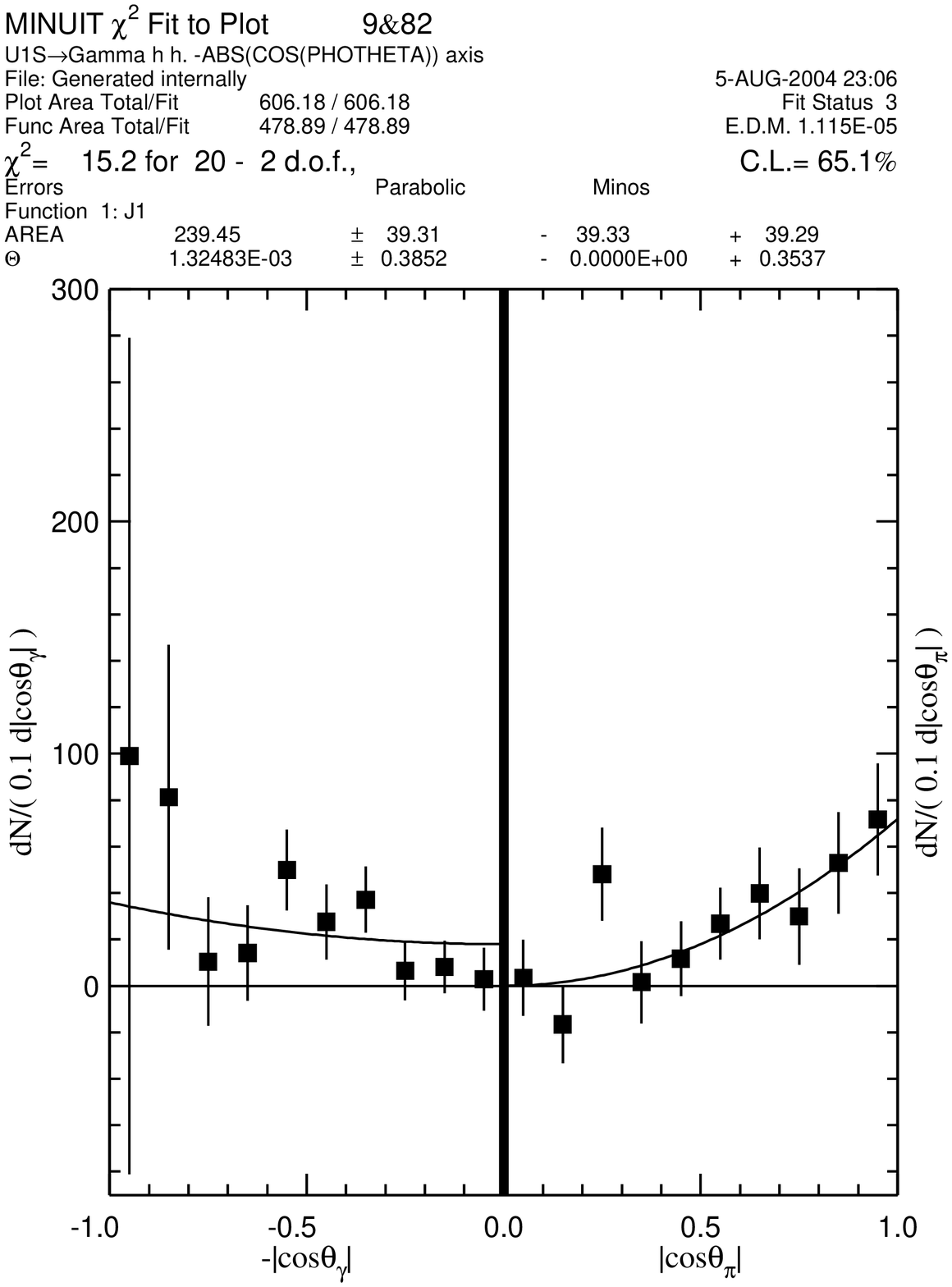}
\end{center}
\caption{Angular distribution for the excess events in the
$f_0(980)$ mass region. The fit corresponds to J =
1.}\label{Figure:pipiplot_ct_cg_1.ps}   
\end{figure}

\begin{figure}[ht]
\begin{center}
\epsfig{bbllx=66,bblly=100,bburx=580,bbury=775,width=4.5in,clip=,file=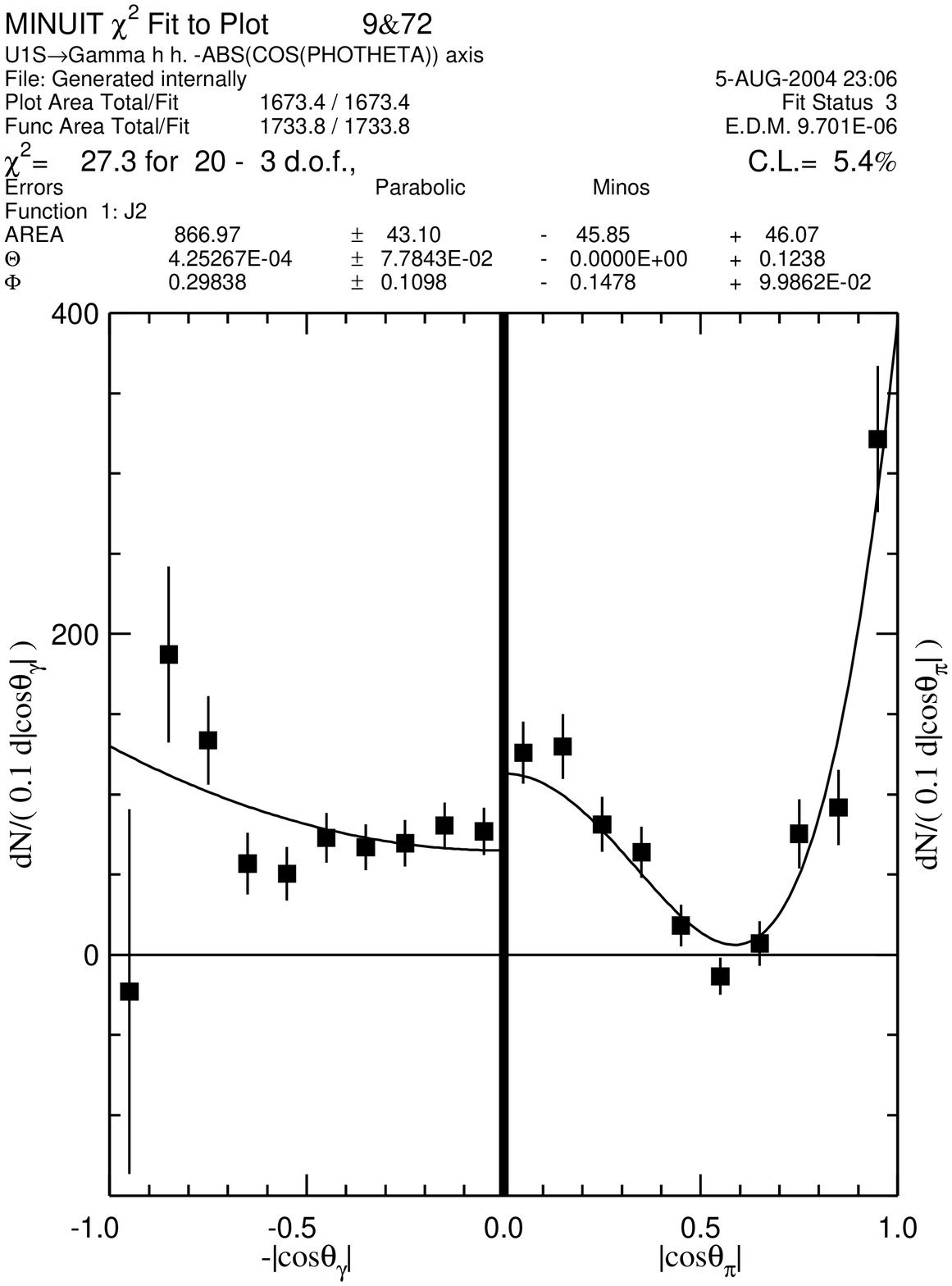}
\end{center}
\caption{Angular distribution for the excess events in the
$f_2(1270)$ mass region. The fit corresponds to J =
2.}\label{Figure:pipiplot_ct_cg_2.ps}   
\end{figure}

\begin{figure}[ht]
\begin{center}
\epsfig{bbllx=66,bblly=100,bburx=580,bbury=775,width=4.5in,clip=,file=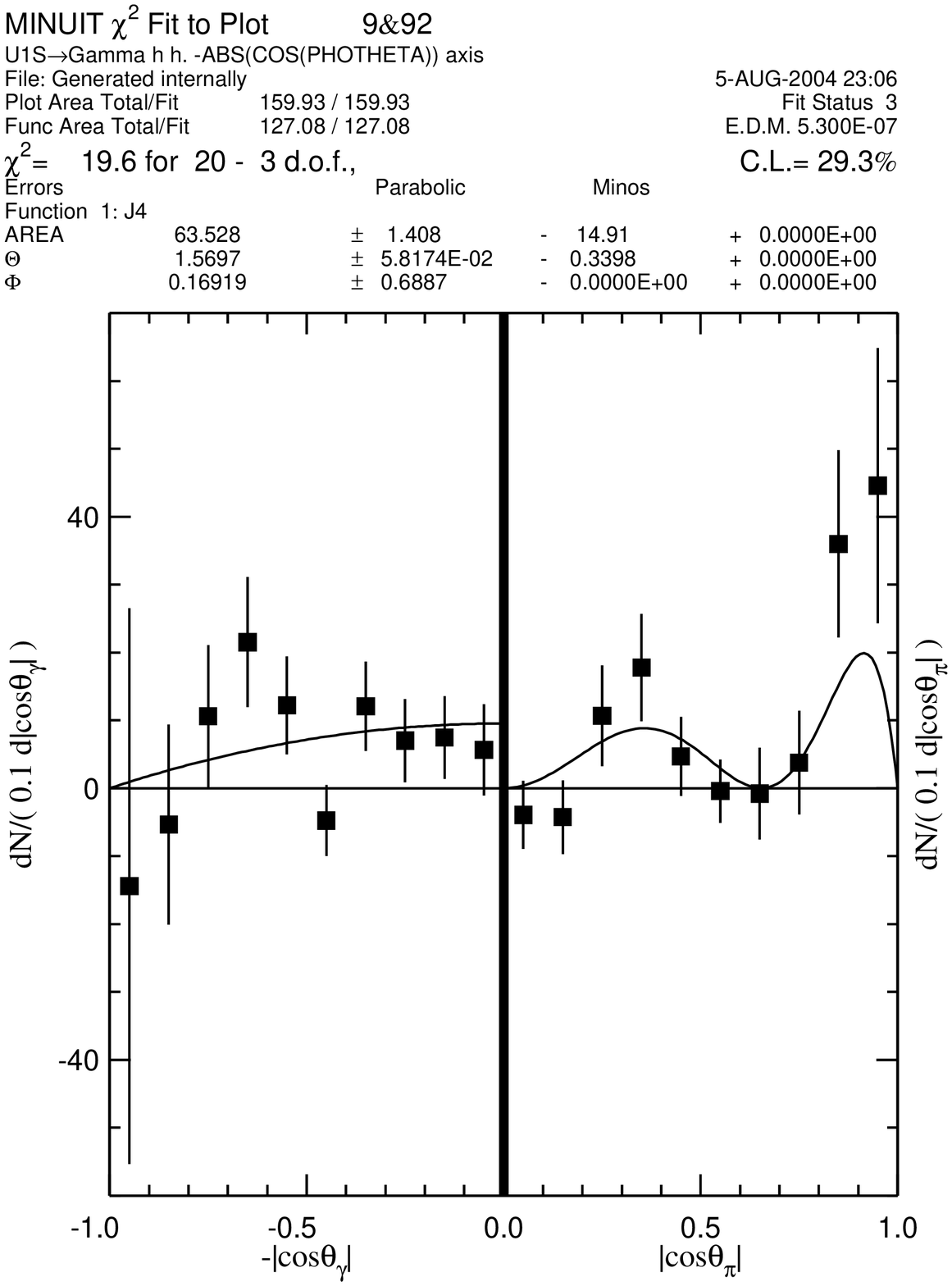}
\end{center}
\caption{Angular distribution for the excess events in the
$f_4(2250)$ mass region. The fit corresponds to J =
4.}\label{Figure:pipiplot_ct_cg_4.ps}   
\end{figure}

For each resonance, we fit the data to the simultaneous
$\cos\theta_{\gamma}$ and $\cos\theta_{\pi}$ helicity angle
distribution obtained from data using Equation~\ref{Equation:solutionlala}
to the helicity formalism prediction, 
Equations~\ref{Equation::J0}-\ref{Equation::J4}, projected on each
angle and folded in opposite directions 
around their symmetry axis in order to show both
distributions on the same plot. The corresponding J value of the best fit for
each resonance, shown in
Figures~\ref{Figure:pipiplot_ct_cg_1.ps}
through~\ref{Figure:pipiplot_ct_cg_4.ps}, 
is defined as the J assignment ($J_a$). We obtain $J_a = 1$ for the $f_0(980)$
which is inconsistent with the known spin of the $f_0(980)$ which is $J
= 0$. For the $f_2(1270)$ and the $f_4(2050)$ we obtain $J_a =2$ and
$J_a = 4$ respectively, which is consistent with their known spins.

\begin{figure}[ht]
\begin{center}
\epsfig{bbllx=0,bblly=100,bburx=580,bbury=775,width=4.5in,clip=,file=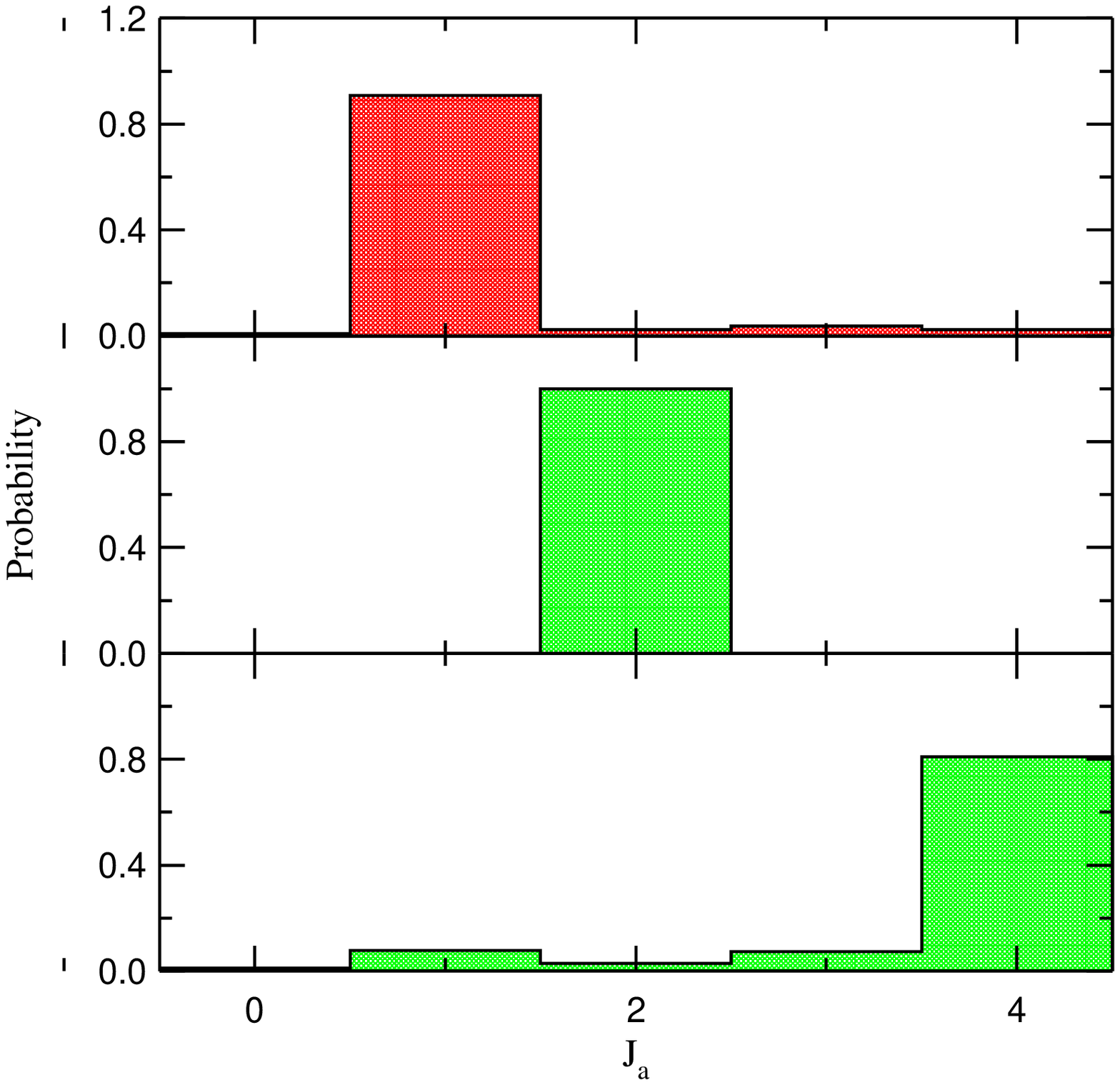}
\end{center}
\caption{$J_a$ probability distribution for resonances in the
$f_0(980)$ (top),
$f_2(1270)$ (middle), and $f_4(2050)$ (bottom) invariant mass region
when only the 
hypotheses $J_a$ = 0, 1, 2, 3, 4 are considered.}\label{assignedj.ps} 
\end{figure}

To have an idea of how well the angular distribution determines J
among the hypotheses  J = 0, 1, 2, 3, 4, we do a statistical fit for
each hypothesis and assign each one a
probability proportional to $e^{-(\chi^2+d.o.f.)/2}$ where $d.o.f.$ are
the degrees of freedom in the fit. The resulting normalized
probability distributions give an idea of the assigned J significance 
and are shown in Figure~\ref{assignedj.ps}. In particular, this figure shows
that the $J_a = 1$ for the $f_0(980)$ inconsistency can be due to a
statistical fluctuation.

\begin{figure}[ht]
\begin{center}
\epsfig{bbllx=66,bblly=100,bburx=580,bbury=775,width=4.5in,clip=,file=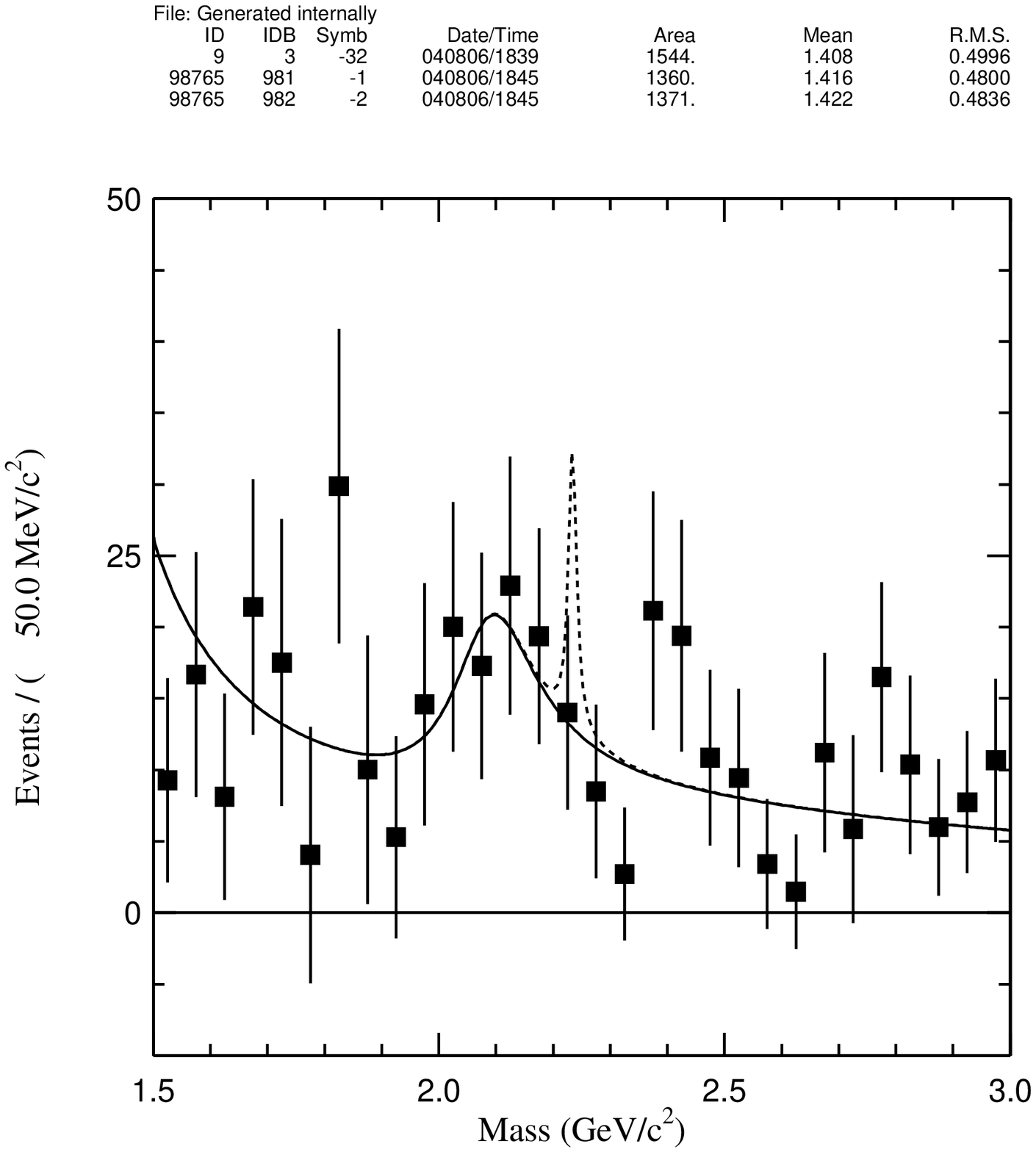}
\end{center}
\caption{Blow up of the invariant mass of $\pi^+ \pi^-$ in the
$f_4(2050)$ mass region, including the upper limit on the
$f_J(2220)$ shown as a dashed line.}\label{Figure:pipiplot_2.ps}
\end{figure}

\begin{figure}[ht]
\begin{center}
\epsfig{bbllx=66,bblly=100,bburx=580,bbury=775,width=4.5in,clip=,file=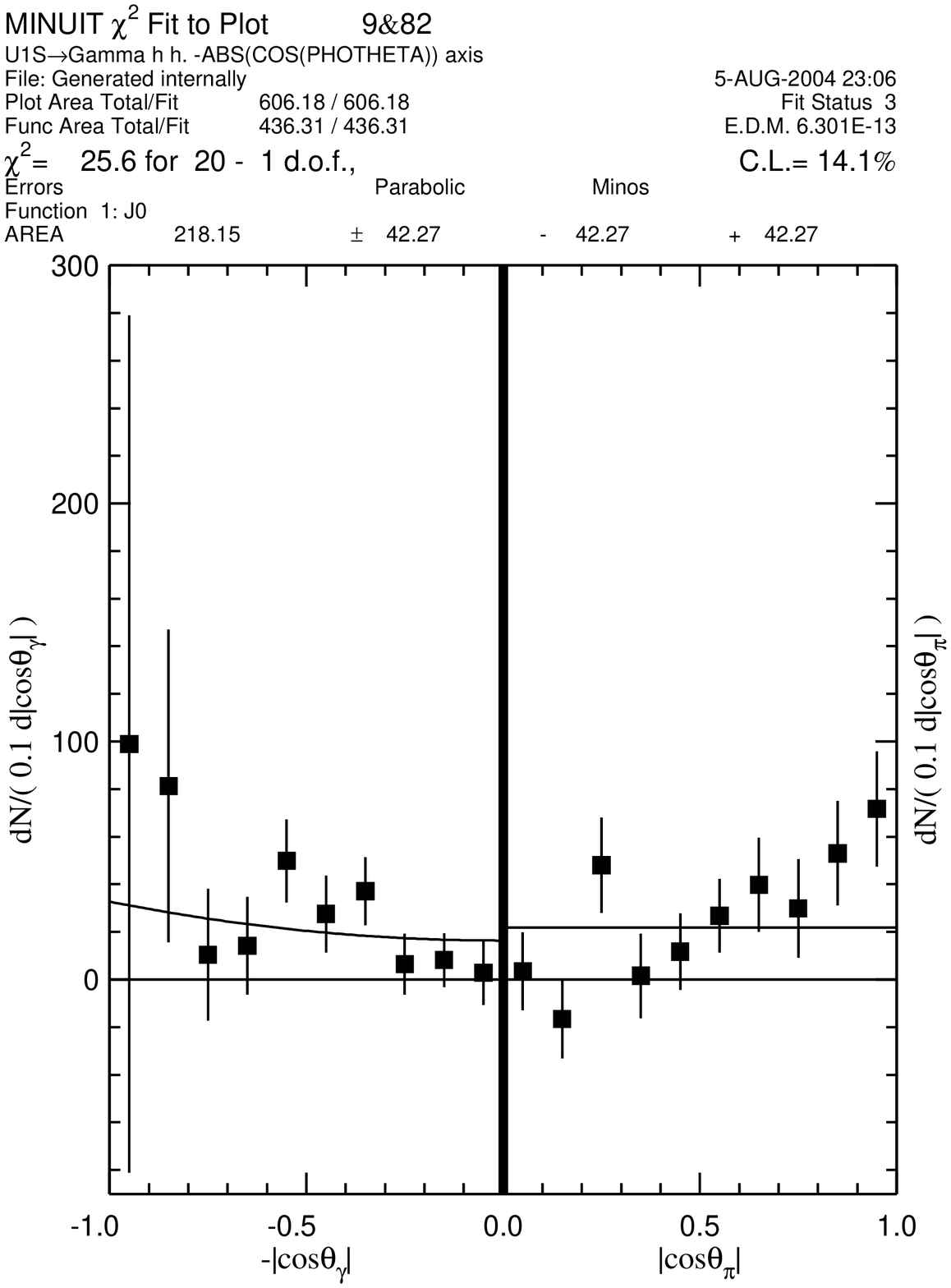}
\end{center}
\caption{Angular distribution for the excess events in the
$f_0(980)$ mass region. The fit corresponds to J =
0.}\label{Figure:pipiplot_ct_cg_0.ps}  
\end{figure}

\clearpage
\chapter{EXCLUSIVE RADIATIVE DECAY $\Upsilon(1S) \rightarrow \gamma
K^+K^-$}
\begin{figure}[ht]
\begin{center}
\epsfig{bbllx=66,bblly=100,bburx=580,bbury=775,width=4.5in,clip=,file=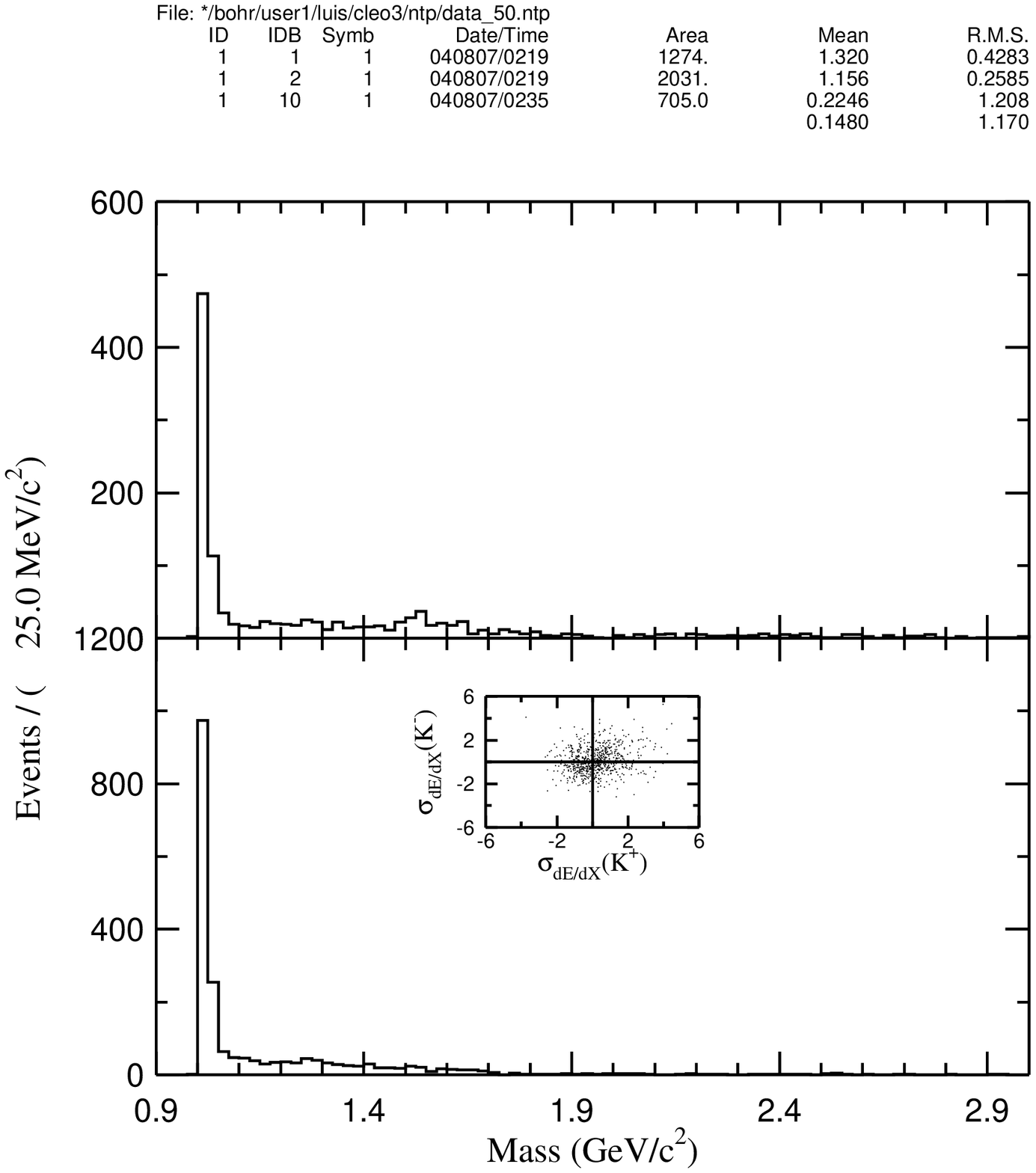}
\end{center}
\caption{Invariant mass of $K^+ K^-$ for 1S (top) and 4S (bottom)
data. For the 4S data the inset shows the
$\sigma_{dE/dX}(K)$ for both tracks for events in the
$1.1-3\mass$ mass region. This inset indicates that most of the events are
constant with having two kaons.}\label{Figure:plot_mass_kk.ps}
\end{figure}

In figure \ref{Figure:plot_mass_kk.ps} the $K^+K^-$ invariant mass
plot is shown for both
resonance and continuum running. This figure also has an inset showing the
$\sigma_{dE/dX}(K)$ for both tracks for events in the $1.1-3\mass$
region. This inset indicates that most of the events in this mass region
have indeed two kaons and that the amount of $\rho$ reflection is
small. Furthermore, when the $\pi^+\pi^-$ invariant mass is plotted for
these events only 40 out of 700 can be fit under a $\rho$ peak (this
fit is what we use to calculate the pion faking kaon rate in
Table~\ref{Table:table_ef}).

Figure~\ref{Figure:kkplot_1.ps} shows our fit to the $K^+K^-$ invariant
mass continuum subtracted plot as defined is Section 3.2.1 with the most
likely statistical fit overlayed (which is described in the next
section). The number of
events near the
$\phi$ region ($1.01-1.03\mass$) left after the continuum
subtraction is $50\pm70$.

\section{Statistical Fit of the Invariant Mass Distribution}

The results of this section are summarized in
Table~\ref{Table:table_kk}.

From the measurement of the previous section we expect a small contribution
($\approx 50$ events assuming no interference) from $f_2(1270) \to
K^+ K^-$. We do find some evidence for $f_2(1270) \to K^+ K^-$ events
($110\pm40$) in the fit. We also find strong evidence for 
the resonance $f_2'(1525)$ and weak evidence for the $f_0(1710)$
resonance. The $f_2(1270)$, and $f_0(1710)$ are fitted with their
widths fixed to their PDG  
values\cite{pdg} because they have large errors if allowed to
float. The rest of the resonances parameters are consistent with their
PDG values\cite{pdg}, which are $m_{f_2'(1525)} = 1525 \pm 5 \miss$,
$\Gamma_{f_2'(1525)} = 76 \pm 10  \miss$, $m_{f_0(1710)} = 1715 \pm
6 \miss$, and  
$\Gamma_{f_0(1710)} = 125 \pm 10  \miss$. We also observe
an excess of events in the $2-3\mass$ region which we
can't attribute to any known resonances.

Significances of the signals of the identified resonances in the fit
are calculated as described in the 
previous section.  The $\Upsilon(1S) \to \gamma
f_J(2220)$, $f_J(2220) \to K^+ K^-$ upper limit is also calculated
using multiple fits, 
except that this time the events under the 
$f_J(2220)$ are of unknown origin, so we use a first order polynomial
allowed to 
float. The significance of the excess of events in the
$2-3\mass$ invariant mass region is calculated assuming a normal
distribution; we simply add up 
the number of events in each bin along with its error. Results are
shown in Table~\ref{Table:table_kk}.

\begin{landscape}
\vspace*{\fill}
 \begin{table}[ht]
 \begin{center}
 \caption{Results for $\Upsilon(1S) \rightarrow \gamma K^+K^-$}
 \begin{tabular}{c|c|c|c}
 \hline
 \hline
 Mode & Area & Branching Fraction $(10^{{-5}})$ & Significance \\
 \hline
 $ \gamma f_2(1270)$&
 $   110\pm40$& $    23\pm8$ &
 $5.4\times10^{-4}(3.3\sigma)$ \\
 $ \gamma f^{'}_2(1525)$&
 $   360^{+    80}_{    -70}$& $    3.9^{+   0.9}_{   -0.7}$ &
 $<10^{-45}(>14\sigma)$ \\
 $ \gamma f_0(1710), f_0(1710) \rightarrow K^+K^-$&
 $   75\pm30$& $   0.35\pm0.14$ & $7.5\times10^{-4}(3.2\sigma)$\\
 $ \gamma K^+ K^- (2-3\mass)$&
 $   220\pm    20$ & $    1.03\pm   0.12$ & $    8.8\sigma$ \\
 $ \gamma f_J(2220), f_J(2220) \rightarrow K^+K^-$& $<      10$ & $ <5\times10^{-2}$ & - \\
 \hline
 \end{tabular}
 \label{Table:table_kk}
 \end{center}
 \end{table}
\vfill
\end{landscape}

\section{Angular Distribution of The Signal}

In this section we adapt the ideas presented in Section 3.4.3 to the
$\Upsilon(1S) \to \gamma K^+ K^-$ situation. 

\begin{figure}[ht]
\begin{center}
\epsfig{bbllx=66,bblly=100,bburx=580,bbury=775,width=4.5in,clip=,file=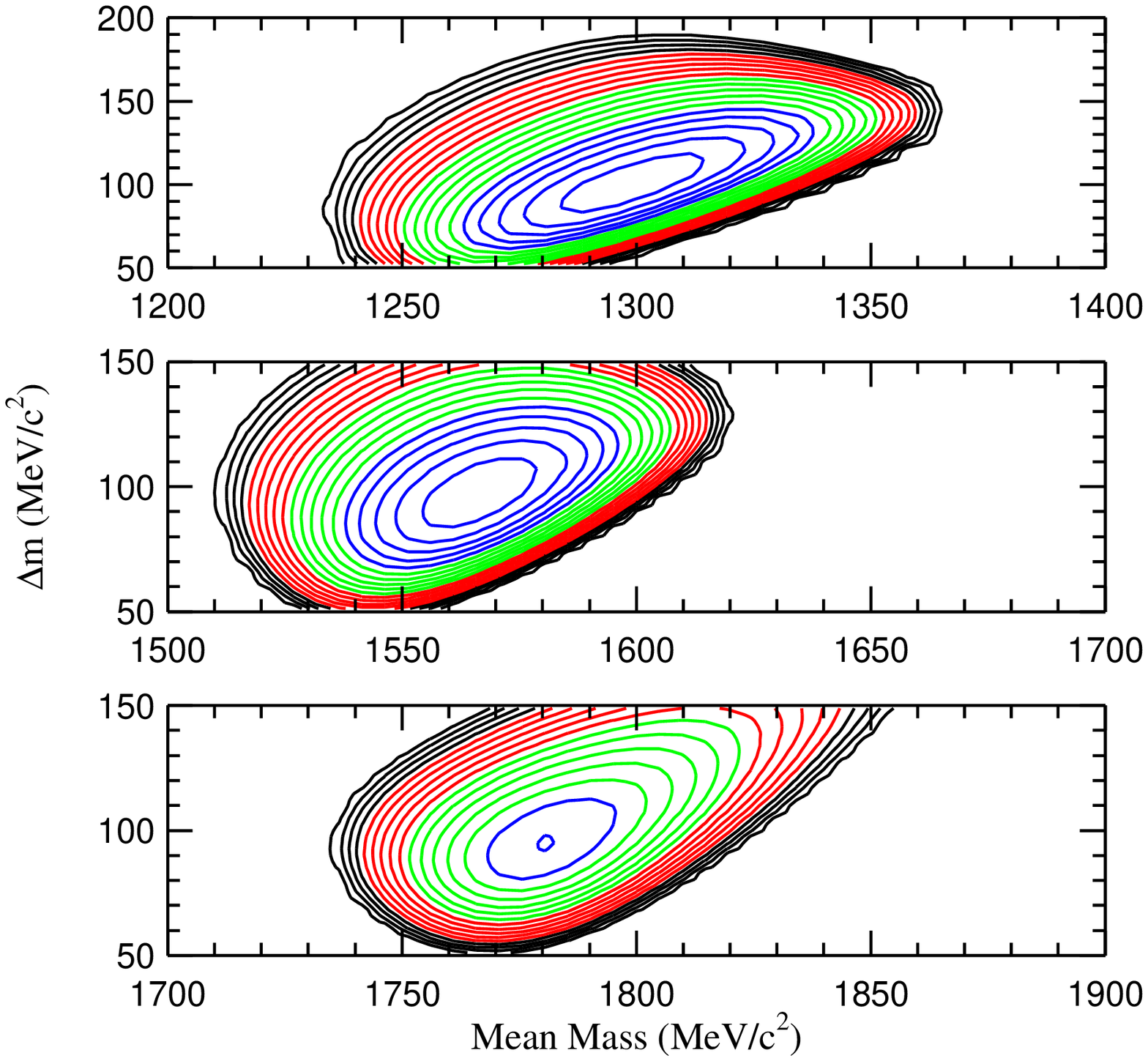}
\end{center}
\caption{Contour plot of the inverse
of the average relative bin-error from background subtractions in the
$f_2(1270)$ (top), $f_2^{'}(1525)$ (middle), and $f_0(1710)$ (bottom)
angular distributions.}\label{Figure:mbardmkk.ps}
\end{figure}

The derived statistical errors from the signal and
background subtractions can be used to calculate the mass interval which
best represents the helicity angular distribution. The inverse of the
expected average relative bin error as a function of the mass interval
is shown in Figure~\ref{Figure:mbardmkk.ps}, and the mass interval
that maximizes it are tabulated in Table~\ref{Table:intervalresults2}.

\begin{landscape}
\vspace*{\fill}
 \begin{table}[ht]
 \begin{center}
 \caption{Mean masses, widths in$mass$ and inverse of the average bin
 relative error (F) from background subtractions for the angular
 distribution of different resonances. Standard mean masses and 
 widths, corresponding to 1$\Gamma$, are taken from the fit in
 Figure~\ref{Figure:kkplot_1.ps}, and are labeled with the 
 subscript ``s'', while those that maximize
 F are labeled with the 
 subscript ``m''. The last column shows the factor by which the
 effective statistics increase.} 
 \begin{tabular}{c||c|c||c|c||c}
 \hline
 \hline 	
 Resonance & $(\bar{m}_s,\Delta m_s)$ & $F(m_s,\Delta m_s)$ &
 $(\bar{m}_m,\Delta  m_m)$ & $F(m_m,\Delta m_m)$ &
 $\frac{F^2(m_m,\Delta m_m)}{F^2(m_s,\Delta m_s)}$ \\ 
 \hline
$f_2(1270)$ & $(1.276,0.185)$ & 3.5 & $(1.300,0.100)$ & 4.1 & 1.37  \\ 
$f_2'(1525)$ & $(1.540,0.085)$ & 9.6 & $(1.565,0.100)$ & 9.8 & 1.04
 \\ 
$f_0(1710)$ & $(1.760,0.125)$ & 3.1 & $(1.780,0.095)$ & 3.3 & 1.13 \\ 
 \hline
 \hline
 \end{tabular}
\label{Table:intervalresults2} 
 \end{center}
 \end{table}
\vfill
\end{landscape}

The tails from the resonances contribute to the continuum subtracted
helicity distributions,
\begin{equation} 
\begin{cases}
c_{f_2(1270)} = t_{f_2(1270)} + \beta t_{f_2'(1525)}  \\
c_{f_2'(1525)} = t_{f_2'(1525)} + \gamma t_{f_2(1270)} + \delta
t_{f_0(1710)}  \\ 
c_{f_0(1710)} = t_{f_0(1710)} + \epsilon t_{f_2'(1525)}\label{Equation:system}
\end{cases}
\end{equation}

\begin{figure}[ht]
\begin{center}
\epsfig{bbllx=66,bblly=100,bburx=580,bbury=775,width=4.5in,clip=,file=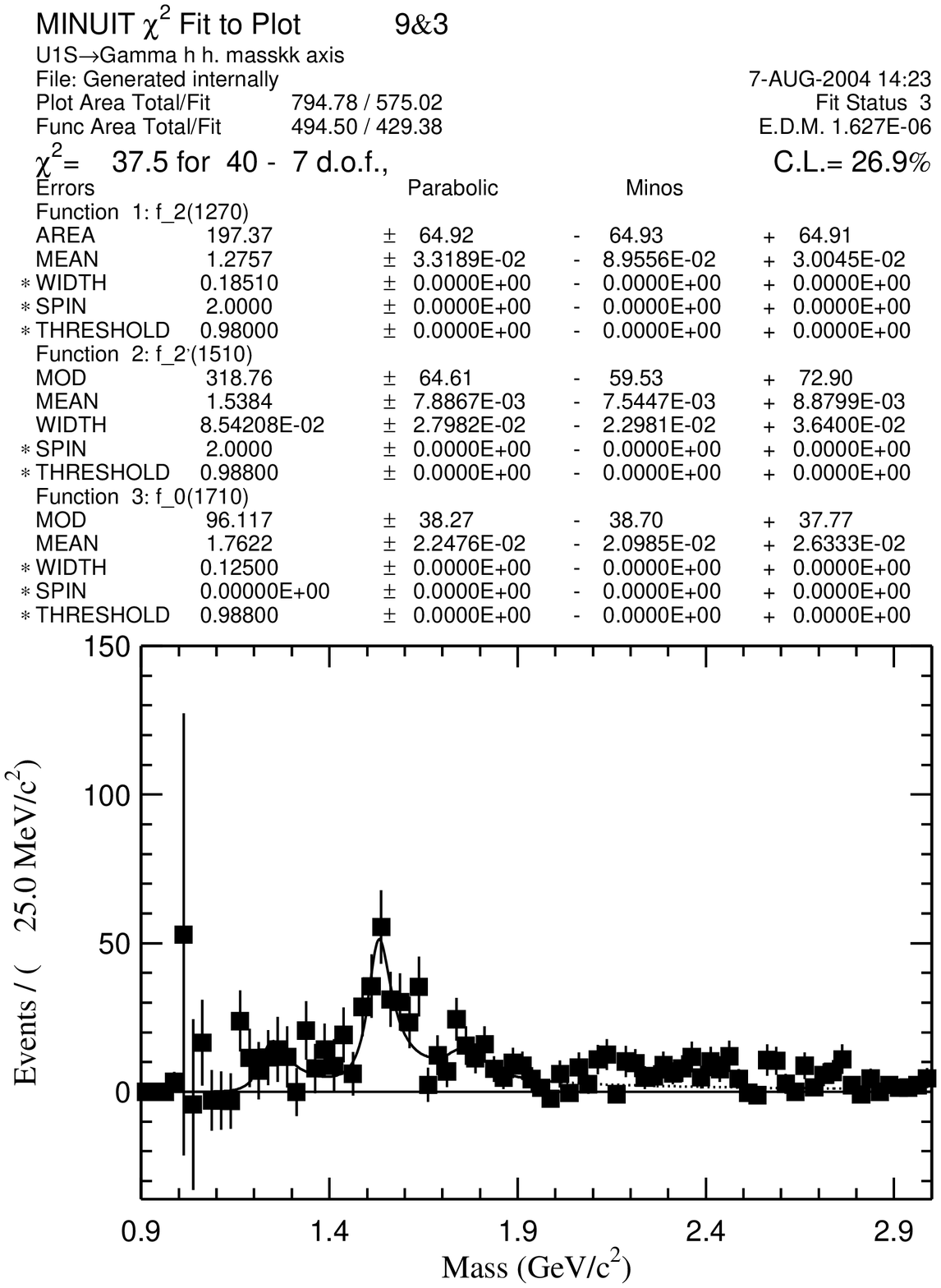}
\end{center}
\caption{Continuum subtracted invariant mass of $K^+ K^-$.}\label{Figure:kkplot_1.ps}
\end{figure}

Where again the small numbers $ \beta,\ \gamma,\ \delta$, and $\epsilon$ are
the ratios of the number of events from a resonance in the mass
interval where the contamination is taking place to the number of
events from the same resonance in the mass interval used to select
it. Using the $(m_m, \Delta m_m)$ values in
Table~\ref{Table:intervalresults2} and the fit in
Figure~\ref{Figure:kkplot_1.ps} we obtain $\beta = 1.3\times10^{-2}$,
$\gamma = 0.14$, $\delta = 0.11$, and
$\epsilon = 0.19$.

The background subtracted helicity angular distributions are,
\begin{equation} 
\begin{cases}
t_{f_2'(1525)} = \frac{1}{1-\beta \gamma - \delta \epsilon}
(c_{f_2'(1525)} - \gamma c_{f_2(1270)} - \delta c_{f_0(1710)})  \\ 
t_{f_2(1270)} = c_{f_2(1270)} - \beta t_{f_2'(1525)}  \\
t_{f_0(1710)} = c_{f_0(1710)} - \epsilon t_{f_2'(1525)}\label{Equation:solution}
\end{cases}
\end{equation}

\begin{figure}[ht]
\begin{center}
\epsfig{bbllx=66,bblly=100,bburx=580,bbury=775,width=4.5in,clip=,file=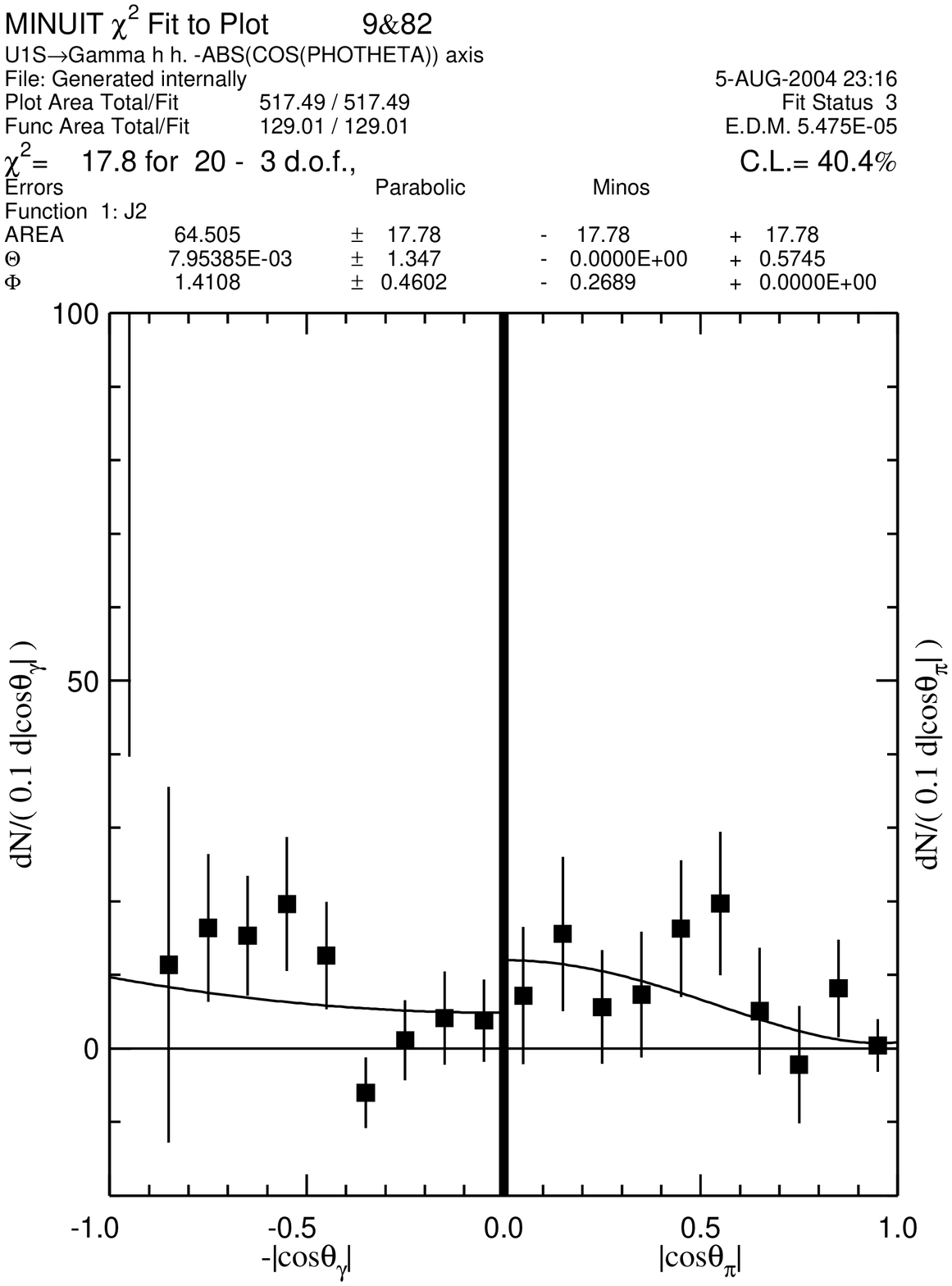}
\end{center}
\caption{Background subtracted $K^+K^-$ angular
distribution in the $f_2(1270)$ mass region as defined in the
text. The fit corresponds to J = 2.}\label{Figure:kkplot_ct_1.ps}
\end{figure}

\begin{figure}[ht]
\begin{center}
\epsfig{bbllx=66,bblly=100,bburx=580,bbury=775,width=4.5in,clip=,file=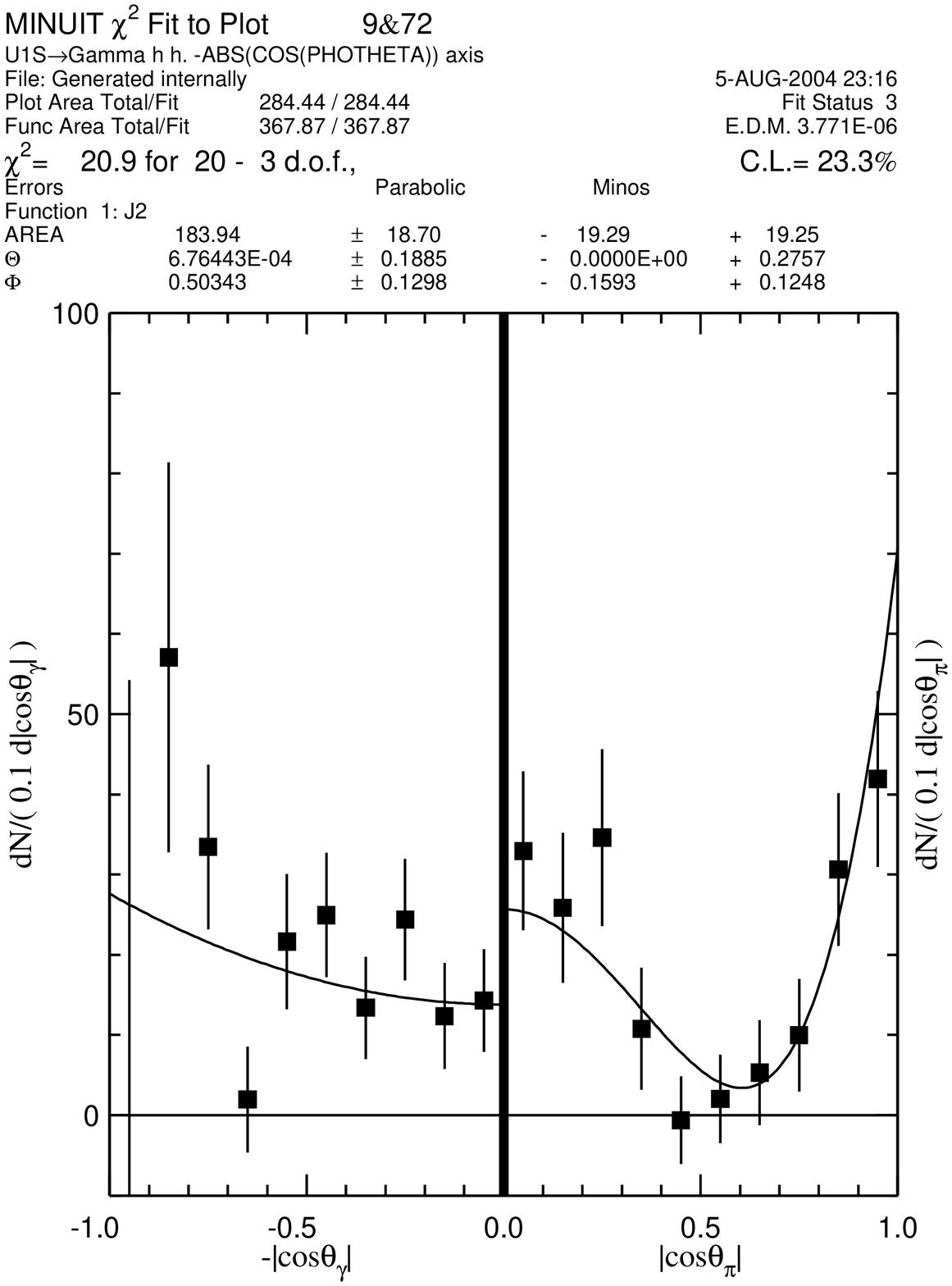}
\end{center}
\caption{Background subtracted $K^+K^-$ angular
distribution in the $f_2^{'}(1525)$ mass region as defined in the
text. The fit corresponds to J = 2.}\label{Figure:kkplot_ct_2.ps}
\end{figure}

\begin{figure}
\begin{center}
\epsfig{bbllx=66,bblly=100,bburx=580,bbury=775,width=4.5in,clip=,file=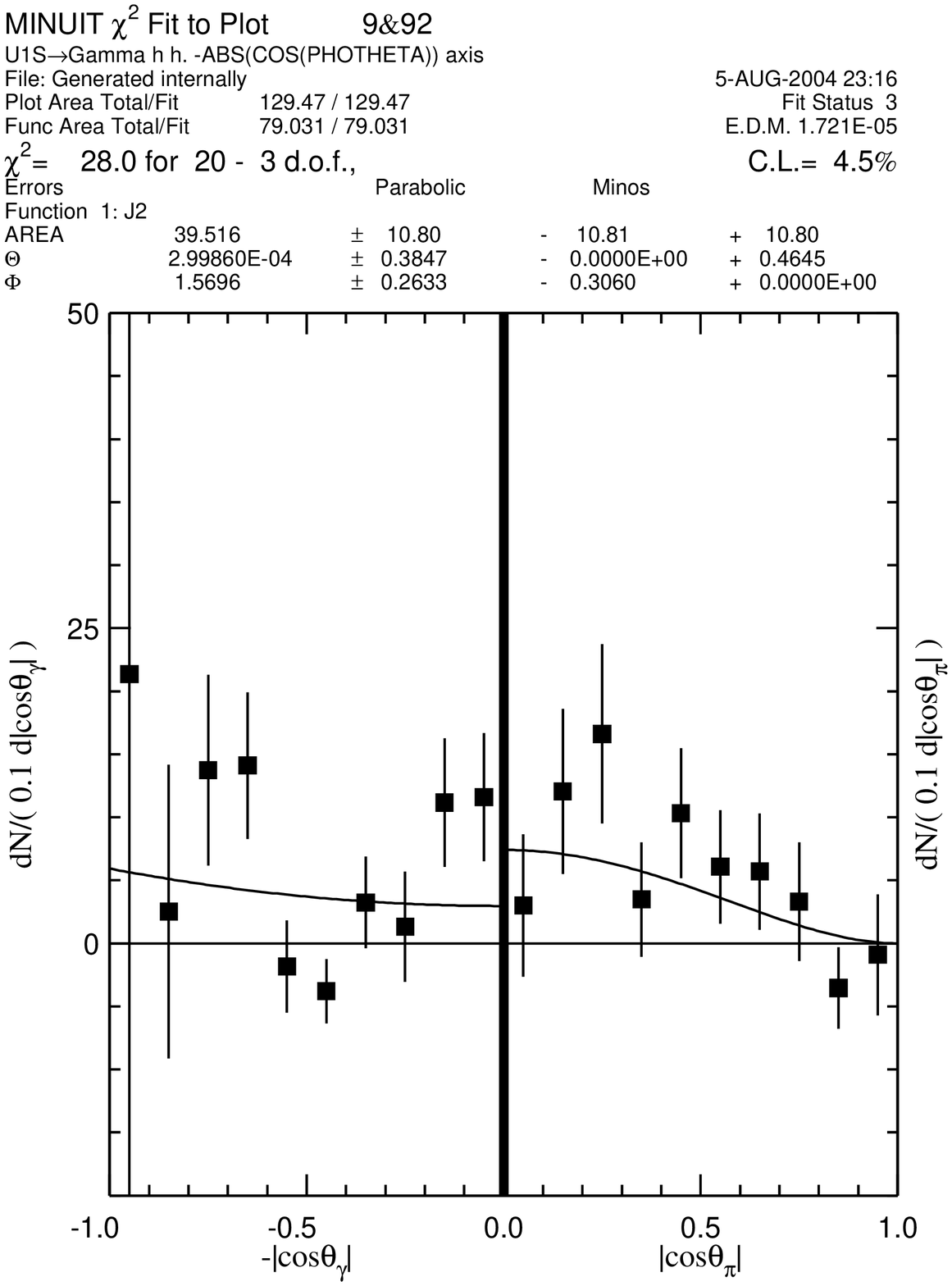}
\end{center}
\caption{Background subtracted $K^+K^-$ angular
distribution in the $f_0(1710)$ mass region as defined in the
text. The fit corresponds to J = 2.}\label{Figure:kkplot_ct_3.ps}
\end{figure}

\begin{figure}
\begin{center}
\epsfig{bbllx=66,bblly=100,bburx=580,bbury=775,width=4.5in,clip=,file=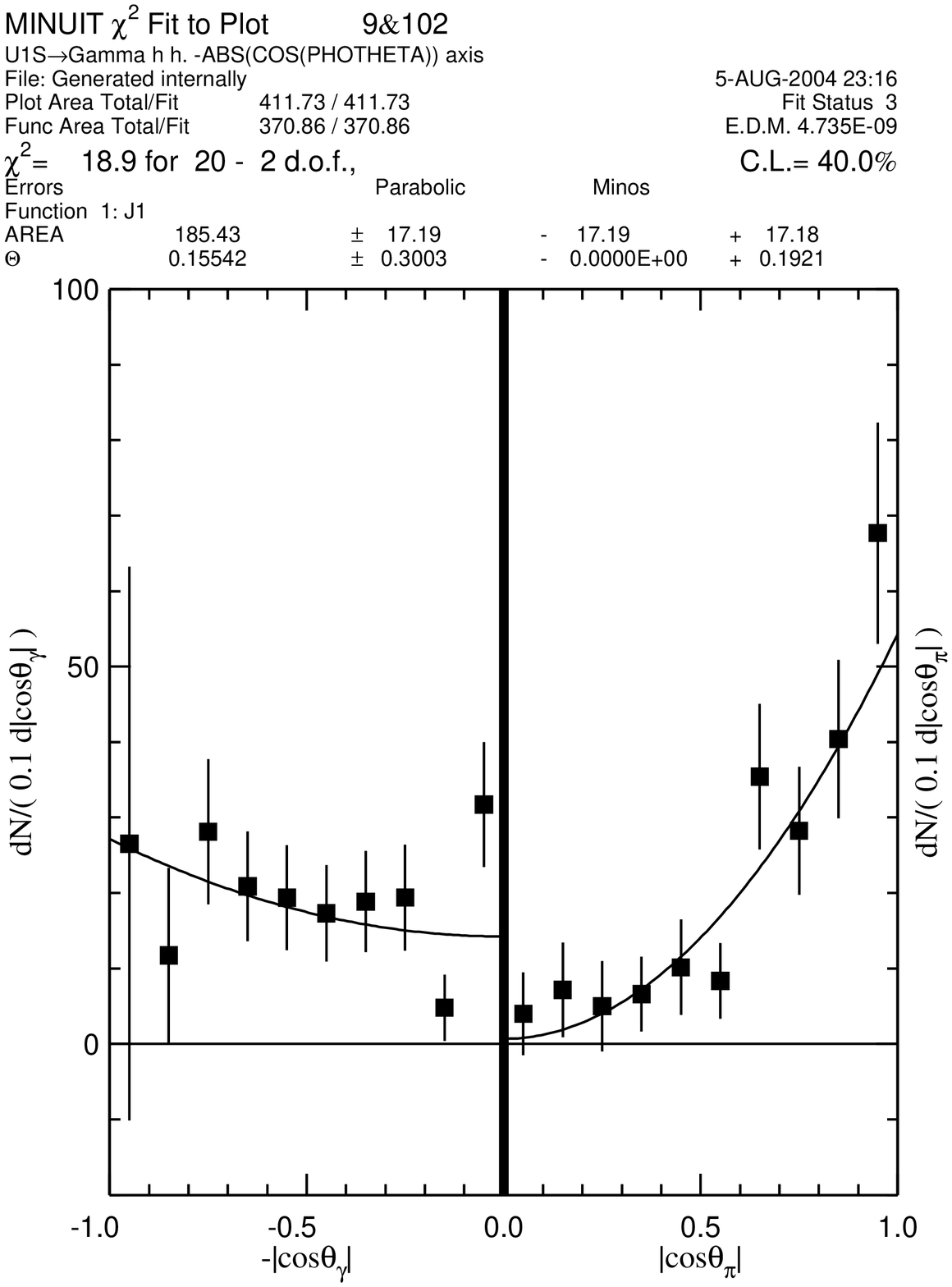}
\end{center}
\caption{$K^+K^-$ angular
distribution for events within the $2-3\mass$.}\label{Figure:kkplot_ct_5.ps} 
\end{figure}

The the best fit for
each resonance and the excess of events in the $2-3\mass$ are shown in
Figures~\ref{Figure:kkplot_ct_1.ps}
through~\ref{Figure:kkplot_ct_5.ps}. The best spin assignment for the  
$f_2(1270)$ is $J_a = 2$, for the $f^{'}_2(1525)$ it is $J_a = 2$, for the
it is $f_0(1710)$, and it is $J_a = 1$ for the excess of events in the
$2-3\mass$ mass region. The $J_a = 2$ value for the $f_0(1710)$ is inconsistent
with 
its known spin. Also, examination of the normalized helicity
amplitudes for the $f_2(1270)$ reveals that they are inconsistent with
those obtained for the $f_2(1270)$ in the $\pi^+ \pi^-$ mode.

\begin{figure}[ht]
\begin{center}
\epsfig{bbllx=66,bblly=100,bburx=580,bbury=775,width=4.5in,clip=,file=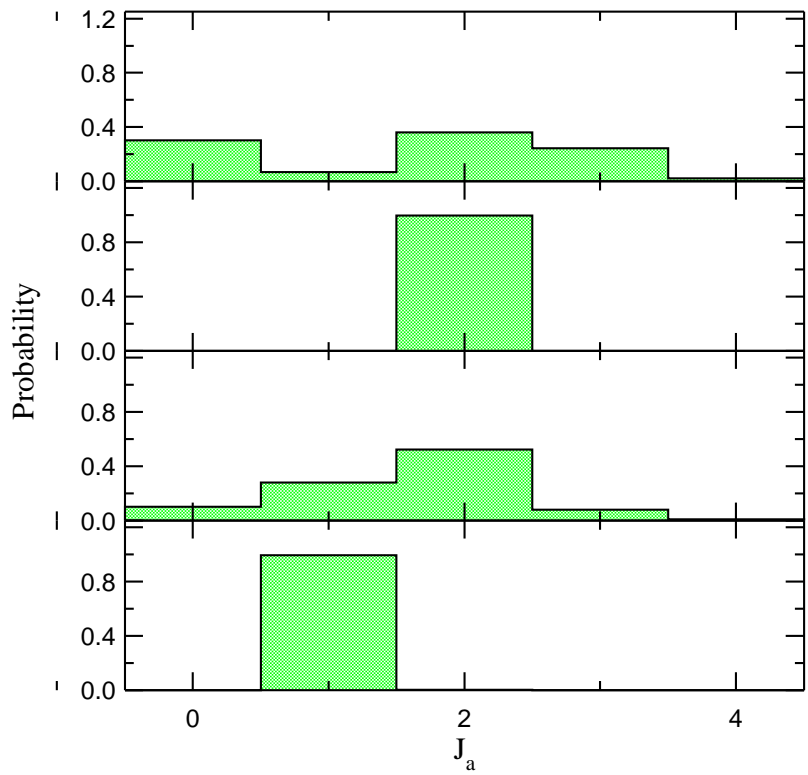}
\end{center}
\caption{$J_a$ probability distribution for resonances in the $f_2(1270)$(top),
$f_2'(1525)$ (middle-top), $f_0(1710)$ (middle-bottom) invariant mass region and the excess of
events in the $2-3\mass$ (bottom)  region when only the
hypotheses $J_a$ = 0, 1, 2, 3, 4 are considered.}\label{assignedj2.ps} 
\end{figure}

The assigned J probability distributions are shown in
Figure~\ref{assignedj2.ps}. They reveal that the inconsistencies in
the $f_2(1270)$ and $f_0(1710)$ are not significant and can be
attributed to the statistical uncertainty.

\begin{figure}[ht]
\begin{center}
\epsfig{bbllx=66,bblly=100,bburx=580,bbury=775,width=4.5in,clip=,file=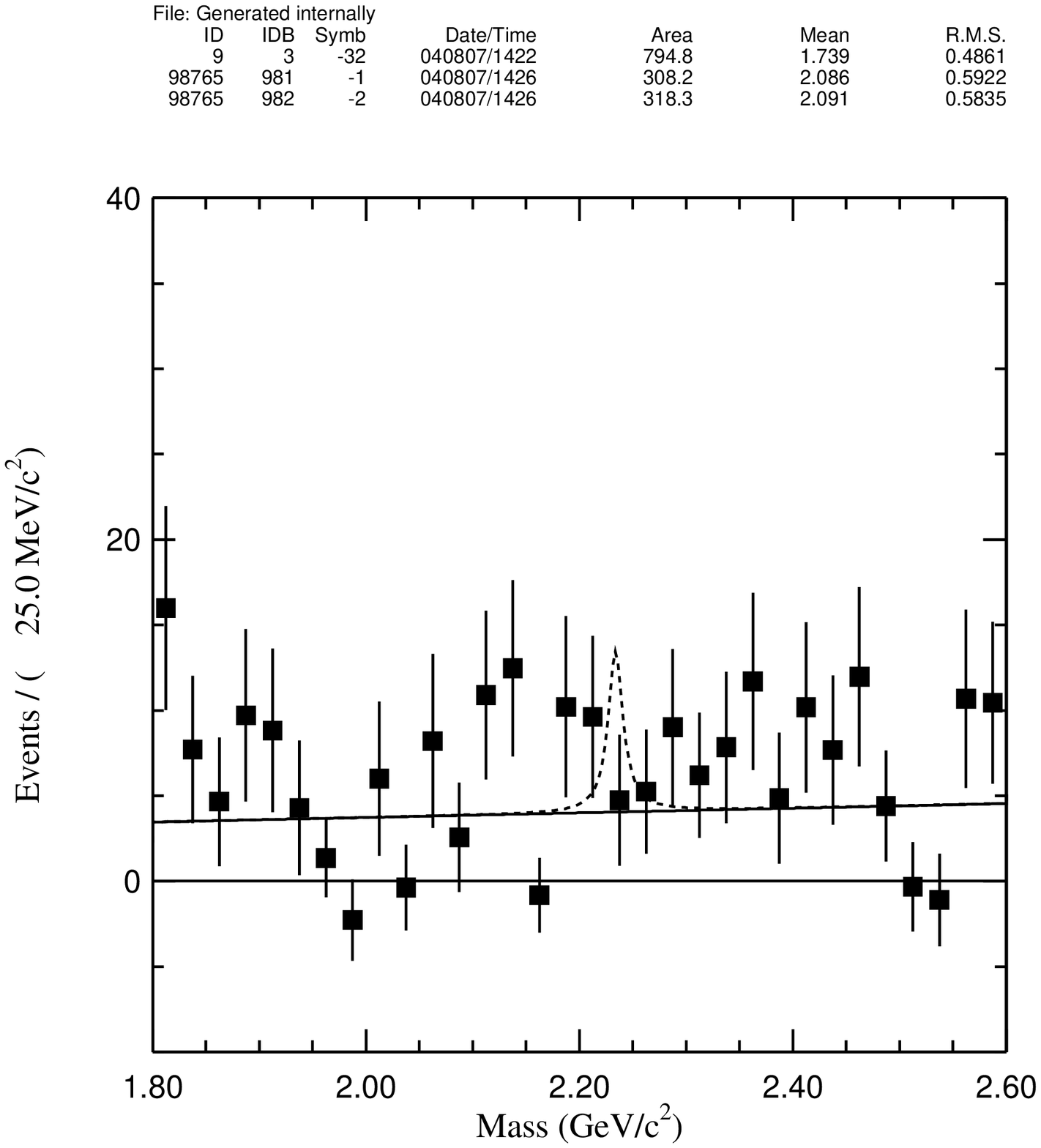}
\end{center}
\caption{Blow up of the $f_J(2220)$ region, with the 90\% CL upper
limit overlaid. The mass and width are taken to be $m_{f_J(2220)} =
2.234\mass$ and $\Gamma_{f_J(2220)} = 17\miss$ as
in~\cite{glueball}.}\label{Figure:kkplot_3.ps} 
\end{figure}

\begin{figure}[ht]
\begin{center}
\epsfig{bbllx=66,bblly=100,bburx=580,bbury=775,width=4.5in,clip=,file=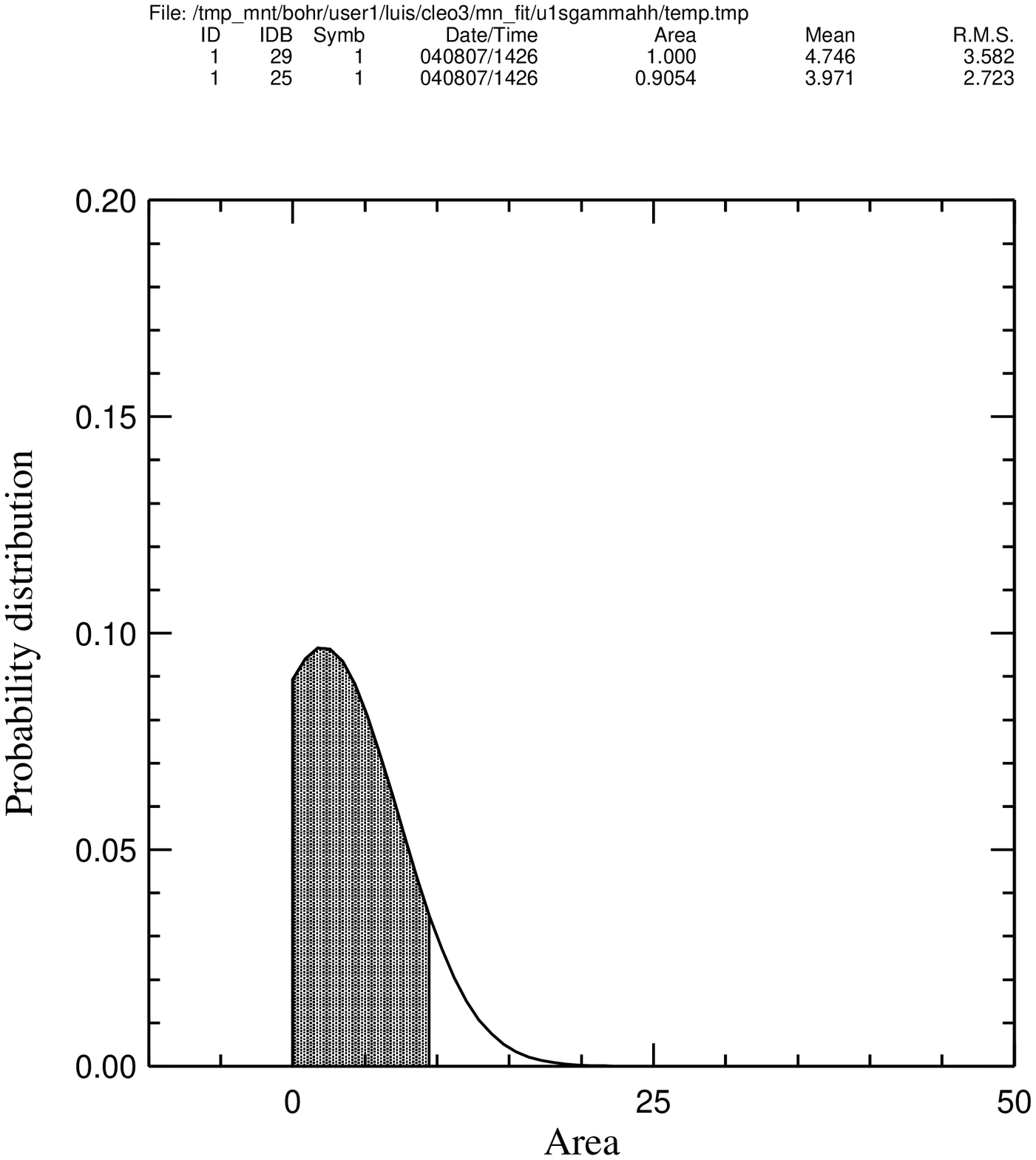}
\end{center}
\caption{Normalized probability distribution for different $f_J(2220)
\to K^+ K^-$
signal areas. The shaded area spans 90\% of the
probability.}\label{Figure:fj_prob2.ps} 
\end{figure}

\begin{figure}
\begin{center}
\epsfig{bbllx=66,bblly=100,bburx=580,bbury=775,width=4.5in,clip=,file=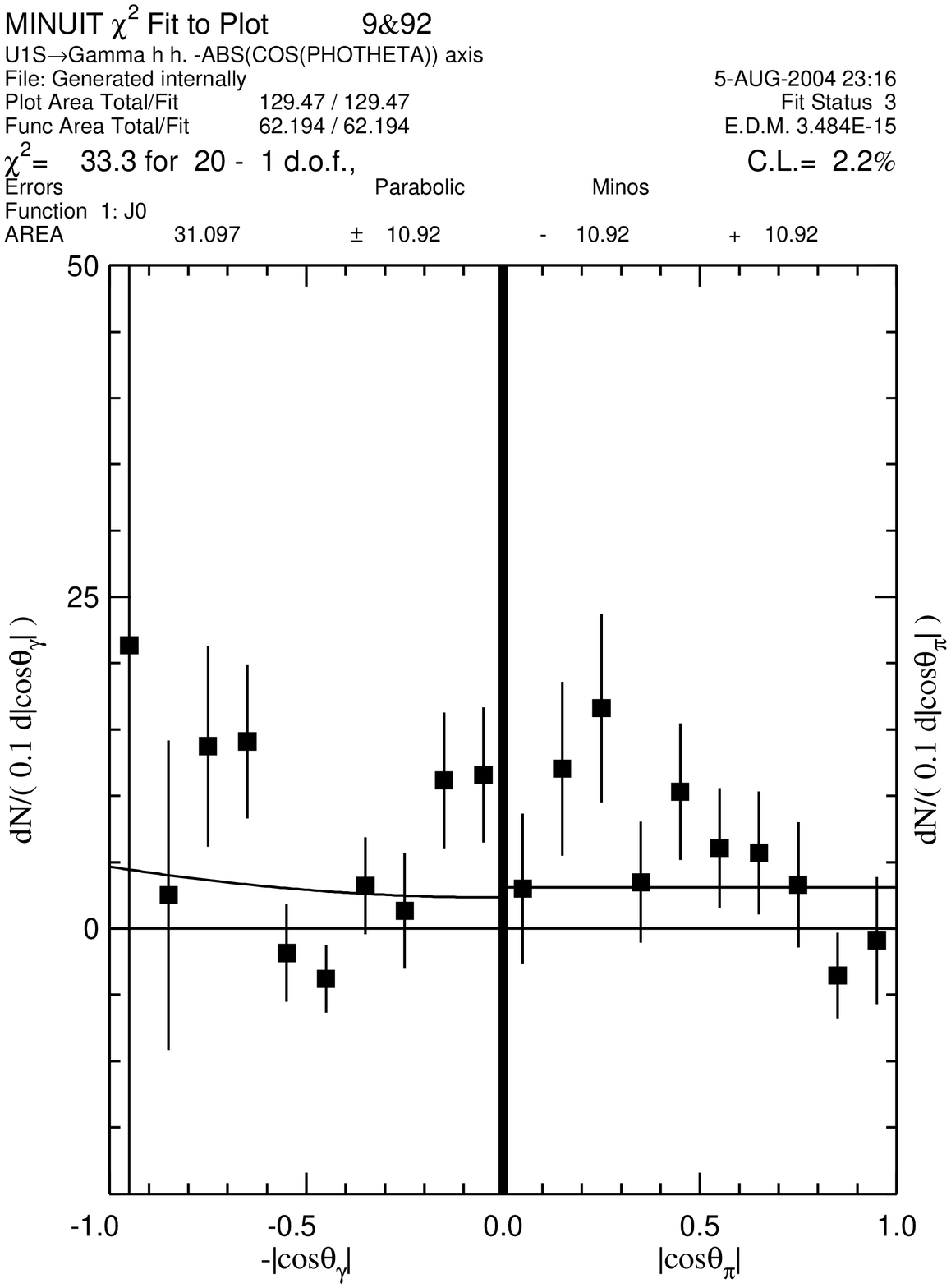}
\end{center}
\caption{Background subtracted $K^+K^-$ angular
distribution in the $f_0(1710)$ mass region as defined in the
text. The fit corresponds to J = 0.}\label{Figure:kkplot_ct_4.ps}
\end{figure}

\clearpage
\chapter{EXCLUSIVE RADIATIVE DECAY $\Upsilon(1S) \rightarrow \gamma
p \bar{p}$}
\begin{figure}[ht]
\begin{center}
\epsfig{bbllx=66,bblly=100,bburx=580,bbury=775,width=4.5in,clip=,file=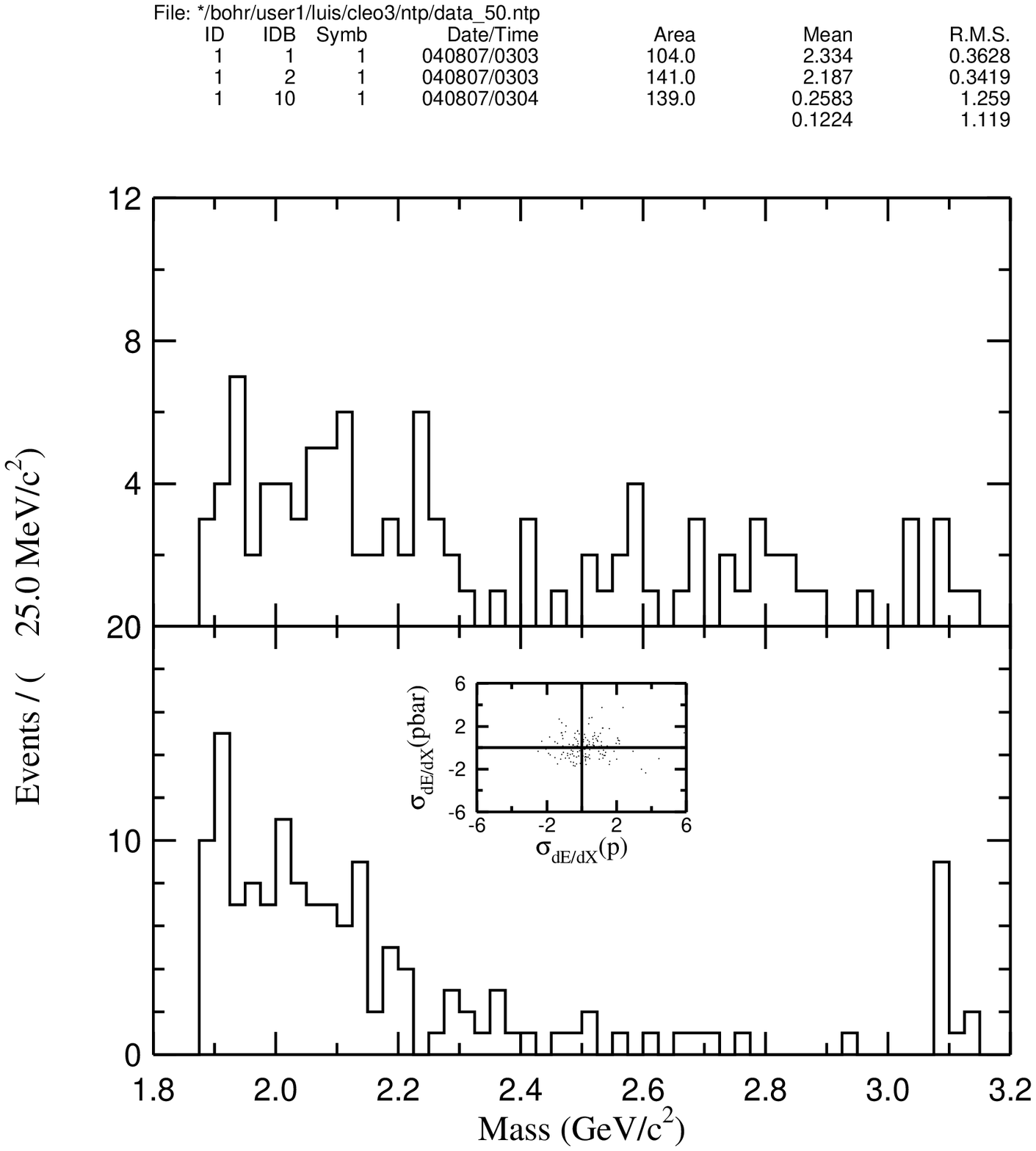}
\end{center}
\caption{Invariant mass of $p \overline{p}$ for 1S (top) and 4S
(bottom) running. For the 4S running the inset is consistent with the
events having a proton and an
anti-proton.}\label{Figure:plot_mass_pp.ps} 
\end{figure}

In figure \ref{Figure:plot_mass_pp.ps} the $p\bar{p}$ invariant mass
plot is shown for both 1S and 4S data, with an inset showing that
most of the events indeed have a proton and an anti-proton and that
the $\rho$ and $\phi$ reflections are small. The
enhancement at 3.1 $\mass$ in the 4S plot corresponds to the
process $e^+e^- \to \gamma J/\psi$ with $J/\psi \to p \bar{p}$. This
enhancement is not as pronounced in the 1S plot because the 1S data
only has 32.4\% of the luminosity of the 4S data. The number of events
in the $J/\psi$ invariant mass region after continuum subtraction is
$6\pm3$ and is consistent with 0.

\begin{figure}[ht]
\begin{center}
\epsfig{bbllx=66,bblly=100,bburx=580,bbury=775,width=4.5in,clip=,file=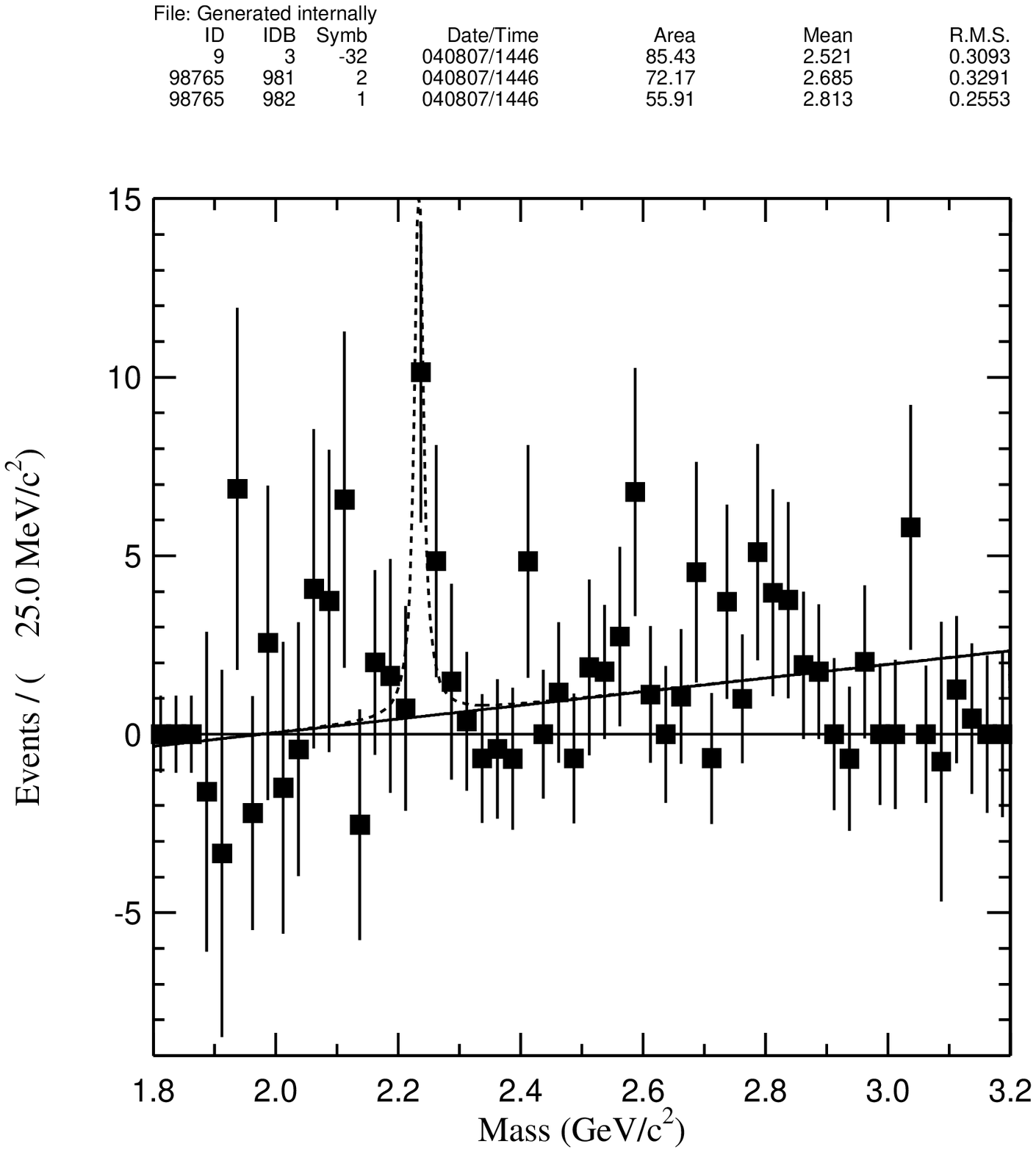}
\end{center}
\caption{Invariant mass of $p \bar{p}$. The plot is continuum subtracted
and efficiency corrected. An overlay with the 90\% confidence level upper limit
for $f_J(2220)$ is shown. The mass and width are taken to be
$m_{f_J(2220)} = 2.234 \mass$ and $\Gamma_{f_J(2220)} = 17 \miss$
as in~\cite{glueball}. }\label{Figure:ppplot_1.ps}
\end{figure}

Figure
\ref{Figure:ppplot_1.ps} shows the continuum subtracted invariant
$p\bar{p}$ mass distribution (as defined in Section 3.2.1) with a 90\%
confidence level upper limit
for $f_J(2220)$ overlaid. A direct fit to the $f_J(2220)$ yields
$12\pm5$ events.

There is an excess of events in the continuum subtracted invariant
mass plot. We measure this excess and the upper limit
of $e^+e^- \to \gamma f_J(2220)$, $f_J(2220) \to p \bar{p}$
the same way we measured the excess of events inside $2 \mass $ $<
m(K^+K^-) < 3 \mass$ region and the $f_J(2220)$
upper limit in Section 5.1. Results are shown in
Table~\ref{Table:table_pp}.

\begin{figure}[ht]
\begin{center}
\epsfig{bbllx=66,bblly=100,bburx=580,bbury=775,width=4.5in,clip=,file=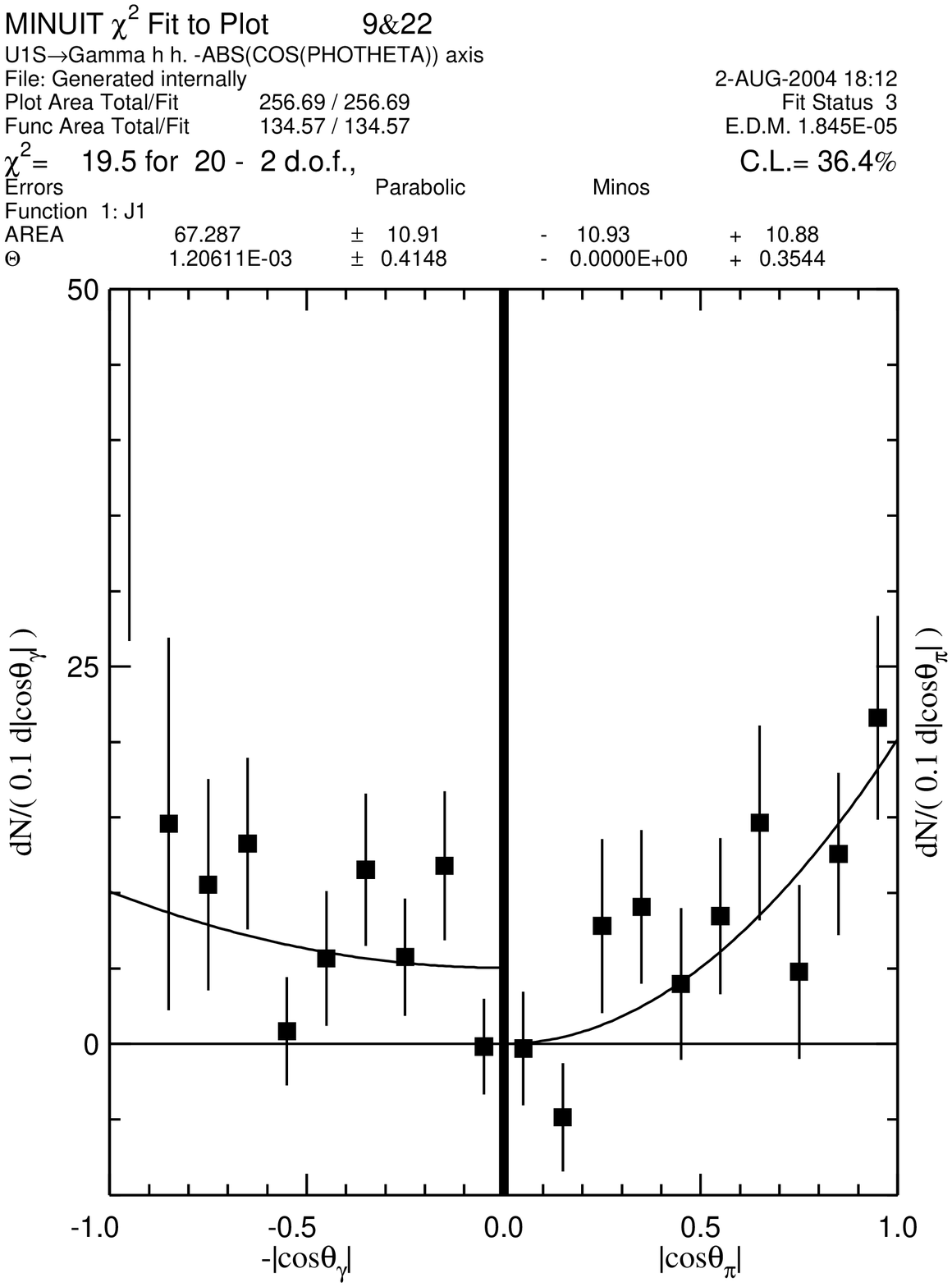}
\end{center}
\caption{$p \bar{p}$ angular
distribution for the excess of events in the mass range $2 \mass$$ <
m(p \bar{p}) < 3 \mass$.}\label{Figure:ppplot_ct_0.ps}
\end{figure}

\begin{figure}[ht]
\begin{center}
\epsfig{bbllx=66,bblly=100,bburx=580,bbury=775,width=4.5in,clip=,file=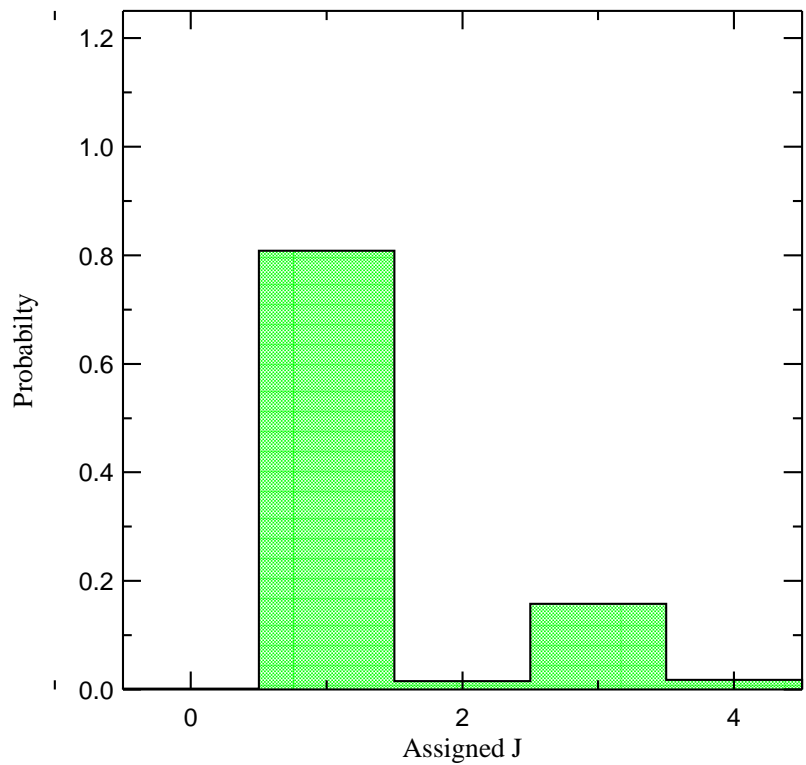}
\end{center}
\caption{$J_a$ probability distribution for the excess of events in the
mass range $2 \mass$$ < m(p \bar{p}) < 3 \mass$  
when only the hypotheses $J_a$ = 0, 1, 2, 3, 4 are
considered.}\label{assigned3.ps}
\end{figure}

The $p \bar{p}$ angular distribution for the mass range $2 \mass$$ <
m(p \bar{p}) < 3 \mass$ is shown in Figure
\ref{Figure:ppplot_ct_0.ps} with the most likely $J$ assignment, $J_a$
= 1, fit overlayed. The probability distribution for $J_a$ is shown in
Figure~\ref{assigned3.ps}.

 \begin{table}[ht]
 \begin{center}
 \caption{Results for $\Upsilon(1S) \rightarrow \gamma p \overline{p}$.}
 \begin{tabular}{c|c|c|c}
 \hline
 \hline
 Mode & Area & B.F. or 90 \% U.L. $(10^{{-5}})$ & Significance \\
 \hline
 $\ \gamma p \bar{p}\ (2-3\mass)$&
 $   85\pm    18$ & $   0.41\pm   0.08$ & $    4.85\sigma$ \\
 $\ \gamma f_J(2220), f_J(2220) \rightarrow p \bar{p}$& $<      20$ &
 $ <     9\times10^{-2}$ & - \\ 
 \hline
 \end{tabular}
 \label{Table:table_pp}
 \end{center}
 \end{table}

\begin{figure}[ht]
\begin{center}
\epsfig{bbllx=66,bblly=100,bburx=580,bbury=775,width=4.5in,clip=,file=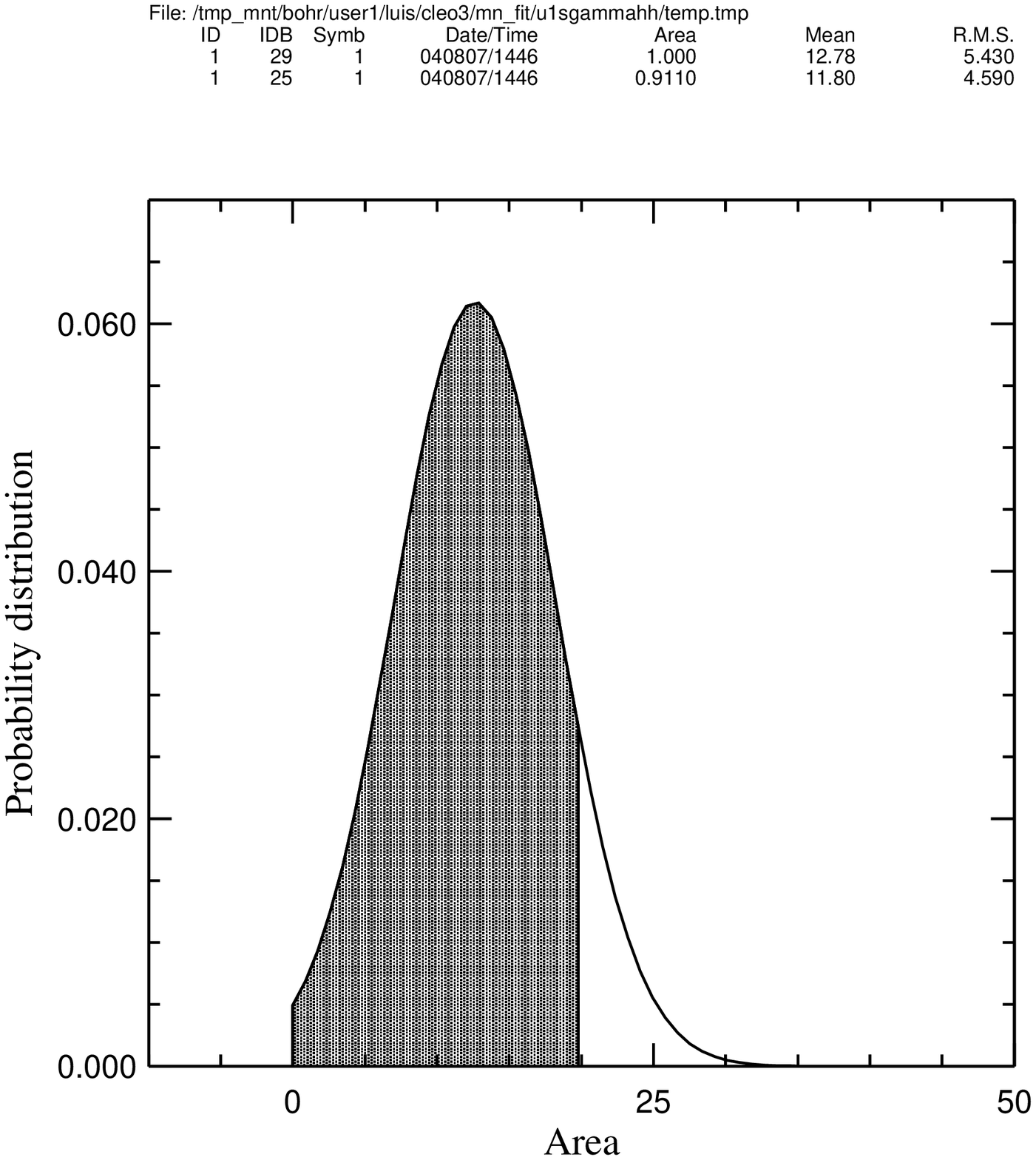}
\end{center}
\caption{Normalized probability distribution for different $f_J(2220)
\to p \bar{p}$
signal areas. The shaded area spans 90\% of the
probability.}\label{Figure:fj_prob3.ps} 
\end{figure}

\clearpage
\chapter{SYSTEMATIC UNCERTAINTIES}
Systematic uncertainties are any sources of experimental uncertainty
other than the statistical ones. Limits on the accuracy of our detector
simulation and any physical processes that interferes with the experimental 
measurement are typical examples of systematic uncertainties. These
uncertainties need to be identified, quantified, and if possible,
corrected.





\section{Cuts}

From now on we report individual cut efficiencies relative to the
events that survive all the other cuts.

According to our MC, most of the skim cuts, except for the hardGam
requirement, are
nearly 100\% efficient. Therefore, such cuts should
not be a source  of systematic uncertainty. The only skim cut worth
taking a closer look at is the hardGam cut, which is about 93\% efficient in
MC. Measuring this efficiency in our data is not possible because
events not classified as hardGam are not in the data to begin
with. Closer examination of our MC 
reveals that that most of the 7\% inefficiency in hardGam comes from the
eOverP1 cut (4\%), some from the Sh2 cut (2\%), and the rest (1\%) from the
other cuts present in hardGam. We can measure the efficiency of
eOverP1 in data by looking at the eOverP2 distribution of data tracks
from $\rho$ and $\phi$ decay. We are satisfied by this check on the MC
modeling (see  Tables ~\ref{Table:table_systematics_rho.tex} and ~\ref{Table:table_systematics_phi.tex}),
and we won't measure how well MC models the 
rest of the hardGam cuts, which are 97\% efficient in MC.

To quantify the quality in the MC modeling of our analysis cuts in the $\pi$
and $K$ modes we use the $\rho$ and $\phi$ signals present in
our data\footnote{A study using $K_S^0$ or $\Lambda$
signals from hadronic environments would have larger statistics, but
is problematic because the large number of tracks
and showers artificially decrease the cut efficiency. See~\cite{cleo3} Appendix
A.3 for an example.}. In this case the process is straightforward. We
measure the  $\rho$ and $\phi$ signal 
signals in data and MC over 
a floating background function with all cuts in place, and with all in
place cuts except the one under consideration. From these numbers we
calculate the effective efficiency of the cut. The
differences between data and MC are taken as the systematic
errors which are added in quadrature. Results are shown in Tables.
~\ref{Table:table_systematics_rho.tex}-~\ref{Table:table_systematics_phi.tex}.
The 4-momentum cut does not appear because its efficiency is close to
100\%.

 \begin{table}[ht]
 \begin{center}
 \begin{tabular}{c|c|c||c}
 \hline
 \hline
 Cuts & MC eff. & Data eff. & Systematic Error \\
 \hline
 eOverP2 & 96.2 & 96.6 & 0.4 \\
 QED $e^+e^- \to \gamma e^+e^-$ suppresion &     93.1 &      94.2 &       1.2 \\
 QED $\mu^+\mu^- \to \gamma \mu^+\mu^-$ suppresion  &     75.3 &     77.7 &      3.2 \\
 Hadron separation &     96.9 &      96.8 &     -0.1\\
 \hline
 \hline
 Overall analysis cut systematic error &&&       $\pm3$ \\
 \hline
 \end{tabular}
 \caption{Efficiencies for $\Upsilon(1S) \to \gamma \pi^+ \pi^-$
 for flat signal MC, efficiencies
 from data ($\rho$), and the derived systematic error in \%.
 Efficiencies are reported as the number of signal events after all cuts
 divided by the number of signal events with all cuts except the one
 under consideration.
 Statisticall errors are 0.1\% or less.}\label{Table:table_systematics_rho.tex}
 \end{center}
 \end{table}

 \begin{table}[ht]
 \begin{center}
 \begin{tabular}{c|c|c||c}
 \hline
 \hline
 Cuts & MC Eff. & Data Eff. & Systematic Error \\
 \hline
 eOverP2 & 98.5 & 100 & 1.5 \\
 QED $e^+e^- \to \gamma e^+e^-$ suppresion &     98.1 &      99.3 &       1.3 \\
 QED $\mu^+\mu^- \to \gamma \mu^+\mu^-$ suppresion  &     93.5 &     96.4 &      3.1 \\
 Hadron separation &     88.4 &      82.2 &      -7.0\\
 \hline
 \hline
 Overall analysis cut systematic error &&&      $\pm8$  \\
 \hline
 \end{tabular}
 \caption{Efficiencies for $\Upsilon(1S) \to \gamma K^+ K^-$
 for flat signal MC, efficiencies
 from data ($\phi$), and the derived systematic error in \%.
 Efficiencies are reported as the number of signal events after all cuts
 divided by the number of signal events with all cuts except the one
 under consideration.
 Statisticall errors are 0.1\% or less.}\label{Table:table_systematics_phi.tex}
 \end{center}
 \end{table}

For the proton case we don't have a clean sample with high statistics
of $e^+e^- \to \gamma p \overline{p}$ events in data. By extension we
take the systematic error in this mode to be 10\%.

\subsection{Justification of the DPTHMU Cut}

The reason we prefer to use $DPTHMU < 5$, instead of the more traditional
(see ~\cite{cleo1}) $DPTHMU <3$ used in CLEO II, is that, for some
unknown reason, our CLEO III MC has too many
$\pi$ tracks with $ 3 < DPTHMU < 5$.

\begin{figure}[ht]
\begin{center}
\epsfig{bbllx=66,bblly=150,bburx=580,bbury=700,width=4.5in,clip=,file=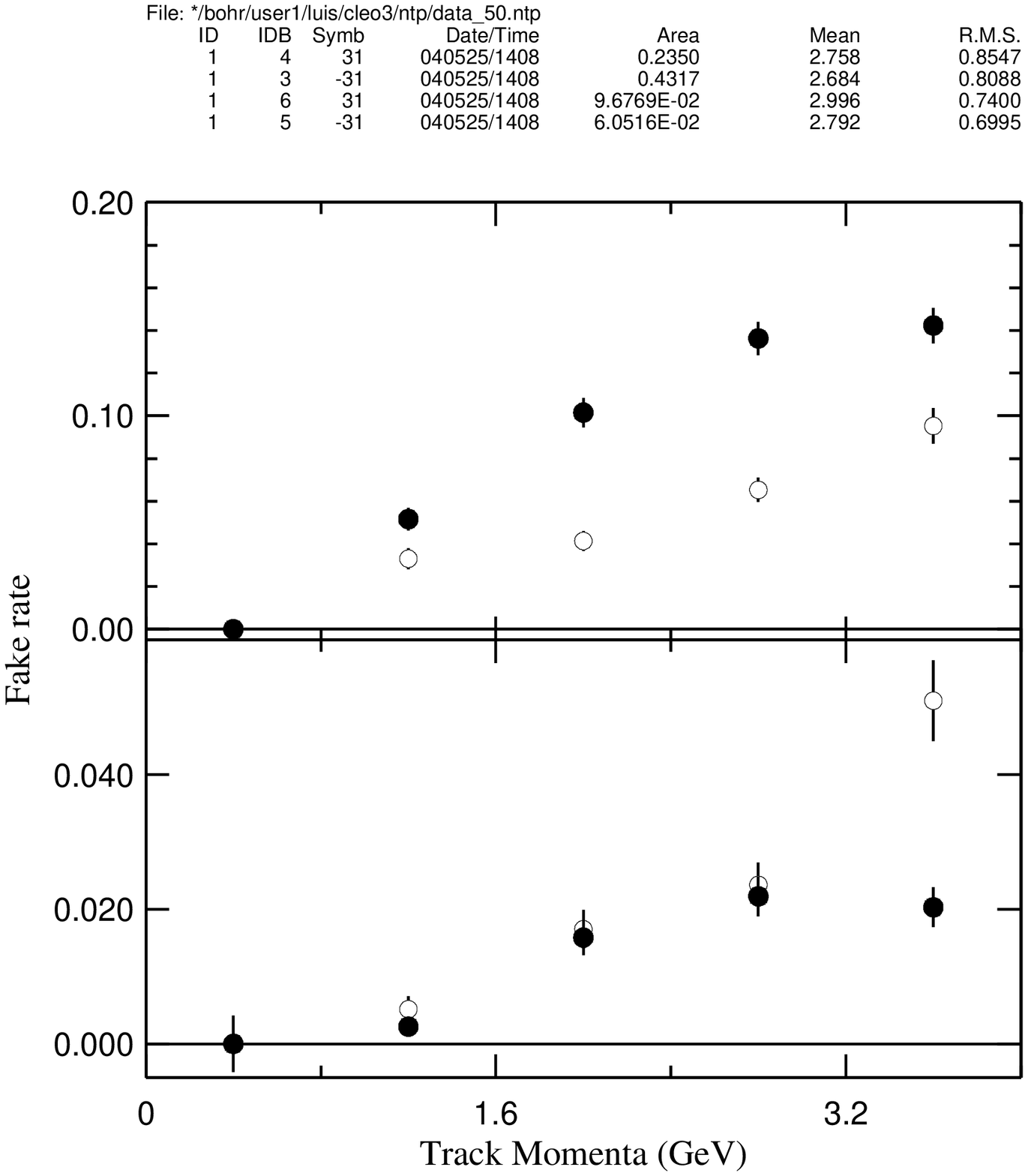}
\end{center}
\caption{Pion faking muon fake rates for a cut on $DPTHMU < 3$ (top) and
$DPTHMU < 5$ (bottom). MC is shown as solid circles while $\rho$ from
data is shown as hollow circles. Fake rates are reported relative to
all events that pass all cuts, except for the $DPTHMU$ cut for one of
the tracks.}\label{Figure:dpthmu.ps}
\end{figure}

To observe this fact we first select
a relatively clean sample of pion tracks by requiring the event to
have a $\pi^+ \pi^-$ invariant mass consistent with the $\rho$ mass,
and to pass all our analysis cuts except for the cut on $DPTHMU$ on one
track. For such events we plot the rate as a function of momentum, at
which the $\pi$ track whose DPTHMU cut we released has
$DPTHMU > 3$ and $DPTHMU > 5$. To increase our statistics we do this
procedure twice, once for each track, and average the fake
rate. Figure~\ref{Figure:dpthmu.ps} shows the results. Clearly the 
CLEO III MC we are using has some problem modeling the $DPTHMU <3$ cut. 

\begin{figure}[ht]
\begin{center}
\epsfig{bbllx=66,bblly=150,bburx=580,bbury=700,width=4.5in,clip=,file=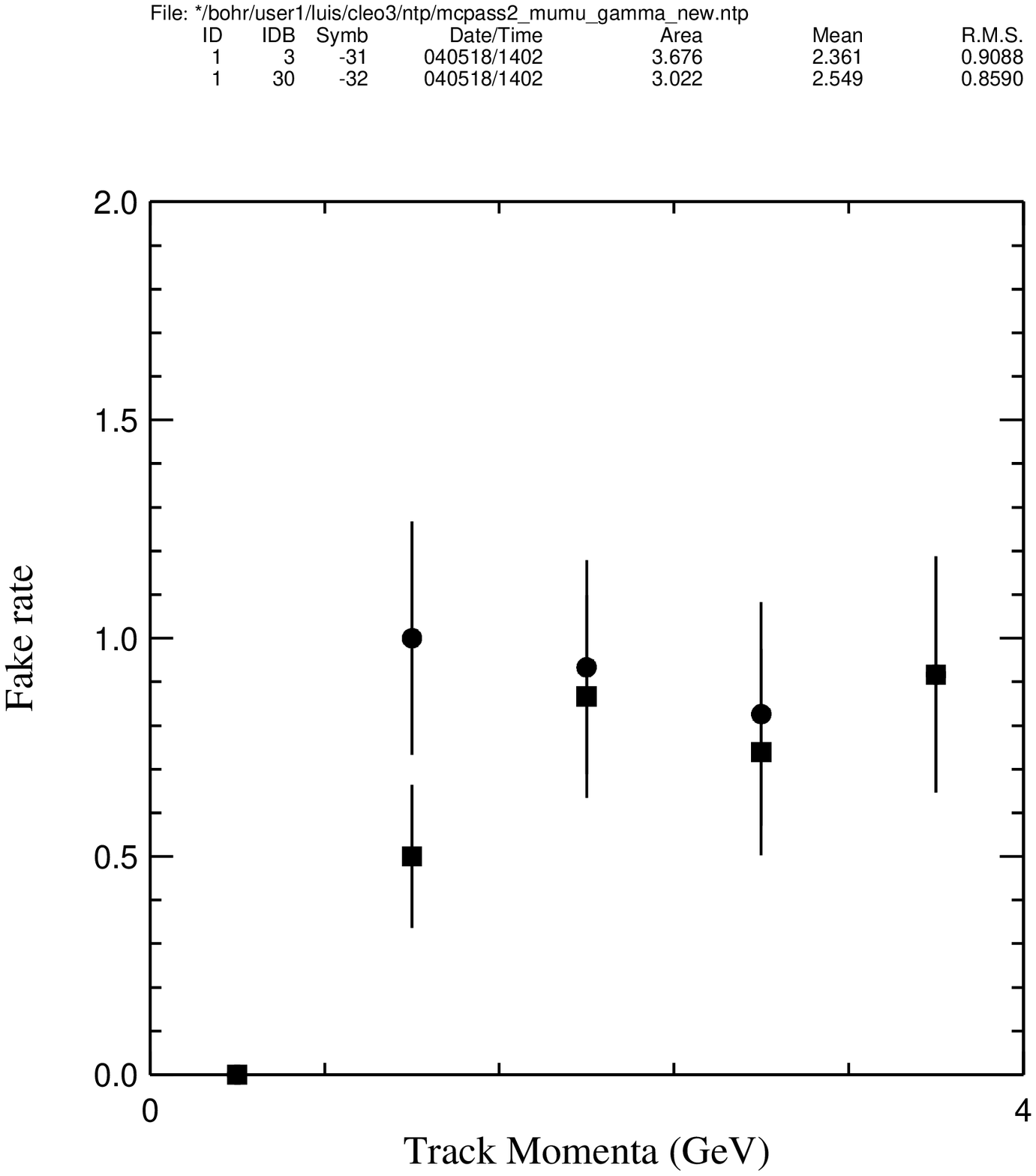}
\end{center}
\caption{Muon faking pion fake rate (in \%) for $DPTHMU < 3$ (solid
squares) and $DPTHMU < 5$ (solid circles). There is a complete overlap
in the last bin.}\label{Figure:mufakes.ps}
\end{figure}

To keep the systematic error low
we choose a cut at $DPTHMU < 5$ ($DPTHMU < 3$ gives a systematic
error of about 20\%). This does not change the efficiency
in data very much, but it increases the efficiency reported by MC,
bringing it closer to reality. The increase in $\mu$ fakes after
loosening the cut is estimated to be low using QED MC (see
Figure~\ref{Figure:mufakes.ps}).

\section{Angular Distribution of Signal}

The photon and tracks from the process $\Upsilon(1S) \to
\gamma X$ with $X \to h^+ h^-$ have a different angular distribution
than that of flat MC. Examples of possible angular distributions are
shown in the appendix. 

Figure ~\ref{Figure:systematics_helicity.ps} shows the efficiency in
flat MC as a function of $\cos\theta_{\gamma}$ and
$\cos\theta_{h^+}$\footnote{ $\theta_{\gamma}$ and $\theta_{h^+}$ are the
helicity angles of the sequential decay, defined in the appendix  (see
Figure ~\ref{Figure:drawing2.eps}).}. Note that the $K$
and $p$ modes are nearly insensitive to the track angular
distribution, while the $\pi$ mode is more sensitive to
$\theta_{h^+}$. This happens because of the stronger muon rejection cut
in the $\pi$ mode.

\begin{figure}[ht]
\begin{center}
\epsfig{bbllx=66,bblly=150,bburx=580,bbury=700,width=4.5in,clip=,file=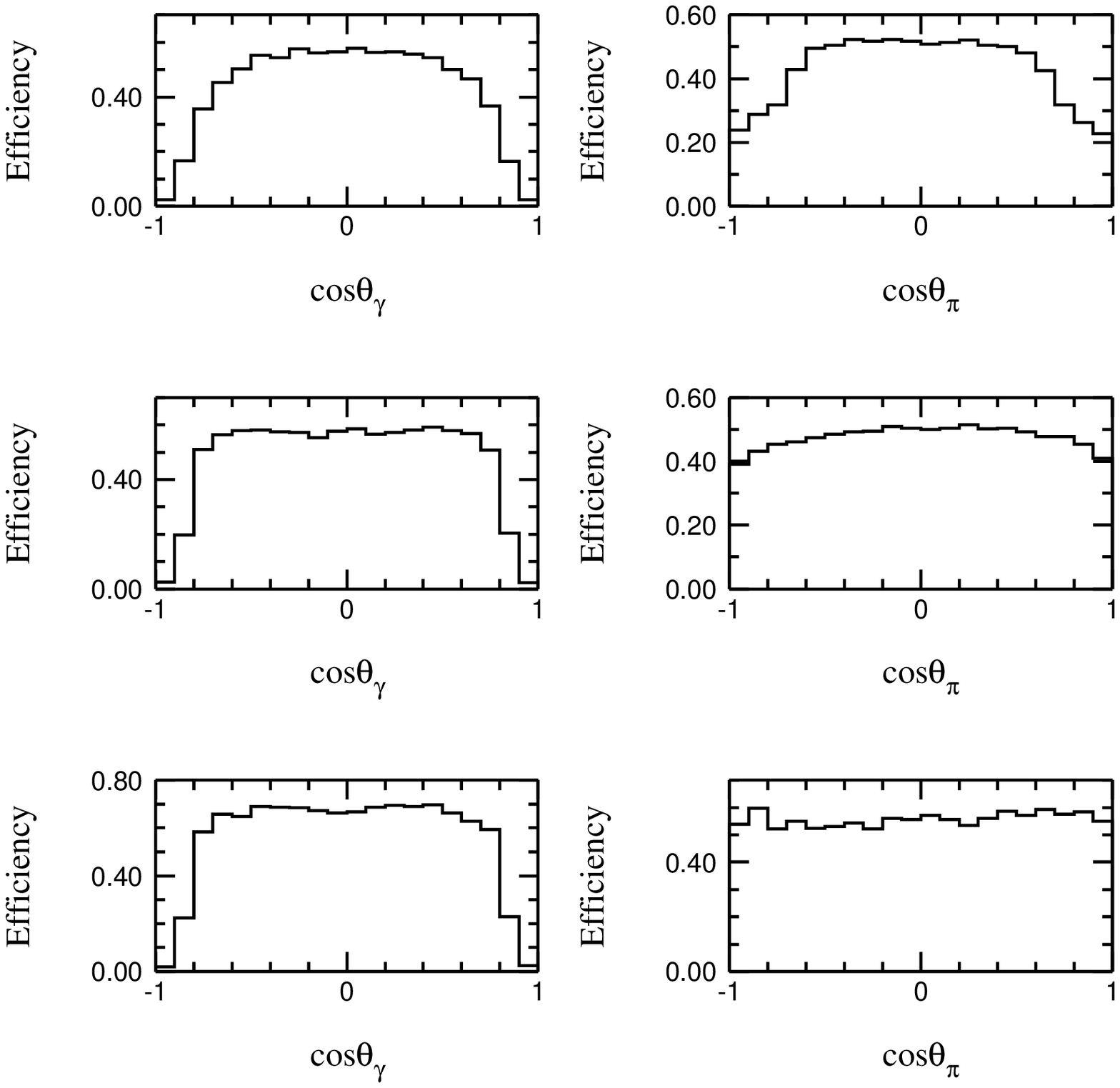} 
\end{center}
\caption{Flat MC efficiency as a function of $cos\theta_{\gamma}$
(left column) and as a function of $cos\theta_{h^+}$ (right
column). Top row corresponds to the pion mode, middle row to the
kaon mode and bottom row to the proton mode.}\label{Figure:systematics_helicity.ps}
\end{figure}

We measure the systematic effects of flat MC efficiency by convoluting each
plot in Figure~\ref{Figure:systematics_helicity.ps} with different
possible angular distributions calculated in the appendix. Tables
\ref{Table:table_systematics_helicity1.tex}-\ref{Table:table_systematics_helicity2.tex}  
show the necessary correction factors relative to flat MC efficiency
for decays with definite $\gamma$ and $X$ helicities $\lambda_{\gamma}$ and
$\lambda_X$, due to the non-flat photon and hadron distributions. We call
these factors $\epsilon^{J_X}_{\lambda_{\gamma}\lambda_{X}}(\gamma)$
and $\epsilon^{J_X}_{\lambda_{\gamma}\lambda_{X}}(h)$ respectively.

 \begin{table}[ht]
 \begin{center}
 \begin{tabular}{c||c|c}
 \hline
 \hline
 Mode & $a_{10}$ or $a_{12}$ & $a_{11}$ \\
 & $\Theta = 0$ & $\Theta = \frac{\pi}{2} $ \\
 \hline
 \hline
 $\pi$ &    0.910 &     1.18 \\
 K &    0.925 &     1.15 \\
 p &    0.922 &     1.16 \\
 \hline
 \hline
 \end{tabular}
 \caption{$\epsilon^{J_X}_{\lambda_{\gamma}\lambda_{X}}(\gamma)$,
 efficienciy correction with respect to flat MC factors due to the
 non-flat angular distribution of the photon
 in $\Upsilon(1S) \to \gamma X$ for different X
 spin values ($J_{X}$)
 and $\gamma$, $X$ helicities ($\lambda_{\gamma},\lambda_X$).}
 \label{Table:table_systematics_helicity1.tex}
 \end{center}
 \end{table}
 \begin{table}[ht]
 \begin{center}
 \begin{tabular}{c||c||c|c|c}
 \hline
 \hline
 Mode & $J_{X}$ & $a_{10}$ & $a_{11}$ & $a_{12}$ \\
 & & $\Theta = 0$, $\Phi = 0$ & $\Theta = \frac{\pi}{2}$ & $\Theta = 0$, $ \Phi = \frac{\pi}{2} $ \\
 \hline
 \hline
 & 0 & 1.00 & - & - \\
 & 1 &    0.785 &     1.11 &  - \\
 $X \to \pi^+\pi^-$ & 2 &    0.829 &    0.934 &      1.15\\
 & 3 &    0.898 &    0.853 &      1.03\\
 & 4 &    0.899 &    0.863 &     0.922\\
 \hline
 & 0 & 1.00 & - & - \\
 & 1 &    0.944 &     1.03 &  - \\
 $X \to K^+K^-$ & 2 &    0.950 &    0.988 &      1.04\\
 & 3 &    0.957 &    0.971 &      1.01\\
 & 4 &    0.965 &    0.962 &     0.992\\
 \hline
 \hline
 \end{tabular}
 \caption{$\epsilon^{J_X}_{\lambda_{\gamma}\lambda_{X}}(h)$,
 efficienciy correction factors with respect to non-flat MC due to the
 track angular distribution
 in $\Upsilon(1S) \to \gamma X$, with $X \to h^+ h^-$ for different X
 spin values ($J_{X}$)
 and $\gamma$, $X$ helicities ($\lambda_{\gamma},\lambda_X$).}
 \label{Table:table_systematics_helicity2.tex}
 \end{center}
 \end{table}

The efficiency of a decay with definite $\lambda_{\gamma}$,
$\lambda_X$ can be obtained using the flat MC efficiency corrected by
a factor $\epsilon^{J_X}_{\lambda_{\gamma}\lambda_{X}}$ =
$\epsilon^{J_X}_{\lambda_{\gamma}\lambda_{X}}(\gamma)\times
\epsilon^{J_X}_{\lambda_{\gamma}\lambda_{X}}(h)$. In general, the
final state is a mixture of all possible $\lambda_{\gamma}$,
$\lambda_X$ pairs, and the efficiency correction factor is,
\begin{equation}
\epsilon^{J_X} =
\cos^2\Theta\cos^2\Phi \epsilon^{J_X}_{10}
+  \sin^2\Theta \epsilon^{J_X}_{11}
+  \cos^2\Theta\sin^2\Phi \epsilon^{J_X}_{12}.
\label{Equation::ecorrect}
\end{equation}

The fits in
Figures~\ref{Figure:pipiplot_ct_cg_1.ps} and \ref{Figure:pipiplot_ct_cg_4.ps}
and~\ref{Figure:kkplot_ct_1.ps}-\ref{Figure:kkplot_ct_3.ps} measure
the pair $(\Theta,\Phi)$. These values are summarized in Table
~\ref{Table:sys_angle}, where they are used to obtain $\epsilon^{J_X}$
for each mode. 

\begin{figure}[ht]
\begin{center}
\epsfig{bbllx=66,bblly=150,bburx=580,bbury=700,width=4.5in,clip=,file=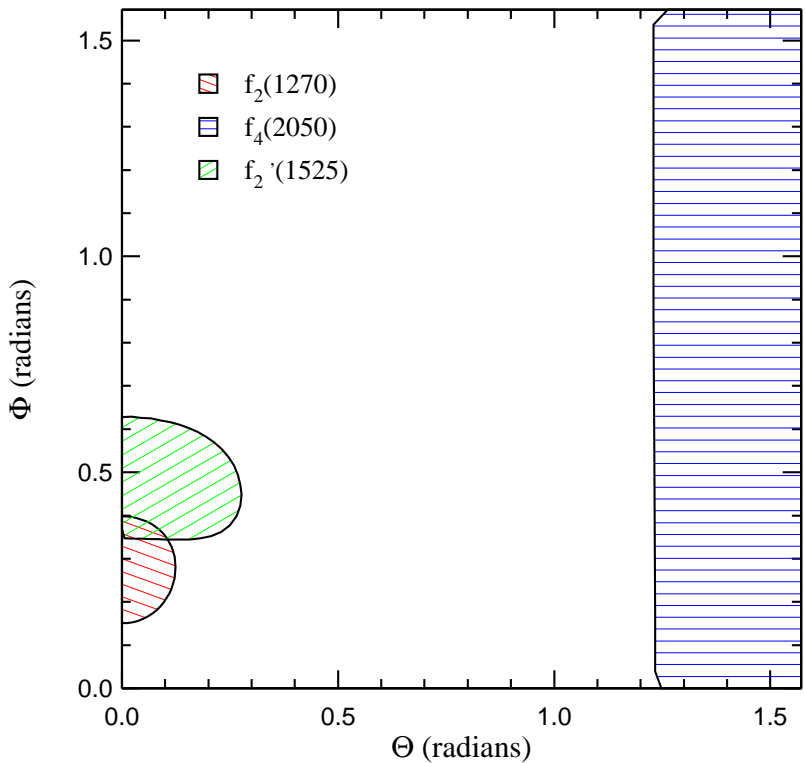}
\end{center}
\caption{Surfaces in the $(\Theta,\Phi)$ plane used to determine the
systematic uncertainty in the efficiency correction factor for the
modes with $f_2(1270)$, $f_4(2050)$, and $f_2^{'}(1525)$. The
$f_4(2050)$ surface may seem large, but when drawn in spherical
coordinates it is a small ``north pole cap''.}\label{Figure:elipse.ps}
\end{figure}

The pair  $(\Theta,\Phi)$ carry an
error which is a source of systematic uncertainty. We 
calculate this systematic uncertainty by inspecting the
differences in efficiency when $(\Theta,\Phi)$ move away
from the value which gives the minimum chi-squared,
$\chi^2_{min}$, under the condition $\chi^2 <
\chi_{min}^2+1$~\footnote{In a two dimensional linear problem such a
set of points defines the surface of the standard error
ellipse.}. For $J_X > 1$ both $(\Theta,\Phi)$ are free to move,
defining a surface in the $(\Theta,\Phi)$ plane. These surfaces are
shown in Figure~\ref{Figure:elipse.ps} for the $f_2(1270)$,
$f_4(2050)$, and $f_2^{'}(1525)$.

\begin{figure}[ht]
\begin{center}
\epsfig{bbllx=66,bblly=150,bburx=580,bbury=700,width=4.5in,clip=,file=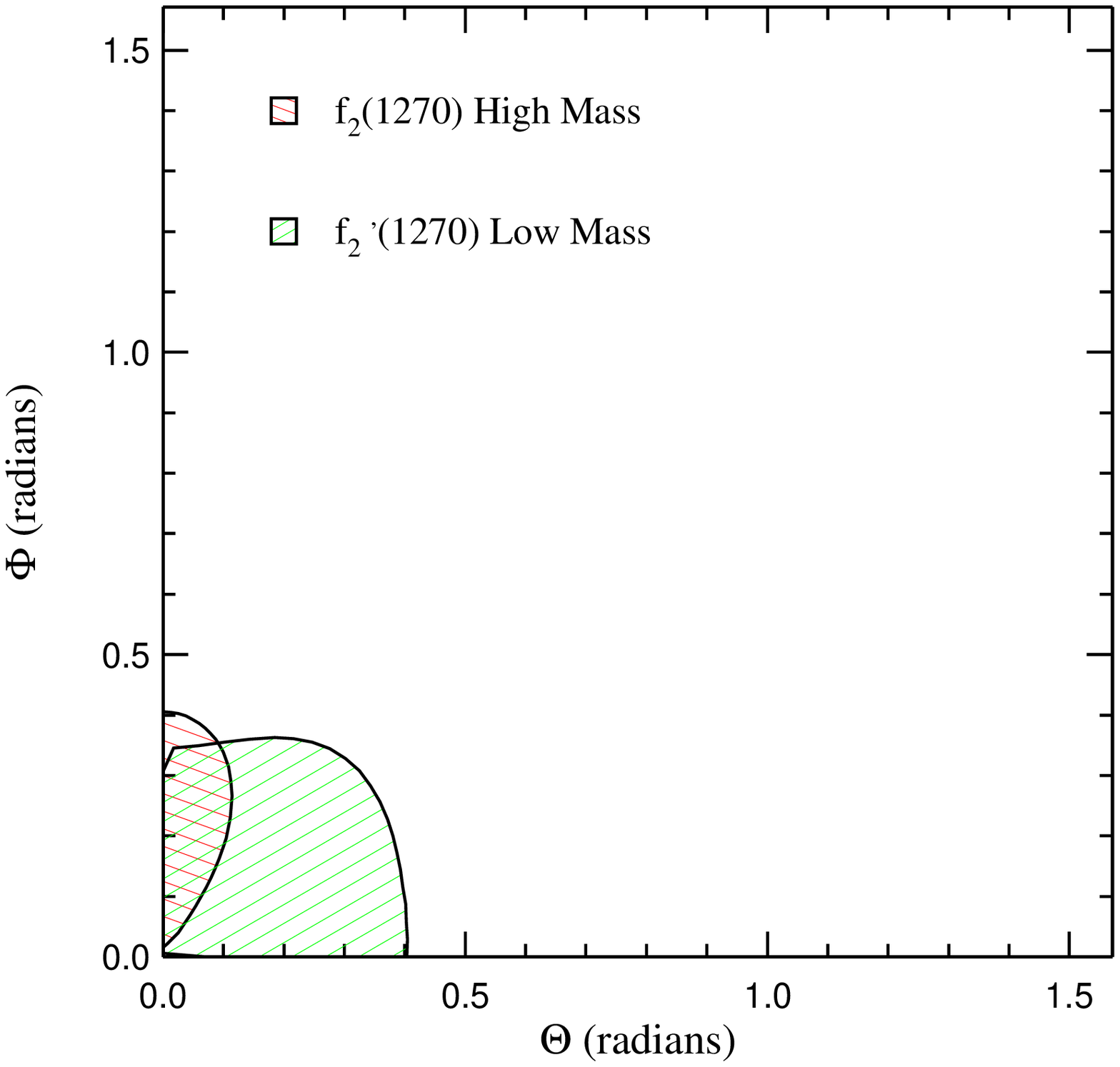}
\end{center}
\caption{Measured $(\Theta,\Phi)$ surfaces for the $f_2(1270)$ high
and low mass regions.}\label{Figure:elipse2.ps}
\end{figure}

\begin{figure}[ht]
\begin{center}
\epsfig{bbllx=66,bblly=150,bburx=580,bbury=700,width=4.5in,clip=,file=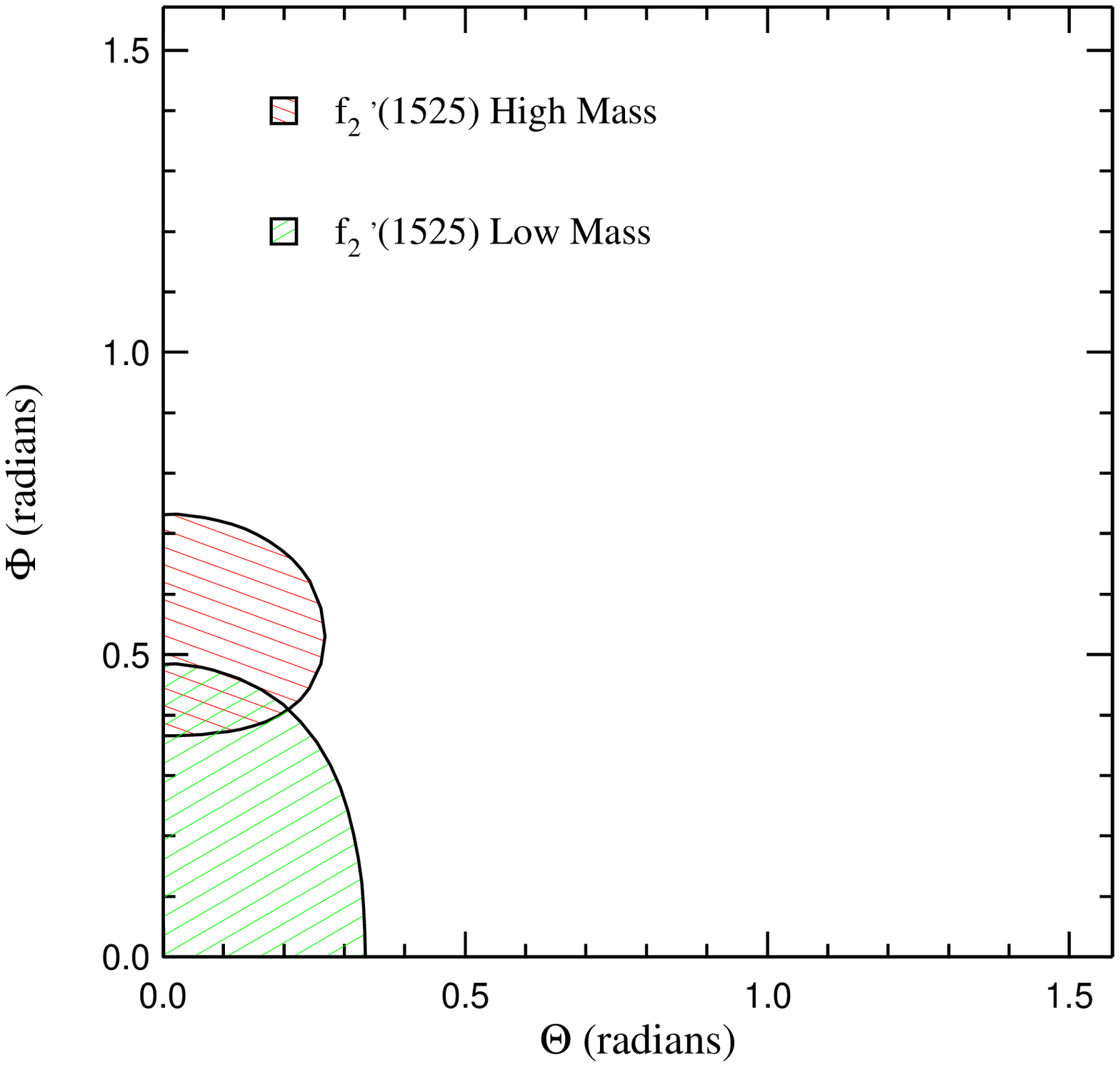}
\end{center}
\caption{Measured $(\Theta,\Phi)$ surfaces for the $f_2^{'}(1525)$ high
and low mass regions.}\label{Figure:elipse3.ps}
\end{figure}

At this point we can check whether $(\Theta,\Phi)$ depend on the mass
of the decay. We split the $f_2(1270)$ and the $f_2^{'}(1525)$ into a
high mass and a low mass region. The plots of the error
surfaces of the measured $(\Theta,\Phi)$ show no significant
separation for the different mass regions (see
Figures~\ref{Figure:elipse2.ps} and \ref{Figure:elipse3.ps}).

Results for the correction factor and its systematic error are shown
in Table~\ref{Table:sys_angle}.

Upper limits on $\Upsilon(1S) \to \gamma  f_2(2220)$, $f_2(2220) \to h^+
 h^-$  are changed to include the angular distribution's effect on
 efficiency. Since we can't measure the helicity amplitudes in this
 case, we choose the worst possible case where the corrected efficiency is
 lowest. This always corresponds to $\Theta=\Phi=0$. Results are in
 Table~\ref{Table:sys_angle2}.

\begin{landscape}
\vspace*{\fill}
 \begin{table}[ht]
 \begin{center}
 \caption{Measured $(\Theta,\Phi)$, calculated $\epsilon^{J_{X}}$,
 efficiency correction factor interval when
 $(\Theta,\Phi)$ move away from their measured value under the
 condition that $\chi^2$ stay within one unit of its minimum value,
 and the systematic  uncertainty on the correction factor.}
 \begin{tabular}{c||c|c|c|c||c|c}
 \hline
 Mode & $J_X$ & $\Theta$ & $\Phi$ & $\epsilon^{J_{X}}$ & Interval & Sys. Err. (\%) \\
 \hline
 $ \gamma f_0(980)$ & 0 & 0 & 0 & 0.91 & 0.91 - 0.91 &
 0 \\ 
 $ \gamma f_0(980)$ & 1 & $0.0^{+0.4}_{-0.0}$ & 0 & 0.71 &
0.71 - 0.80 & +13 \\
 $ \gamma f_2(1270)$ & 2 & 
 $0.00^{+0.08}_{-0.00}$ & $0.30^{+0.15}_{-0.10}$ & 0.78 & 0.76 - 0.80
 & $\pm3$ \\
 $ \gamma f_4(2050)$ & 4 & 
 $1.57^{+0.00}_{-0.34}$ & $0.2\pm0.7$ & 1.02 & 1.00 - 1.02
 & $-2$ \\
 $ \gamma f'_2(1525)$ & 2 &
 $0.00^{+0.27}_{-0.00}$ & $0.50^{+0.12}_{-0.16}$ & 0.90 & 0.89 - 0.91 &
 $\pm1$ \\ 
 $ \gamma f_0(1710)$ & 0 & 0 & 0 & 0.93 & 0.93-0.93
 & 0 \\
 $\gamma K^+ K^-$, $ 2\mass < m(K^+K^-) < 3
\mass$  & 1 & 
 $0.16^{+0.19}_{-0.16}$ & 0 & 0.88 & 0.87 - 0.91
 & $^{+3}_{-1}$ \\
 $\gamma p\bar{p}$, $ 2\mass < m(p\bar{p}) < 3
\mass$  & 1 & 
 $0.00^{+0.35}_{-0.00}$ & 0 & 0.92 & 0.92 - 0.95
 & $+3$ \\
 \hline
 \end{tabular}
 \label{Table:sys_angle}
 \end{center}
 \end{table}

 \begin{table}[ht]
 \begin{center}
 \caption{Worst case efficiency correction factors and their
 effect on upper
 limits of $\Upsilon(1S) \to \gamma  f_J(2220)$, $f_J(2220) \to h^+
 h^-$ decays.}
 \begin{tabular}{c|c|c}
 \hline
 Mode & Worst case correction & Upper limit increase (\%) \\
 \hline
 $f_J(2220) \to \pi^+ \pi^-$ & 0.75 & +33  \\
 $f_J(2220) \to K^+ K^-$ & 0.88 & +14 \\
 $f_J(2220) \to p \bar{p}$ & 0.92 & +9 \\
 \hline
 \end{tabular}
 \label{Table:sys_angle2}
 \end{center}
 \end{table}
\vfill
\end{landscape}

\section{Different Hadronic Fake Rates Between 1S and 4S}

Differences in the fake rates of hadronic events between the $1S$ and
$4S$ data are another potential source of systematic
errors when doing a continuum subtraction. 

In principle this is a second order effect since the contamination is
proportional to the difference of two small numbers. However since we
have such a large number of $\rho$ and $\phi$ events in the underlying
1S continuum, we should quantify any possible signal
contamination. Section 3.2.2
outlines the three cases we need to worry about; 
$\rho$ events contaminating the $K^+K^-$ invariant mass plot,
$\rho$ contaminating the $p\bar{p}$ invariant mass plot, and $\phi$ events
contaminating the $p\bar{p}$ invariant mass plot. 

It is reasonable to assume that the 1S and 4S fake rates
stay within 50\% of each other (see Section 3.3.2 for some examples from
MC measurements). With this assumption, taking $n_{\rho} \approx
20000$ and $n_{\phi} \approx 2000$ in the 1S data, and the fake rates
measured in data (see
Table~\ref{Table:table_ef}) we can calculate the 
systematic uncertainties. 
We expect $\pm27$ events from $\rho$ to
contaminate the $K^+K^-$ invariant mass plot. However, these events can
be ignored as source of systemtic uncertnty because 90\% of them
fall in the $1-1.5 \mass$ region, below the $f_2^{'}(1525)$ peak. %
We also expect $\pm 7$ from $\rho$ and  $\pm 3$
events from $\phi$ contaminating the $p \bar{p}$ invariant mass
plot. This represents a $\pm 8\%$ systematic error for the
$\Upsilon(1S) \to \gamma p \bar{p}$ mode.

\section{Other Systematic Sources}

Besides the systematic uncertainties described above, we also add a 2\%
systematic effect from track finding (1\% per track) and 5\% from the
number of $\Upsilon(1S)$.

We find no evidence for non-resonant hadron pairs that could interfere
with the resonances. There is a possible interference between
$f_2(1275)$ and $f_2^{'}(1525)$ because they both are in the J = 2
state. From the $f_2(1275)$ measurement in the $\pi^+\pi^-$ mode we expect 3
events from this source in the $f_2^{'}(1525)$ mass region $1.45-1.6\
\mass$. In the most extreme cases we have, $N \propto
||A_{f_2^{'}(1525)}||^2+||A_{f_2(1270)}||^2
\pm2||A_{f_2^{'}(1525)}||||A_{f_2(1270)}||$, 
where $A_{f_2^{'}(1525)}$ and $A_{f_2(1270)}$ are the amplitudes
associated with $f_2^{'}(1525)$ and $f_2(1270)$ respectively. From the
data we have $ N \approx 300$, and using the fact that
$||A_{f_2(1270)}||^2$ contributes with 3 events, the contribution from
$||A_{f_2^{'}(1525)}||^2$ can range from 400 to 200 events. This
represents a 30\% variation from the central value. Since this is the extreme
interference case, we take half of this value, 15\%, as a reasonable
systematic uncertainty from possible interference in the
$f_2^{'}(1525)$.

The possible decay $\Upsilon(1S) \to \rho \pi$ could contaminate our
$\pi^+$ $\pi^-$ invariant mass plot, since a $\pi^0$ with a momentum
greater than $4 \mass$ looks very much like a photon. Using phase
space MC and isospin symmetry, we estimate that about 40\% of possible
$\Upsilon(1S) \to \rho \pi$ have such a $\pi^0$. The latest measured
upper limit 
~\cite{cleo4} for this 
decay mode is $1.9 \times 10^{-6}$. This gives an upper limit
of 16 events in the efficiency corrected $\pi^+$ $\pi^-$ invariant mass
plot, of which 10 events fall in the $\rho$ mass region, behind the
$f_0(980)$ mass region. The remaining potential 6 events could
contaminate the rest of the invariant mass region. None of the signal
regions is affected by more than 1\%.

To test the robustness of the statistical fits, we redo them with all masses
and widths of possible resonances required to stay within $1\sigma$ of
their nominal PDG values
~\cite{pdg}. The results from fitting the data in this way are within the
statistical errors when compared to the fit results
summarized in Table~\ref{Table:table_pipi} and
~\ref{Table:table_kk}. We do not observe a systematic difference
between these two fitting techniques.

The branching ratios for $f_2(1270) \to \pi^+\pi^-$, $f_4(2050) \to
\pi^+\pi^-$, and $f_2^{'}(1525) \to K^{+}K^{-}$ are taken from the
PDG~\cite{pdg} and contribute to the systematic uncertainty with
$^{+2}_{-3}$, $\pm9$, and $\pm3$ percent respectively.

\section{Overall Systematic Uncertainties}

We combine the systematic errors of each section in quadrature.
Tables~\ref{Table:final_sys}-~\ref{Table:final_sys2} summarize the
systematic errors.

 \begin{table}[ht]
 \begin{center}
 \caption{Systematic errors expressed as \% from the different sources
 described in the text. Not 
 shown in the table, but included in the total, are a systematic uncertainty
 of 5\% from the number of $\Upsilon(1S)$, a 2\% systematic uncertanty
 from MC
 tracking in all the modes, a 15\% systematic uncertainty in the
 $\gamma f'_2(1525)$ from possible interference with $\gamma
 f_2(1270)$, an 8\% systematic uncertainty in the  $ \gamma p
 \bar{p}$ mode from possible hadronic contamination, and the
 systematic uncertanties in the $f_2(1270)$, $f_4(2050)$, and
 $f_2^{'}(1525)$ hadronic branching fractions.} 
 \begin{tabular}{c||p{2.cm}|p{2.2cm}|p{2.62cm}p{2.4cm}c}
 \hline
 \hline
 Mode & Analysis Cuts & Angular Distribution & Total \\
 \hline
 \hline 
 $ \gamma f_0(980)$ & $\pm3$ & 0 & $\pm6$ \\
 $ \gamma f_2(1270)$ & $\pm3$ & $\pm3$ &  $\pm7$ \\ 
 $ \gamma f_4(2050)$ & $\pm3$ & $+2$ & $\pm7$ \\
 \hline
 $ \gamma f'_2(1525)$ & $\pm8$ & $\pm1$ & $\pm20$ \\
 $ \gamma f_0(1710)$ & $\pm8$ & 0 & $\pm10$ \\
 $ \gamma K^+K^- (2-3 \mass)$ & $\pm8$ & $^{+1}_{-3}$ & $\pm10$ \\     
 \hline
 $ \gamma p \bar{p}(2-3 \mass)$ & $\pm10$ & $-3$ & $\pm14$ \\     
 \hline
 \hline
 \end{tabular}
 \label{Table:final_sys}
 \end{center}
 \end{table}

 \begin{table}[ht]
 \begin{center}
 \caption{Increase in the upper limits of $\Upsilon(1S) \to \gamma
 f_J(2220)$ for different $f_J(2220)$ decay modes due to the possible
 systematic effects described in the text added in quadrature. Not
 shown in the table, but 
 included in the total, are a  
 5\% contribution from the number of $\Upsilon(1S)$ and of 2\%
 contribution from MC tracking.} 
 \begin{tabular}{c||p{2.cm}|p{2.2cm}|p{2.62cm}p{2.4cm}c}
 \hline
 \hline
 Mode & Analysis Cuts & Angular Distribution & Total \\
 \hline
 \hline
 $ f_J(2220) \to \pi^+ \pi^-$ & $\pm3\%$ & +33\% & +34\% \\
 $ f_J(2220) \to K^+ K^-$ & $\pm8\%$ & +14\% &  +17\% \\
 $ f_J(2220) \to p \bar{p}$ & $\pm10\%$ & +9\% & +14\% \\
 \hline
 \hline
 \end{tabular}
 \label{Table:final_sys2}
 \end{center}
 \end{table}

\clearpage

\clearpage
\chapter{RESULTS AND CONCLUSION}

We report on a new search for two-body radiative $\Upsilon(1S)$
decays. We place stringent
limits on the production of the $f_J(2220)$ particle in the pion and
kaon modes, and a 
less stringent limit in the proton mode where some excess of events is
observed in the region of interest.

In the decay channel $\Upsilon(1S) \to \gamma\pi^+\pi^-$ we find clear
evidence for the resonance $f_2(1270)$  
and measure a branching fraction of  $(13.3 \pm 1.4) \times 10^{-5}$, which is
consistent with the 
earlier CLEO measurement of $(7.4^{+2.7}_{-1.8}) \times 10^{-5}$
\cite{cleo1}. The angular 
distributions of the photon and tracks strongly indicate that the
hadron pairs we assign to the $f_2(1270)$ are indeed in a $J =
2$ state, and that nature prefers to produce the $f_2(1270)$ with 0
helicity. In contrast, for $J/\psi \to \gamma f_2(1270)$ it was
found~\cite{oldy} 
that the $f_2(1270)$ is produced at equal rates with both helicity 0
and helicity 1, and at a rate consistent with 0 for helicity 2, but
more recently~\cite{newestBES} measured helicity 0 dominance for this
same mode.  

There is a barely significant ($4.3\sigma$) excess of events in the
$f_0(980) \to \pi^+\pi^-$ invariant mass region. The angular 
distributions of the  photon and tracks indicate that this excess of
events is in a $J=1$ 
state,  rather than $J=0$ which would be the
case if the excess were due to $f_0(980)$ decays. It may be that this
excess is due to non-resonant $\gamma \pi^+\pi^-$ production. We conclude that
more data is needed to resolve this situation. 


We find weak evidence for the production of the resonance
$f_4(2050)$ in the $\Upsilon(1S) \to \gamma\pi^+\pi^-$ decay channel.

In the decay channel $\Upsilon(1S) \to \gamma K^+ K^-$ we find strong
evidence for the production of the resonance $f_2'(1525)$
and weak evidence for the production of $f_0(1710)$. The
photon and track angular distributions show that the two tracks
attributed to the $f_2'(1525)$ are
indeed in a $J = 2$ state and that the $f_2'(1525)$ is produced mostly
with helicity 0. This is consistent with Kramer's theoretical
prediction~\cite{very_old} and similar measurents done in the
$J/\psi$ system~\cite{kkBes}.

There is also 
evidence of an excess of events in the $2-3\mass$
$K^{+}K^{-}$ invariant mass region, which we cannot attribute to any
known resonances. 

Finally, we also find evidence of an excess of events in the
channel $\Upsilon(1S) \to \gamma p \bar{p}$, which we cannot attribute
to any known resonances. We find no evidence of an enhancement near
$p\bar{p}$ threshold. Such an enhancement was recently reported
by~\cite{ppbar} in the $J/\psi$ system, and is currently being
interpreted (See e.g.~\cite{ppt1},~\cite{ppt2},~\cite{ppt3}).

When comparing the radiative $\Upsilon(1S)$ decays studied in this
analysis to radiative $J/\psi$ decays we observe a suppression ratio
of 0.09 that is in reasonable agreement with with the
naive expectation of 0.025 obtained from scaling
arguments.

Table \ref{Table:brs} summarizes our results.

While we find hadronic resonances at a level consistent with the
estimate of Section 1.2, we do not observe any glueball states. This
is not the result we would expect from a naive interpretation of QCD,
as the two gluons from a \Uis radiative decay can form a bound state
directly and need to go through at least two strong interaction
vertexes to form a hadron. 

Internal consistency in QCD predicts the existence of glueballs. Even
though theoretical calculations of the glueball mass spectrum exist, a clear
glueball observation has not been made yet. We conclude that more glueball
searches are necessary to clarify this situation.

\begin{landscape}
\vspace*{\fill}
\begin{table}[ht]
\begin{center}
\caption{Final measured branching ratios, measured branching ratios reported
relative to $J/\psi$ branching ratios, and statistical significance for
each decay channel. The measured branching ratios have been corrected
by the factors calculated in Appendix A2. For the branching ratios the
first uncertainty is statistical and the second is
systematic.}\label{Table:brs} 
\begin{tabular}{c|c|c|c}
\hline
\hline
Channel & Branching Fraction $\times(10^-5)$
&$\frac{\Upsilon(1S)}{J/\psi}$ & Significance \\
\hline 
\hline 
$ \gamma f_0(980),\ f_0(980) \to \pi^+ \pi^-$ & 
$1.8^{+0.8}_{-0.7} \pm0.1$ & - & $4.3\sigma$ \\
$ \gamma f_2(1270)$  & $13.3 \pm 1.0\pm0.9 $
& $0.09\pm0.01$ &  $> 14 \sigma $\\
$ \gamma f_4(2050)$ & $3.5\pm1.3\pm0.2$ &
$0.019 \pm 0.006$ & $2.6\sigma$ \\ 
\hline 
$ \gamma f_2'(1525)$ & $4.3^{+1.0}_{-0.8}\pm0.9$ &
$0.09^{+0.05}_{-0.04}$ & $> 14\sigma$\\ 
$ \gamma f_0(1710)$, $f_0(1710) \to K^+ K^-$ & 
$0.38\pm0.16\pm0.04$ & $0.007^{+0.004}_{-0.005}$ & $3.2\sigma$\\
$ \gamma K^{+} K^{-}$ & $1.14\pm0.14\pm0.11$ &
- & $9.1\sigma$ \\
\hline 
$ \gamma p \bar{p} $ & $0.41\pm0.08\pm0.10$ &
$0.011\pm0.005$ & $4.8\sigma$ \\
\hline 
$ \gamma f_J(2220)$, $f_J(2220) \to \pi^+ \pi^-$ &
$<0.08$ & - & - \\ 
$ \gamma f_J(2220)$, $f_J(2220) \to K^+ K^-$ &
$<0.06$ & - & - \\ 
$ \gamma f_J(2220)$, $f_J(2220) \to p \bar{p}$ &
$<0.11$ & - & - \\ 
\hline
\hline
\end{tabular}
\end{center}
\end{table}
\vfill
\end{landscape}

\clearpage

\clearpage
\appendix %

\addcontentsline{toc}{extrachapter}{APPENDIX \protect\hspace*{1.0em}%
 HELICITY FORMALISM FOR TWO BODY DECAYS}%

\chapter*{APPENDIX \\ HELICITY FORMALISM FOR TWO BODY DECAYS}%
\newcommand{\calA}{\mathcal{A}\mathrm{}}
\newcommand{\calC}{\mathcal{C}\mathrm{}}
\newcommand{\calAC}{\mathcal{AC}\mathrm{}}

\begin{figure}[t]
\begin{center}
\epsfig{clip=,file=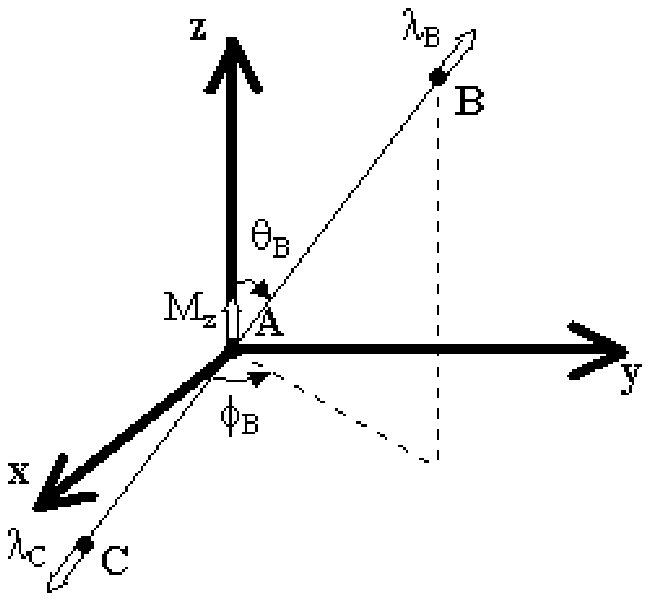}
\end{center}
\caption{$A \to B + C$.}\label{Figure:drawing.eps}
\end{figure}

Helicity formalism can be used to obtain the angular distribution of
a decay process. In this appendix we restrict ourselves to the case of
two body decays. The situation is as follows: particle A  with
spin $J_A$, and z-axis spin projection $M_A$ decays to two particles, $A
\to B + C$ with definite helicity $\lambda_B$ and $\lambda_C$. Let the
direction of the of the decay be $\hat{n}(\theta,\phi) = \hat{n}(\theta_B,\phi_B) =
\hat{p}_B = -\hat{p}_C$ (see Figure ~\ref{Figure:drawing.eps}). The
amplitude for this decay is,
\begin{equation}
        \calA_{\lambda_B\lambda_C}^{M_A} = \mathnormal{<\theta\phi\lambda_B\lambda_C|H|J_AM_A>.}
\label{Equation::basic}
\end{equation}
 
We have little chance of directly
calculating~\ref{Equation::basic}. But we can exploit the fact that $H$ conserves angular
momentum by considering the basis
$\{<jm\lambda_B\lambda_C|\}$, where j is the total angular
momentum of $B+C$ and m its projection along the z axis\footnote{The
rotational invariance of helicities is a necessary condition to construct
$\{<jm\lambda_B\lambda_C|\}$ as a basis for the two-particle
state $B+C$. Rotational invariance is one of two key ideas in helicity
formalism. The other key idea, which is boost invariance, is
useful in sequential decays.}. Inserting this basis
into Equation~\ref{Equation::basic},

\begin{eqnarray*}
\calA_{\lambda_B\lambda_C}^{M_A}
&=&\sum_{jm}<\theta\phi\lambda_B\lambda_C|jm\lambda_B\lambda_C><jm\lambda_B\lambda_C|H|J_AM_A>
\nonumber \\
&=&\sum_{jm}<\theta\phi\lambda_B\lambda_C|jm\lambda_B\lambda_C>\delta_{jJ_A}\delta_{mM_A}A_{\lambda_B\lambda_C}
\\
&=&<\theta\phi\lambda_B\lambda_C|J_AM_A\lambda_B\lambda_C>A_{\lambda_B\lambda_C}.
\nonumber
\end{eqnarray*}

The above equation contains two factorized terms. The first term is independent
of the interaction $H$ and contains the angular distribution
information we are interested in. The second term,
$A_{\lambda_B\lambda_C}$, is usually called the ``helicity coupling
amplitude''. It is independent of the angular distribution, and contains the
physics of the decay (in particular it depends on $\lambda_B$ and
$\lambda_C$, but not on $M_A$). In the helicity formalism, the helicity coupling
amplitudes are unknown parameters.

For strong and electromagnetic decays there is another invariance of $H$
we can exploit; parity. For these kind of interactions it can be shown
by inserting the parity operator that,
\begin{equation}
A_{-\lambda_B-\lambda_C}=(-1)^{J_A-J_B-J_C}\eta_A\eta_B\eta_C
A_{\lambda_B\lambda_C}
\end{equation}
where $\eta_A\eta_B\eta_C$ is the product of the parity of particles A, B
and C. This useful property of $A_{\lambda_B\lambda_C}$ reduces the number of
unknown parameters in the angular distribution of strong and
electromagnetic decays.

The next step is to derive explicit formulae for
$<\theta\phi\lambda_B\lambda_C|J_AM_A\lambda_B\lambda_C>$ for the
$B+C$ system. Such
derivation, which is out of the scope of this appendix, will be omitted
here\footnote{The interested reader can find this derivation in
~\cite{helicity}}. The result can be expressed in terms of the Wigner
functions,
\begin{equation}
<\theta\phi\lambda_B\lambda_C|J_AM_A\lambda_B\lambda_C> =
\sqrt{\frac{2J+1}{4\pi}}D_{M_A,\lambda}^{*J_A}(\phi,\theta,-\phi)
\end{equation}
where $\lambda = \lambda_B-\lambda_C$ and,
\begin{equation}
D_{M_A,\lambda}^{J_A}(\phi,\theta,-\phi) = e^{-i\phi(M_A-\lambda)}d^{J_A}_{M_A,\lambda}(\theta).
\end{equation}
The functions $d_{M_A,\lambda}^{J_A}(\theta)$ have real values, and
some of them are given in the particle data book.

We have finally arrived to the following important relation,
\begin{equation}
\calA_{\lambda_B\lambda_C}^{M_A}
=\sqrt{\frac{2J_A+1}{4\pi}}D_{M_A,\lambda}^{*J_A}A_{\lambda_B\lambda_C}.
\end{equation}
All helicity distribution calculations are reduced to
handling these basic amplitudes. Here are some of their very important
properties,
\begin{equation}
\calA_{\lambda_B\lambda_C}^{M_A} = 0 ~~~if ~~~|\lambda_B-\lambda_C| > J_A,
\label{Equation::rules}
\end{equation}
\begin{equation}
\parallel \calA_{\lambda_B\lambda_C}^{-M_A} \parallel =
\parallel \calA_{\lambda_B\lambda_C}^{M_A} \parallel,
\label{Equation::norm}
\end{equation}
\begin{equation}
\int_0^{2\pi}d\phi
\calA_{\lambda_B\lambda_C}^{M_A}\calA_{\lambda_B^{'}\lambda_C^{'}}^{*M_A^{'}}
=
2\pi
\delta^{(M_A-M_A^{'})}_{(\lambda-\lambda^{'})}
\calA^{M_A}_{\lambda_B\lambda_C}
\calA^{*M_A^{'}}_{\lambda_B^{'}\lambda_{C}^{'},}\label{Equation::phiprojection}
\end{equation}
\begin{equation}
\int d\Omega
\calA_{\lambda_B\lambda_C}^{M_A}\calA_{\lambda_B^{'}\lambda_C^{'}}^{*M_A^{'}}
=\delta^{(M_A-M_A^{'})}_{(\lambda-\lambda^{'})}\parallel
A_{\lambda_B\lambda_C}
\parallel^2.
\label{Equation:projection}
\end{equation}
Property~\ref{Equation::rules} is a reflection of the
physical fact that the projection of the total angular momentum along
a particular axis cannot be  greater than the total angular momentum itself.

$\parallel \calA_{\lambda_B\lambda_C}^{M_A} \parallel^2$ is the
angular distribution of a final state where $M_A$ and the helicities of
$B$ and $C$ are known (see Equation~\ref{Equation::basic}). However, it
is usually the case that $M_A\lambda_B\lambda_C$ are not known. In such
cases we can only measure the overall angular distribution, which is
simply the average over possible initial states of the
incoherent sum of final states,
\begin{equation}
\frac{dP}{d\Omega_B} = \sum_{M_A\lambda_B\lambda_C}P_{M_A}\frac{dP_{\lambda_B\lambda_C}^{M_A}}{d\Omega_B} = \sum_{M_A\lambda_B\lambda_C}P_{M_A}\parallel
\calA_{\lambda_B\lambda_C}^{M_A} \parallel^2
\end{equation}
where $P_{M_A}$ is the weight of the initial state $M_A$ and $d\Omega_B = d(cos(\theta_B))d\phi_B$.

\begin{figure}
\begin{center}
\epsfig{clip=,file=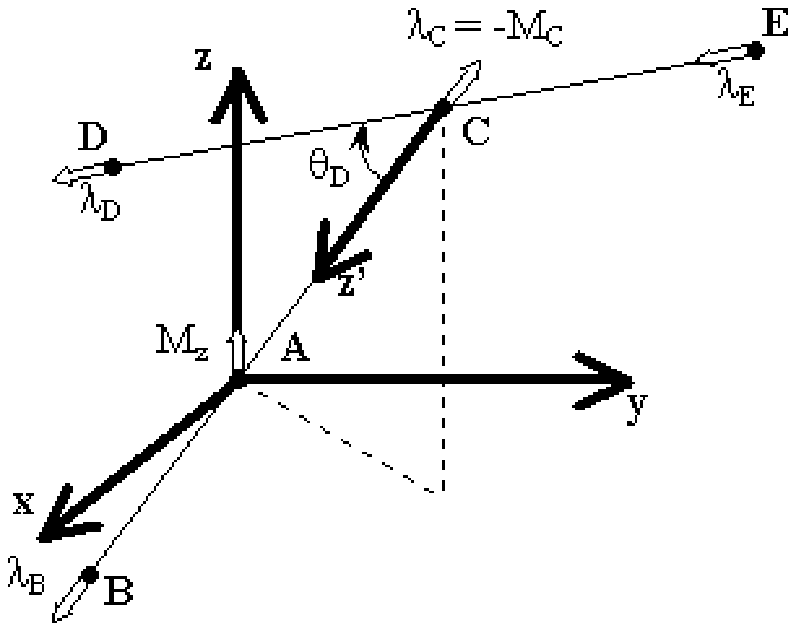}
\end{center}
\caption{$A \to B + C$ and $C \to D + E$. The drawing is peculiar in
that $C$'s daughters are drawn as seen in $C$'s rest
frame (back-to-back).}\label{Figure:drawing2.eps}
\end{figure}

Next, consider a sequential decay, $A \to B+C$ with $C \to
D+E$ (see Figure ~\ref{Figure:drawing2.eps}). The rotational and
boost invariance of helicities makes it easy
to extend the helicity formalism to this case.
To do so we consider a new reference frame, O', at rest with $C$,
such that $\hat{z}'=\hat{n}(\theta_B,\phi_B)$\footnote{This condition
does not unlikely determine O'. We can be more precise by defining
$O' = L(\vec{p}_C)R(\phi_B,\theta_B,0)O$, where $L$ is the Lorenz Boost
operator, $R$ is the rotation operator in terms of the Euler
angles, and $O$ is $A$'s original rest frame.}. All particle
helicities remain the
same in O', and in particular $M_z(C) = -\lambda_C$ (the negative sign arises from
the fact that $\hat{n}(\theta_B,\phi_B)$ is the direction in which $B$ is
traveling).  It is now straightforward to calculate the intermediate
amplitude $\calC_{\lambda_D\lambda_E}^{-\lambda_C}$ at O'. Of course
$\calA_{\lambda_B\lambda_C}^{M_A}$ is still calculated at $A$'s
original rest frame. Since $\lambda_C$ is an intermediate state, the
helicity amplitude for the final state $\lambda_B\lambda_D\lambda_E$
is,

\begin{equation}
\calAC_{\lambda_B\lambda_D\lambda_E}^{M_A} = \sum_{\lambda_C}
\calA_{\lambda_B\lambda_C}^{M_A}\calC_{\lambda_D\lambda_E}^{-\lambda{C}}
\label{Equation::chain1}
\end{equation}
where we have coherently summed over the intermediate state $C$. If
$M_A\lambda_B\lambda_D\lambda_E$ are not known,
\begin{equation}
\frac{dP}{d\Omega_B d\Omega_D} =
\sum_{M_A\lambda_B\lambda_D\lambda_E}P_{M_A}\frac{dP_{\lambda_B\lambda_D\lambda_E}^{M_A}}{d\Omega_B
d\Omega_D}=\sum_{M_A\lambda_B\lambda_D\lambda_E}P_{M_A}\parallel
\calAC_{\lambda_B\lambda_D\lambda_E}^{M_A} \parallel^2.
\label{Equation::chain2}
\end{equation}

The sequential decay amplitudes have the following important
properties,
\begin{equation}
\parallel \calA\calC_{\lambda_B\lambda_D\lambda_E}^{-M_A} \parallel =
\parallel \calA\calC_{\lambda_B\lambda_D\lambda_E}^{M_A} \parallel,
\label{Equation::norm2}
\end{equation}

\begin{equation}
\int_0^{2\pi}d\phi_B\int_0^{2\pi}d\phi_D
\parallel \calA\calC_{\lambda_B\lambda_D\lambda_E}^{M_A} \parallel ^2
=
4\pi^2\sum_{\lambda_C}
\parallel \calA^{M_A}_{\lambda_B\lambda_C} \parallel ^2
\parallel \calC^{\lambda_C}_{\lambda_D\lambda_E} \parallel ^2.
\label{Equation::phiprojection2}
\end{equation}
This kind of logic can be extended to a general two-body decay
chain.

\subsubsection{Example: $\Upsilon(1S) \to \gamma f_2(1270) (f_2(1270)
\to \pi^+ \pi^-)$}

From our previous discussion we can immediately write (see Equation ~\ref{Equation::chain1}),
\begin{eqnarray*}
\calAC_{\lambda_{\gamma}\lambda_{\pi^+}\lambda_{\pi^-}}^{M_{\Upsilon(1S)}} =
\sum_{\lambda_{f_2(1270)}}
\calA_{\lambda_{\gamma}\lambda_{f_2(1270)}}^{M_{\Upsilon(1S)}}
\calC_{\lambda_{\pi^+}\lambda_{\pi^-}}^{-\lambda_{f_2(1270)}}
\end{eqnarray*}

In this case, $\lambda_{\pi^-} = \lambda_{\pi^+} = 0$,
$\lambda_{\gamma} = \pm 1$. In CLEO
$M_{\Upsilon(1S)} = \pm 1$, there is no $M_{\Upsilon(1S)} = 0$ since at
the $\Upsilon(1S)$ energy electrons couple only to positrons of the
opposite helicity, this way $P_1 = P_{-1} = \frac{1}{2}$ and $P_0 =
0$ is an excellent approximation.. There are therefore four amplitudes
we must consider, $\calAC_{100}^{1}$, $\calAC_{100}^{-1}$, $\calAC_{-100}^{1}$ and
$\calAC_{-100}^{-1}$.

The overall angular distribution is (see Equation ~\ref{Equation::chain2}),
\begin{equation}
\begin{split}
\frac{dP}{d\Omega_{\gamma}d\Omega_{\pi^{+}}} &= \frac{1}{2}\parallel \calAC_{100}^{1} \parallel^2 +
\frac{1}{2}\parallel \calAC_{-100}^{1} \parallel^2 +
\frac{1}{2}\parallel \calAC_{100}^{-1} \parallel^2 +
\frac{1}{2}\parallel \calAC_{100}^{-1} \parallel^2 \\
&= \parallel \calAC_{100}^{1} \parallel^2 + \parallel
\calAC_{-100}^{1} \parallel^2
\end{split}
\label{Equation::haha}
\end{equation}
where we have used property ~\ref{Equation::norm2}. The two relevant
amplitudes are,
\begin{eqnarray*}
\calAC_{100}^{1} &=&
\calA_{10}^{1}\calC_{00}^{0}+\calA_{11}^{1}\calC_{00}^{-1}+\calA_{12}^{1}\calC_{00}^{-2}\\
\calAC_{-100}^{1} &=&
\calA_{-10}^{1}\calC_{00}^{0}+\calA_{-1-1}^{1}\calC_{00}^{1}+\calA_{-1-2}^{1}\calC_{00}^{2}
\end{eqnarray*}
where we have used the fact that
$|\lambda_{\gamma}-\lambda_{f_2(1270)}| \leq J_{\Upsilon(1S)} = 1$ to
sum over all possible $\lambda_{f_2(1270)}$.

All that remains is to substitute,
\begin{equation}
\begin{split}
\parallel \calAC_{100}^{1} \parallel^2 = &
\parallel  \calA_{10}^{1}\calC_{00}^{0} \parallel^2 +
\parallel \calA_{11}^{1}\calC_{00}^{-1} \parallel^2 +
\parallel \calA_{12}^{1}\calC_{00}^{-2} \parallel^2 + \\
&  \{\calA_{10}^{1}\calC_{00}^{0}\calA_{11}^{*1}\calC_{00}^{*-1} +
\calA_{10}^{1}\calC_{00}^{0}\calA_{12}^{*1}\calC_{00}^{*-2} +
\calA_{11}^{1}\calC_{00}^{-1}\calA_{12}^{*1}\calC_{00}^{*-2} +
c.c. \}\\
\label{Equation::mess}
\parallel \calAC_{-100}^{1} \parallel^2 = &
\parallel  \calA_{-10}^{1}\calC_{00}^{0} \parallel^2 +
\parallel \calA_{-1-1}^{1}\calC_{00}^{1} \parallel^2 +
\parallel \calA_{-1-2}^{1}\calC_{00}^{2} \parallel^2 + \\
& \{ \calA_{-10}^{1}\calC_{00}^{0}\calA_{-1-1}^{*1}\calC_{00}^{*1} +
\calA_{-10}^{1}\calC_{00}^{0}\calA_{-1-2}^{*1}\calC_{00}^{*2}+
 \calA_{-1-1}^{1}\calC_{00}^{1} \calA_{-1-2}^{*1}\calC_{00}^{*2} + c.c. \}
\end{split}
\end{equation}
into Equation ~\ref{Equation::haha}.

Doing so will give us the full angular distribution of the decay
chain. However, using property ~\ref{Equation::phiprojection}, all the
cross terms inside $\{...\}$ vanish if we integrate over
$\phi_{\pi^{+}}$ or $\phi_{\gamma}$. For the time being we integrate
over both $\phi_{\pi^{+}}$ and $\phi_{\gamma}$. We will return to the
terms inside $\{...\}$ later. 
\begin{eqnarray*}
&&\int_0^{2\pi}d\phi_{\gamma}\int_0^{2\pi}d\phi_{\pi^{+}}
\frac{dP}{d\Omega_{\gamma}d\Omega_{\pi^{+}}} = \\ \\
&&4\pi^2[(\parallel \calA_{10}^{1}\parallel^2 + \parallel \calA_{-10}^{1}\parallel^2)
\parallel \calC_{00}^{0}\parallel^2 +\\
&&\ \ \ \ \ \ \ (\parallel \calA_{11}^{1}\parallel^2 + \parallel \calA_{-1-1}^{1}\parallel^2)
\parallel \calC_{00}^{1}\parallel^2 +\\
&&\ \ \ \ \ \ \ (\parallel \calA_{12}^{1}\parallel^2 + \parallel \calA_{-1-2}^{1}\parallel^2)
\parallel \calC_{00}^{2}\parallel^2]\\
\\
&=&\frac{15}{4}[\parallel A_{10} \parallel^2 \parallel C_{00} \parallel^2
(d^1_{11}(\theta_{\gamma})^2+d^1_{1-1}(\theta_{\gamma})^2)d^2_{00}(\theta_{\pi^{+}})^2+\\
&&\ \ \ \ \parallel A_{11} \parallel^2 \parallel C_{00} \parallel^2
(d^1_{10}(\theta_{\gamma})^2+d^1_{10}(\theta_{\gamma})^{2})d^2_{10}(\theta_{\pi^{+}})^2+ \\
&&\ \ \ \ \parallel A_{12} \parallel^2 \parallel C_{00} \parallel^2
(d^1_{1-1}(\theta_{\gamma})^2+d^1_{1-1}(\theta_{\gamma})^{2})d^2_{20}(\theta_{\pi^{+}})^2] \\
\\
&=&\frac{15}{4}\parallel C_{00} \parallel^2 [
\parallel A_{10} \parallel^2
\frac{1}{2}(1+\cos^2 \theta_{\gamma})(\frac{3}{2} \cos^2
\theta_{\pi^+}-\frac{1}{2})^2+\\
&&\ \ \ \ \ \ \ \ \ \ \ \ \ \ \ \ \parallel A_{11} \parallel^2
\sin^2 \theta_{\gamma}(\sqrt{\frac{3}{2}} \sin \theta_{\pi^+}
\cos \theta_{\pi^+})^2+\\
&&\ \ \ \ \ \ \ \ \ \ \ \ \ \ \ \ \parallel A_{12} \parallel^2
\frac{1}{2}(1+\cos^2 \theta_{\gamma})(\frac{3}{8} \sin^4
\theta_{\pi^+})].
\end{eqnarray*}

At this point it is convenient to define the normalized helicity amplitudes,
\begin{eqnarray*}
a_{10} &=& \frac{A_{10}}{\sqrt{\parallel A_{10} \parallel^2 + \parallel A_{11} \parallel^2 + \parallel A_{12} \parallel^2}} \\
a_{11} &=& \frac{A_{11}}{\sqrt{\parallel A_{10} \parallel^2 + \parallel A_{11} \parallel^2 + \parallel A_{12} \parallel^2}} \\
a_{12} &=& \frac{A_{12}}{\sqrt{\parallel A_{10} \parallel^2 + \parallel A_{11} \parallel^2 + \parallel A_{12} \parallel^2}} \\
\end{eqnarray*}
which satisfy,
\begin{equation}
\parallel a_{10} \parallel^2 + \parallel a_{11} \parallel^2 + \parallel
a_{12} \parallel^2 = 1. ~\label{Equation:norm1}
\end{equation}

The normalized probability distribution,
$\frac{dU}{d(\cos\theta_{\gamma})d(\cos\theta_{\pi^{+}})}$, can be 
obtained by dividing the probability distribution by the branching
fraction of the sequential decay,

\begin{eqnarray*}
P &=&  2\parallel C_{00} \parallel^2 (\parallel A_{10}\parallel ^2 +
\parallel A_{11}\parallel ^2 + \parallel A_{12}\parallel ^2) \\
&=& \parallel C_{00} \parallel^2 \parallel A_{10} \parallel ^2 +
\parallel C_{00} \parallel^2 \parallel A_{11} \parallel ^2 +
\parallel C_{00} \parallel^2 \parallel A_{12} \parallel ^2 +\\
&&
\parallel C_{00} \parallel^2 \parallel A_{-10} \parallel ^2 +
\parallel C_{00} \parallel^2 \parallel A_{-1-1} \parallel ^2 +
\parallel C_{00} \parallel^2 \parallel A_{-1-2} \parallel ^2.
\end{eqnarray*}
This way,
\begin{eqnarray*}
\frac{dU}{d(\cos\theta_{\gamma})d(\cos\theta_{\pi^{+}})} &=&
\parallel a_{10} \parallel^2\frac{3}{8}(1+\cos^2 \theta_{\gamma}) 
        \frac{5}{8}(3\cos^2\theta_{\pi^+}- 1)^2+\\
&&  \parallel a_{11} \parallel^2\frac{3}{4}\sin^2 \theta_{\gamma}
           \frac{15}{4}(\sin \theta_{\pi^+}\cos \theta_{\pi^+})^2+\\
&&  \parallel a_{12} \parallel^2\frac{3}{8}(1+\cos^2 \theta_{\gamma})
           \frac{15}{16} \sin^4\theta_{\pi^+}.
\label{Equation::2D}
\end{eqnarray*}

It is useful to project the angular distribution on $\theta_{\gamma}$
and $\theta_{\pi^+}$,
\begin{equation}
\begin{split}
\frac{dU}{d\cos\theta_{\pi^+}} =&
\parallel a_{10} \parallel^2\frac{5}{8} (3\cos^2\theta_{\pi^+}-1)^2 + 
\parallel a_{11} \parallel^2\frac{15}{4}(\sin \theta_{\pi^+}\cos
\theta_{\pi^+})^2 + \\
&\parallel a_{12} \parallel^2\frac{15}{16} \sin^4\theta_{\pi^+}
\end{split}
\end{equation}
\begin{equation}
\frac{dU}{d\cos\theta_{\gamma}} =
(\parallel a_{10} \parallel^2+\parallel a_{12} \parallel^2)\frac{3}{8}(1+\cos^2 \theta_{\gamma}) + 
           \parallel a_{11} \parallel^2\frac{3}{4}\sin^2 \theta_{\gamma}.
\label{Equation:gamma}
\end{equation}

Let us now return to the terms inside $\{...\}$ in
Equation\ref{Equation::mess}. It turns out that for these terms a
substantial simplification occurs if we integrate over
$\cos\theta_{\pi^+}$. After some algebra, we arrive at
the following expression, 
\begin{equation}
\frac{dU}{d(\phi_{\gamma}-\phi_{\pi^+})} =
\frac{1}{2\pi}(1-\frac{1}{\sqrt{6}}\{{a_{10}a_{12}^*+a_{10}^*a_{12}}\}\cos[2(\phi_{\gamma}-\phi_{\pi^+})]).\label{Equation:fun2}
\end{equation}

\subsubsection{Example: $\Upsilon(1S) \to \gamma f_4(2050) (f_4(2050)
\to \pi^+ \pi^-)$}

For the case $f_4(2050)$ the calculation is very similar to the
$f_2(1270)$ case discussed above. The
allowed helicities for $f_4(2050)$ are the same as the allowed
helicities for $f_2(1270)$. This makes the angular distribution
calculation very similar to the $f_2(1270)$ case, the only difference
being that the functions $d^2_{\lambda_{f_22(1270)}0}$ are replaced by
$d^4_{\lambda_{f_4(2050)}0}$. 

The photon's angular distribution comes out to be the same. The result
for the pion's normalized angular distribution is,
\begin{eqnarray*}
\frac{dU}{d\cos\theta_{\pi^+}} &=&
\parallel a_{10} \parallel^2\frac{9}{128} (35\cos^4\theta_{\pi^+} - 30\cos^2\theta_{\pi^+}
+ 3)^2 + \\
&& \parallel a_{11} \parallel^2\frac{45}{32}(\sin \theta_{\pi^+}\cos
\theta_{\pi^+})^2(7\cos^2\theta_{\pi^+}-3)^2 + \\
&& \parallel a_{12} \parallel^2\frac{45}{64} \sin^4\theta_{\pi^+}(7\cos^2\theta_{\pi^+}-1)^2.
\end{eqnarray*}
\begin{equation}
\end{equation}

\subsubsection{Fit of the Helicity Angular Distribution Obtained from Data}

In practice, a particular bin in the helicity angular
distribution obtained from data has bins with 
$dN = NdU$ events, where $N$ is the total number of events (which is
allowed to float during the fit), and $dU$ is the portion of the
normalized probability assigned to the bin by the helicity formalism
prediction. 

To incorporate Equation~\ref{Equation:norm1} directly into the fit we
make the following change of variables, 
\begin{equation}
\parallel a_{10} \parallel = \cos\Theta\cos\Phi; \
\parallel a_{12} \parallel = \cos\Theta\sin\Phi; \
\parallel a_{11} \parallel = \sin\Theta. \label{Equation:change}
\end{equation}

Because of this definition, both $\Theta$ and $\Phi$ should be required to be
within the interval $(0,\frac{\pi}{2})$ during the fit. The new
variables have the convenient property that for J=0, $\Phi$ = $\Theta$ = 0, and for $J=1$, $\Phi$=0. In general there
are 1, 2 and 3 degrees 
of freedom for the fits with $J=0$, $J=1$, and $J>1$ respectively. One degree
of freedom always corresponds to the number of events, $N$, and the
other two correspond to $\Theta$ and $\Phi$, which determine
the magnitude of the normalized helicity amplitudes through
Equation~\ref{Equation:change}~\footnote{Obtaining the error of the
normalized helicity amplitudes requires some matrix
multiplication. For
compactness let us define 
$\vec{a} = \begin{pmatrix} a_{10} \\ a_{12} \\ a_{11}\end{pmatrix}$ and
$\vec{\omega} = \begin{pmatrix} \Theta \\ \Phi \end{pmatrix}$. The error matrix for the
normalized helicity amplitudes is,
\begin{equation*}
<\delta\vec{a}\delta\vec{a}^T> =
\frac{\partial\vec{a}}{\partial\vec{\omega}}<\delta\vec{\omega}
\delta\vec{\omega}^T>\frac{\partial\vec{a}}{\partial\vec{\omega}}^T
\end{equation*}
where $<\delta\vec{\omega}\delta\vec{\omega}^T>$ is the error
matrix obtained from the fit.}.

When fitting, it is preferable to choose the full 2D helicity angular
distribution which is a function of $\theta_{\gamma}$ and
$\theta_{\pi^+}$ instead of its one dimensional projections. The 2D
choice is better not only because the 2D distribution carries the
maximum amount of information, but because in general the efficiency depends
on both the pion and photon helicity angles, which can lead to
complications when efficiency correcting the one-dimensional projections.
Building on the two previous examples the 2D helicity angular
distributions for $J = 0,\ 1,\ 2,\ 3,\ 4$ are (for compactness we define
$x=\cos\theta_{\pi^+}$ and $y=\cos\theta_{\gamma}$),
\begin{equation}
\frac{dN_{J=0}}{dxdy} =
N_{J=0}\frac{3}{8}(1+y^2)\frac{1}{2}\label{Equation::J0}
\end{equation}
 \begin{equation}
\begin{split}
\frac{dN_{J=1}}{dxdy} =
N_{J=1}(&\cos^2\Theta\ \frac{3}{8}(1+y^2)\frac{3}{2}x^2\ + \\
&\sin^2\Theta\ \frac{3}{4}(1-y^2)\frac{3}{4}(1-x^2))\label{Equation::J1}
\end{split}
\end{equation}
\begin{equation}
\begin{split}
\frac{dN_{J=2}}{dxdy} = N_{J=2}(&
\cos^2\Theta\cos^2\Phi\ \frac{3}{8}(1+y^2)\frac{5}{8}(3x^2-1)^2+\\ 
&  \sin^2\Theta\ \ \ \ \ \ \ \ \ \ \frac{3}{4}(1-y^2)
           \frac{15}{4}x^2(1-x^2)+\\
&  \cos^2\Theta\sin^2\Phi\ \frac{3}{8}(1+y^2)
           \frac{15}{16} (1-x^2)^2)\label{Equation::J2}
\end{split}
\end{equation}
\begin{equation}
\begin{split}
\frac{dN_{J=3}}{dxdy} = N_{J=3}(&
\cos^2\Theta\cos^2\Phi\ \frac{3}{8}(1+y^2)\frac{7}{8}(5x^3-3x)^2+\\ 
&  \sin^2\Theta\ \ \ \ \ \ \ \ \ \ \frac{3}{4}(1-y^2)
           \frac{21}{32}(1-x^2)(5x^2-1)^2+\\
&  \cos^2\Theta\sin^2\Phi\ \frac{3}{8}(1+y^2)
           \frac{105}{16}x^2(1-x^2)^2)\label{Equation::J3}
\end{split}
\end{equation}
\begin{equation}
\begin{split}
\frac{dN_{J=4}}{dxdy} = N_{J=4}(&
\cos^2\Theta\cos^2\Phi\ \frac{3}{8}(1+y^2)\frac{9}{128}(35x^4-30x^2+3)^2+\\ 
&  \sin^2\Theta\ \ \ \ \ \ \ \ \ \ \frac{3}{4}(1-y^2)
           \frac{45}{32}x^2(1-x^2)(7x^2-3)^2+\\
&  \cos^2\Theta\sin^2\Phi\ \frac{3}{8}(1+y^2)
           \frac{45}{64}(1-x^2)^2(7x^2-1)^2).\label{Equation::J4}
\end{split}
\end{equation}
  
In this analysis, because of our low statistics (and by low statistics we mean
that there are numerous bins with $<1$ events in the 2D plots), we project
Equations~\ref{Equation::J0}-\ref{Equation::J4} on $x = \cos\theta_{\pi}$,
and $y = \cos\theta_{\gamma}$. We also fold each distribution around its
symmetry axis $x = 0$ and $y = 0$ to increase each bin's
statistics. Finally, we do a simultaneous fit to both distributions
and obtain $N,\ \Theta,$ and $\Phi$.

Admittedly, this procedure has the
problems described previously when 2D fits where advocated, which
can give rise to systematic effects in the measured helicity
amplitudes. In essence, we need to use a unitary weight function proportional
to the efficiency distribution (see Figure
~\ref{Figure:systematics_helicity.ps}) in the integrals used to make
the projections. This affects the relative strength of each helicity
amplitude and the overall efficiency, but does not change each
projected shape. However, the helicity amplitudes themselves are only used
to calculate systematic effects on the efficiency of resonances, so by
simultaneously fitting the unweighted 1D projections we are just ignoring the
systematics on the systematics.

\begin{figure}[ht]
\begin{center}
\epsfig{bbllx=50,bblly=50,bburx=580,bbury=755,width=4.5in,clip=,file=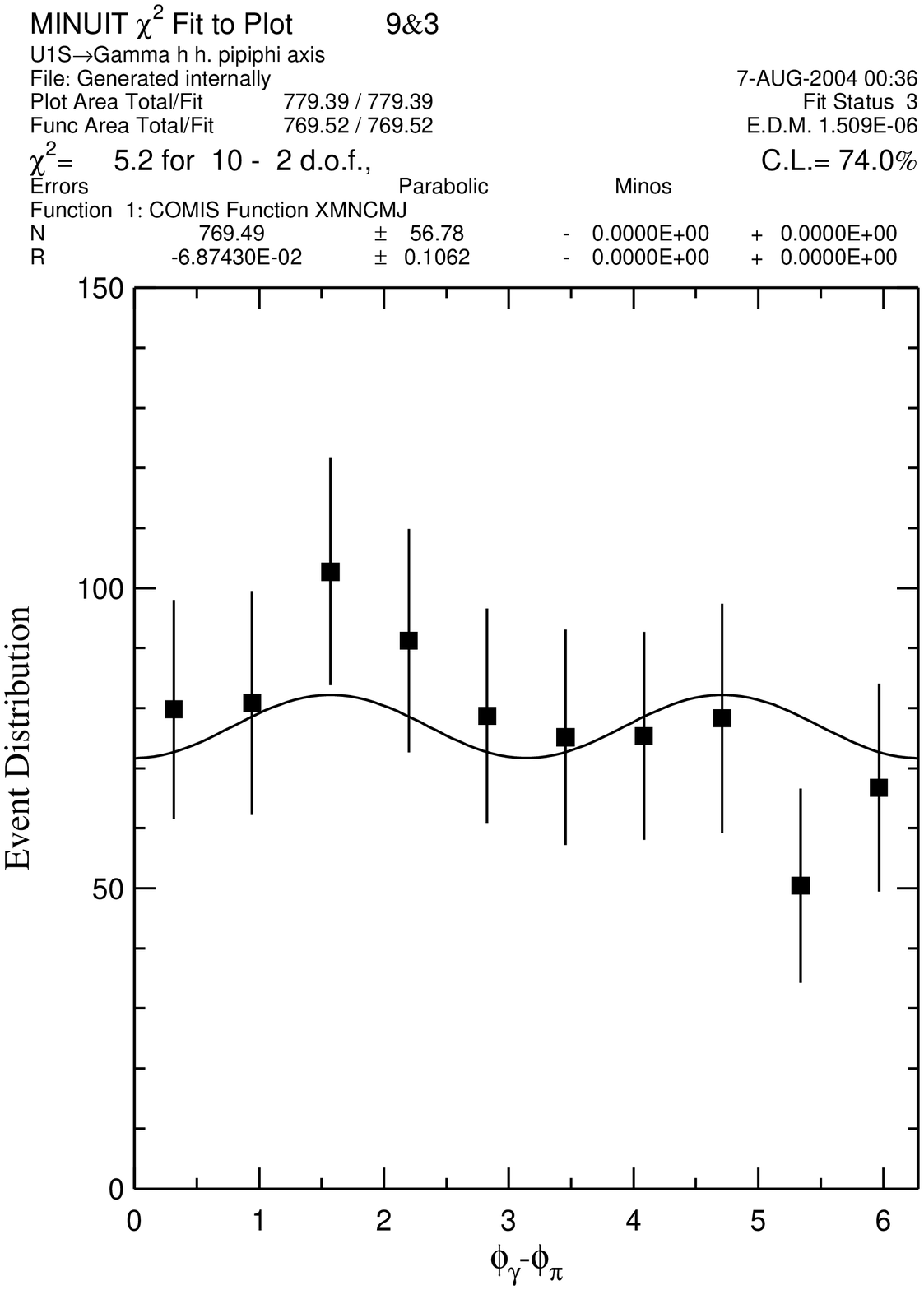}
\end{center}
\caption{Continuum subtracted $\phi_{\gamma}-\phi_{\pi^+}$ distribution for
events near the $f_2(1270)$ resonance with a fit to
Equation~\ref{Equation:fun} overlayed.}\label{Figure:wave.ps}
\end{figure}

Finally, let us consider the $\phi_{\gamma}-\phi_{\pi^+}$ distribution
for $J =2$. Equation~\ref{Equation:fun2} becomes, 
\begin{equation}
\frac{dN_{J=2}}{d(\phi_{\gamma}-\phi_{\pi^+})} =
\frac{N_{J=2}}{2\pi}(1-\frac{1}{\sqrt{6}}\{\cos^2\Theta\sin2\Phi\cos\Delta\}\cos[2(\phi_{\gamma}-\phi_{\pi^+})])\label{Equation:fun}
\end{equation}
where the new degree of freedom $\Delta$ is the relative phase between
$a_{10}$ and $a_{20}$. The deviation from the flat distribution is
characterized by 
\begin{equation}
R =\frac{1}{\sqrt{6}}\{\cos^2\Theta\sin2\Phi\cos\Delta\} \leq
\frac{1}{\sqrt{6}}.
\end{equation} 
The bigger 
this factor is, the easier it is to observe in
data. Figure~\ref{Figure:wave.ps} shows 
the continuum subtracted $\phi_{\gamma}-\phi_{\pi^+}$ distribution for
events near the $f_2(1270)$ resonance. The factor $R$ obtained from
fitting the distribution is non-significant.
It is a bit tedious but straightforward to calculate the
$\phi_{\gamma}-\phi_{\pi^+}$ 
distribution for other $J$ values. We
will not present those results here. They are not very interesting
because we need large statistics to observe deviations from the flat
distribution, and there is no effect on the efficiency from non-flat
$\phi_{\gamma}-\phi_{\pi^+}$ distributions.

\clearpage

\subsubsection{Example: Continuum Events.} 

Our previous examples tackled radiative decays of the
$\Upsilon(1S)$. Now let us consider the continuum ``radiative beam''
process $e^+e^- \to \gamma \pi^+ \pi^-$. The helicity formalism can
be applied at the CM of the $\pi^+ \pi^-$ vertex, where the
initial state has  $J^P = 1^-$. We have  $1^- \to  0^-
0^-$ at the $\pi^+ \pi^-$ vertex and there is only one helicity amplitude
we need to consider,
\begin{equation}
\calA_{00}^{1} = \sqrt{\frac{3}{4\pi}}D^{*1}_{10}A_{00}
\end{equation}
which after a couple simple steps leads to,
\begin{equation}
\frac{dN}{d\cos\theta_{\pi^+}} = N
\frac{3}{4}\sin^2\theta_{\pi^+}.\label{prediction}
\end{equation}
This result is also valid for the kaon case. 

\begin{figure}[ht]
\begin{center}
\epsfig{bbllx=66,bblly=100,bburx=580,bbury=775,width=4.5in,clip=,file=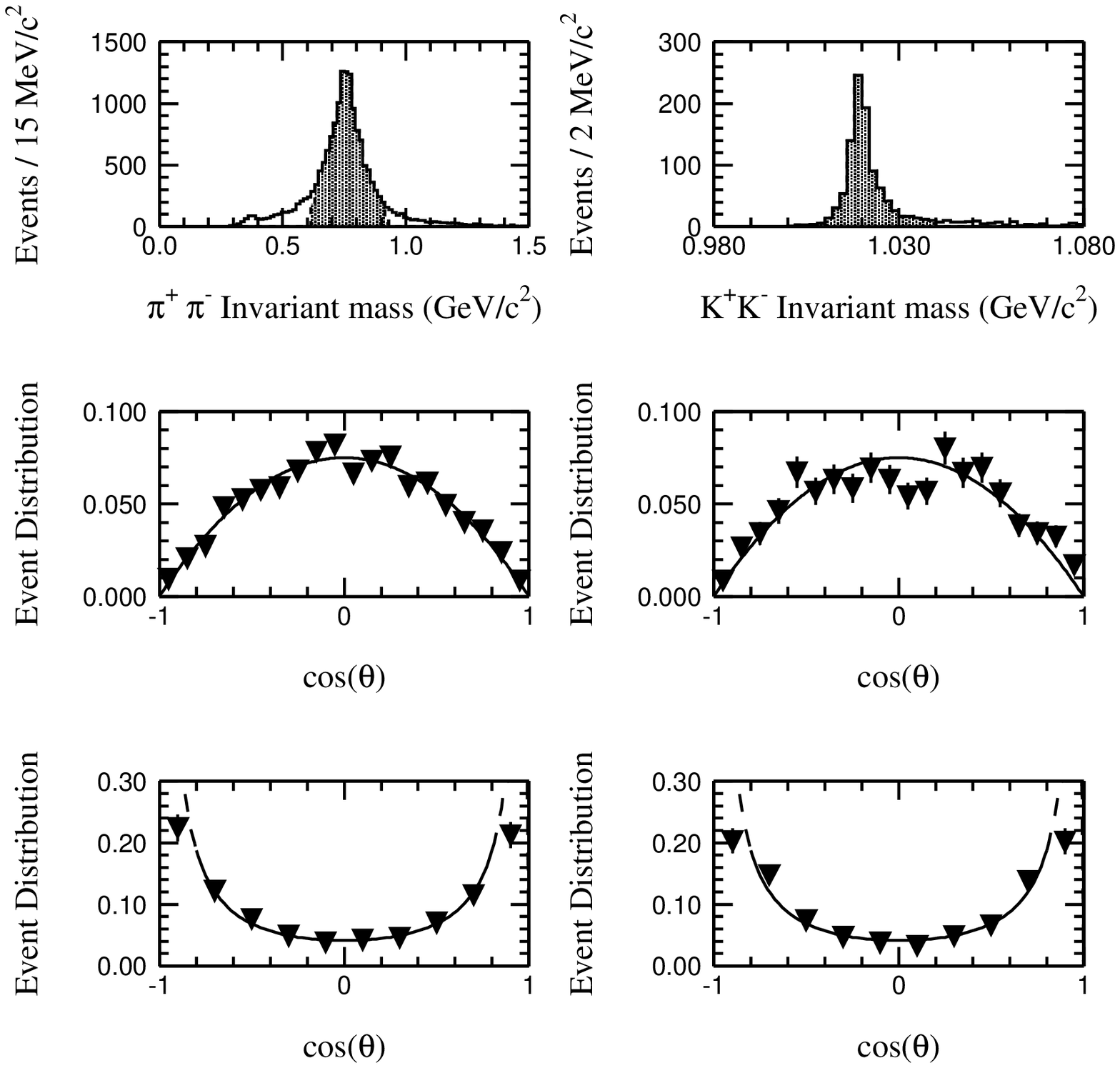}
\end{center}
\caption{Helicity angle distribution for ``radiative beam'' continuum
events $e^+ e^- \to \gamma \rho$ and $e^+ e^- \to \gamma \phi$ in the 4S
data. Top left (right) plot shows in gray the events selected as $\rho
\to \pi^+ \pi^-$ ($\phi \to K^+ K^-$). Middle left (right) plot shows
the $\pi^+ \pi^-$ ($K^+ K^-$) helicity angle distribution for the selected
events, and has the helicity formalism prediction, Equation
~\ref{prediction}, 
overlaid. Bottom left (right) plot shows the $\gamma$ angular
distribution for the photon accompanying the
$\rho$ ($\phi$), and has the first order QED prediction,
Equation~\ref{prediction2}, overlayed.}\label{Figure:helicityproof.ps}
\end{figure}

On the other hand, the photon's angular distribution cannot be
calculated using two-body helicity formalism. A good approximation is
(see~\cite{cleo5} Appendix B),
\begin{equation}
\frac{dP}{d\cos\theta_{\gamma}} \propto
\frac{1+\cos^2\theta_{\gamma}}{1-\cos^2\theta_{\gamma}},\label{prediction2}
\end{equation}
which is not valid at low $\theta$ angles ($\theta \to 0$, $ \theta
\to \pi$) because of the effects on the approximation of the neglected
electron mass.

Figure ~\ref{Figure:helicityproof.ps} shows these angular distributions
in 4S ``radiative beam''
events. The agreement between theory and experiment is excellent. This
is reassuring, because in the main analysis we rely on our angular
distribution measurements to get the correct efficiency in data and
to identify the $J$ value and helicity amplitude distribution of each
resonance.

\clearpage



\backmatter %
\biography{%
Luis Breva-Newell was born in Boulder, Colorado on December 11, 1975. When he
was two years old his family moved to the beautiful Basque Country region of
Spain. He graduated High School from GETXO III and attended the
Universidad del Pais Basco for four years. He moved to Madrid
where he obtained a 5-year degree in Physics from the Universidad
Complutense de Madrid. He was accepted to
attend graduate School at the University of Florida in 1999.}
\end{document}